\documentclass[letterpaper]{article} 
\usepackage{aaai2026}  
\usepackage{times}  
\usepackage{helvet}  
\usepackage{courier}  
\usepackage[hyphens]{xurl}  
\usepackage{graphicx} 
\usepackage{natbib}  
\usepackage{caption} 
\usepackage{algorithm}
\usepackage{algorithmic}

\usepackage{newfloat}
\usepackage{listings}
\DeclareCaptionStyle{ruled}{labelfont=normalfont,labelsep=colon,strut=off} 
\floatstyle{ruled}
\newfloat{listing}{tb}{lst}{}
\floatname{listing}{Listing}
\title{Designing Incident Reporting Systems for Harms from General-Purpose AI}
\author{
    Kevin Wei\textsuperscript{\rm 1, \rm 2},
    Lennart Heim\textsuperscript{\rm 1, \rm 2}
}
\affiliations{
    \textsuperscript{\rm 1} RAND
     \textsuperscript{\rm 2} Centre for the Governance of AI

    kwei@rand.org
}

\usepackage{tabularx}
\usepackage{multirow}
\usepackage{enumitem}
\usepackage{csquotes}
\usepackage{booktabs}
\usepackage{makecell}
\usepackage{array}
\usepackage{tablefootnote}  
\usepackage{epsfig}
\usepackage{caption}
\usepackage[hyperindex, pdfusetitle, linktocpage]{hyperref}
\usepackage[hyphenbreaks]{breakurl}

\newcolumntype{s}{>{\hsize=.25\hsize}X} %
\newcolumntype{Y}{>{\centering\arraybackslash}X} %

\defcitealias{AIA}{EU AI Act 2024}
\defcitealias{republic_of_korea_basic_2024}{AI Basic Act [of South Korea]}

\defcitealias{65_FR_19476}{65 FR 19476}
\defcitealias{79_FR_8832}{79 FR 8832}
\defcitealias{87_FR_61649}{87 FR 61649}
\defcitealias{87_FR_75634}{87 FR 75634}

\defcitealias{10_CFR_20.2201}{10 CFR § 20.2201}
\defcitealias{10_CFR_20.2203}{10 CFR § 20.2203}
\defcitealias{10_CFR_20.2205}{10 CFR § 20.2205}
\defcitealias{10_CFR_30.34}{10 CFR § 30.34}
\defcitealias{10_CFR_36.83}{10 CFR § 36.83}
\defcitealias{10_CFR_50.72}{10 CFR § 50.72}
\defcitealias{10_CFR_50.73}{10 CFR § 50.73}
\defcitealias{10_CFR_70.32}{10 CFR § 70.32}
\defcitealias{10_CFR_70.50}{10 CFR § 70.50}
\defcitealias{10_CFR_72.74}{10 CFR § 72.74}
\defcitealias{10_CFR_72.75}{10 CFR § 72.75}
\defcitealias{10_CFR_73.1200}{10 CFR § 73.1200}
\defcitealias{10_CFR_74.11}{10 CFR § 74.11}
\defcitealias{21_CFR_312.32}{21 CFR § 312.32}
\defcitealias{45_CFR_Part_46}{45 CFR Part 46}
\defcitealias{49_CFR_Part_830}{49 CFR Part 830}

\defcitealias{7_USC_136d}{7 U.S.C. § 136d}
\defcitealias{7_USC_136j}{7 U.S.C. § 136j}
\defcitealias{7_USC_136l}{7 U.S.C. § 136l}
\defcitealias{42_USC_2014}{42 U.S.C. § 2014}
\defcitealias{49_USC_1132}{49 U.S.C. § 1132}
\defcitealias{49_USC_1155}{49 U.S.C. § 1155}

\begin{document}

\maketitle

\pagenumbering{roman}
\setcounter{figure}{0}
\renewcommand{\thefigure}{\Alph{figure}}
\setcounter{table}{0}
\renewcommand{\thetable}{\Alph{table}}

\section*{Executive Summary}

This paper provides a comprehensive overview of the institutional design of AI incident reporting systems. We define ``AI incident reporting systems'' as processes for collecting information about safety- and rights-related events caused by general-purpose AI, which can include systems involving companies, independent organizations, and governments. We define incidents as events that did or could have caused harm to people, property, the environment, legal or human rights, infrastructure, or the public interest.

\begin{figure}[!htb]
    \centering
    \includegraphics[width=0.96\linewidth]{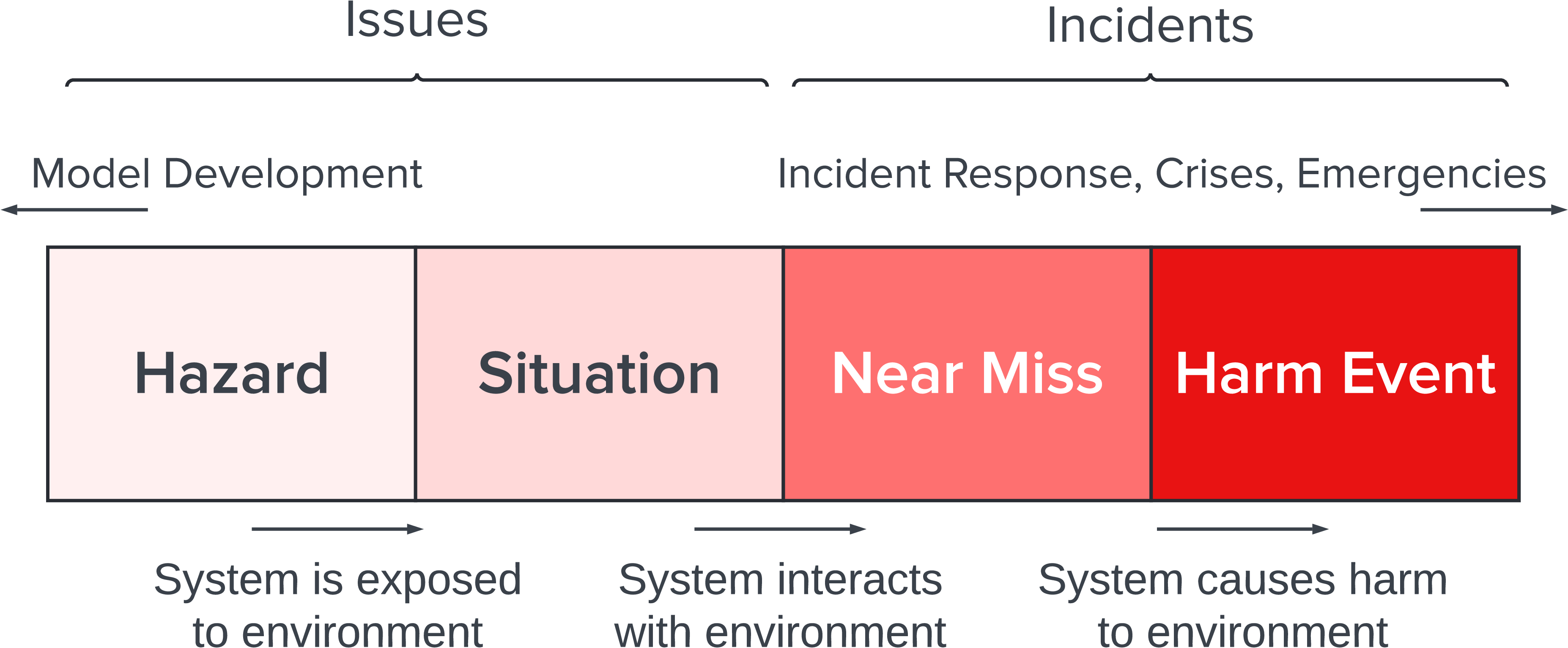}
    \caption{Visualization of the level of risk materialization dimension, i.e., the lifecycle of an (AI) incident. See Section \ref{sec:Framework} for definitions.}
    \label{fig:Overview_Risk_Materialization}
\end{figure}

\begin{figure*}[!hbt]
    \centering
    \includegraphics[width=0.90\textwidth]{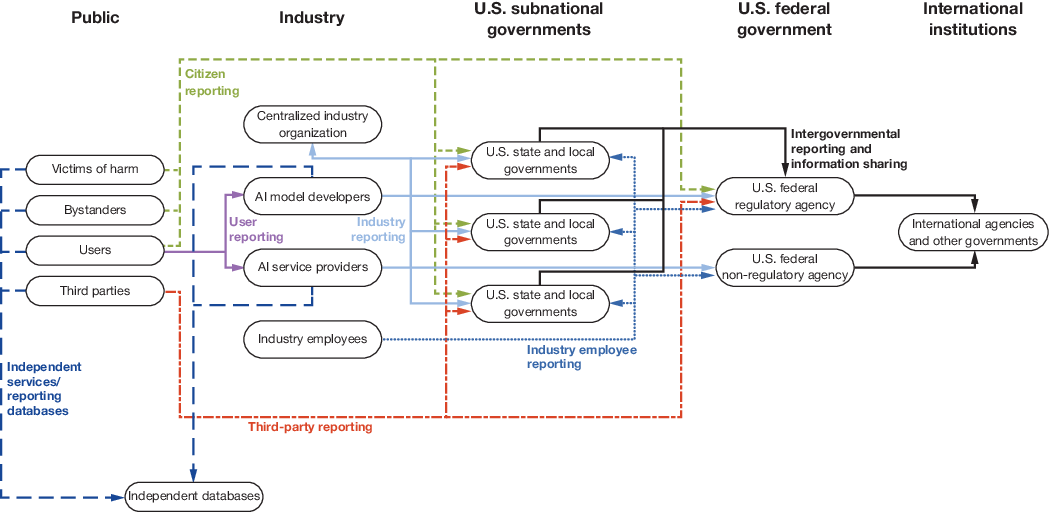}
    \caption[Visualization of the actors dimension]{Visualization of the actors dimension. Colored and dashed lines represent different system types as labeled; dashed lines are for accessibility only and do not represent additional distinctions beyond colors and labels. US state and local governments are repeated to emphasize that there are many such governments, all of which may operate independent reporting systems in the same reporting regime in parallel. See Section \ref{subsec:Framework_Actors} for more and Appendix \ref{subsec:Framework_Actors} for definitions.}
    \label{fig:Overview_Framework_Actors}
\end{figure*}

We first develop an analytical framework for the design of incident reporting systems via a literature review. Our framework contains seven dimensions and is summarized in Table \ref{tab:Overview_Framework_Defs}. We also highlight two dimensions in Figures \ref{fig:Overview_Risk_Materialization} and \ref{fig:Overview_Framework_Actors} that are sometimes confusing in prior work:

\begin{itemize}
    \item Level of risk materialization: Definitions of ``incident'' elsewhere vary according to how close an event is to causing harm. Figure \ref{fig:Overview_Risk_Materialization} visualizes how an AI hazard (a set of conditions or capabilities in a system that enables it to cause harm) eventually materializes into a harm event (an event where a system did cause harm).
    \item Actors submitting and receiving reports: Many different types of systems are often labeled as ``incident reporting systems.'' Figure \ref{fig:Overview_Framework_Actors} visualizes the different types of incident reporting systems involving various sets of actors.
\end{itemize}

\begin{table}[!htb]
    \centering
    \small
    \begin{tabularx}{\linewidth}{ >{\hsize=0.30\hsize}X >{\hsize=0.70\hsize}X }
        \toprule
         \textbf{Dimension} & \textbf{Definition} \\
         \midrule
         
         Policy Goal & The policy aim that the incident reporting system attempts to achieve: safety \textit{learning} or \textit{accountability} for harm. \\ \midrule
         
         Actors Submitting \& Receiving Reports & Possible actors include users, victims of harm, third-party individuals or organizations, companies, industry employees, and governments at various levels. See Figure \ref{fig:Overview_Framework_Actors} \\ \midrule
         
         Type of Incidents \newline Reported & The type of incident reported in the system: \textit{safety}, \textit{rights}, or \textit{security} incidents. \\ \midrule
         
         Level of Risk \newline Materialization & The level of risk materialization reported in the system: \textit{hazards}, \textit{situations}, \textit{near misses}, or \textit{harm events}. See Figure \ref{fig:Overview_Risk_Materialization}. \\ \midrule
         
         Enforcement of \newline Reporting & The procedures that incentivize actors to submit incident reports: \textit{voluntary} or \textit{mandatory} (by law). \\ \midrule
         
         Anonymity of \newline Reporters & The actors who have access to the reporter's identity: \textit{open}, \textit{confidential}, or \textit{anonymous}. \\ \midrule
         
         Post-Reporting \newline Actions & The actions taken by the party receiving incident reports, after reports are received: \textit{information sharing}, \textit{information disclosure}, \textit{audit}, or \textit{regulatory action}. \\
         
        \bottomrule
    \end{tabularx}
    \caption{Seven dimensions of the institutional design of incident reporting systems. Options/categories for each dimension are italicized.}
    \label{tab:Overview_Framework_Defs}
\end{table}

We then examine case studies of incident reporting in nine safety-critical industries in the U.S.: nuclear energy, civilian aviation, agricultural pesticides, pharmaceuticals, cybersecurity, dams, and rail transport. We extract some lessons for incident reporting in general-purpose AI (see full discussion in Section \ref{sec:Recs}), including:

\begin{itemize}
    \item Different policy goals may point toward opposing design choices for incident reporting systems. Achieving different goals may require creating multiple reporting systems.
    \item Existing, third-party AI incident databases lack stakeholder buy-in and often lack information necessary for safety learning. Eight of the nine industries we examined (all except dams) have implemented reporting regimes that go beyond independent databases.
    \item Stakeholders could consider expanding the audiences from whom incident reports are solicited and for whom an incident reporting system is designed, because incident information is often dispersed across actors. 
    \item  Throughout our case study industries, regulators generally oversee reporting by companies, industry employees, citizens, third parties, or product users. Systems where incidents are reported to non-regulatory agencies are generally non-punitive and oriented toward learning.
    \item AI model developers and service providers may benefit from establishing internal systems for accepting and investigating reports of complaints from product users.
    \item Operators of AI incident reporting systems may benefit from recognizing one possible categorization of incidents in terms of safety, security, and rights. See discussion in Section \ref{subsec:Recs_Incident_Type} and definitions in Appendix \ref{subsec:Framework_Incident_Type}.
    \item Governments and industry organizations can consider enabling AI near miss reporting by industry employees, users, third parties, and citizens. Near misses are events that could have but ultimately did not cause harm; they are a valuable source of data for safety learning.
    \item AI issues (see Figure \ref{fig:Overview_Risk_Materialization}) may be useful sources of information for safety learning, in addition to AI incidents.
    \item Mandatory reporting generally applies to (high-severity) harm events (e.g., cases of death or serious injury). Government-mandated reporting thresholds have helped improve regulatory visibility in other industries such as pharmaceuticals and civilian aviation.
    \item Incident reporting systems usually need clear, well-scoped reporting thresholds and definitions to be practically useful.
    \item Mandatory reporting requirements alone may be insufficient to achieve safety learning; voluntary near-miss reporting can be significant for facilitating learning. Some commentators have called for voluntary near-miss reporting systems for AI like those in U.S. aviation, but it is unclear whether aviation is an appropriate model for AI because the same model has failed in the rail industry.
    \item Incident reporting is the first, but not the only step, toward safety learning and accountability. After information is reported, subsequent analysis, mitigations, and monitoring are needed to enable safety learning.
    \item Anonymity for reporters can be warranted in some but not all systems. Anonymity can be warranted to promote safety learning, or when the policy goal is accountability and when reporting parties include those who fall outside the chain of incident responsibility. 
    \item Information must be aggregated and/or routed to relevant stakeholders to facilitate safety or accountability. Successful information sharing requires both identifying the correct actors and ensuring that information can be easily shared (e.g., via standardization and interoperability).
    \item Policymakers seeking to facilitate incident reporting for general-purpose AI can consider legal/regulatory interventions including: clarifying legal frameworks for reporting up front, setting clear reporting incentives, communicating guidelines to potential reporters, and closing legal loopholes that permit avoiding reports.
\end{itemize}

\clearpage

\onecolumn
\tableofcontents
\clearpage

\listoffigures
\listoftables

\clearpage

\pagenumbering{arabic}

\twocolumn

\begin{abstract}
    We introduce a conceptual framework and provide considerations for the institutional design of AI incident reporting systems, i.e., processes for collecting information about safety- and rights-related events caused by general-purpose AI. As general-purpose AI systems are increasingly adopted, they are causing more real-world harms and displaying the potential to cause significantly more dangerous incidents---events that did or could have caused harm to individuals, property, or the environment. Through a literature review, we develop a framework for understanding the institutional design of AI incident reporting systems, which includes seven dimensions: policy goal, actors submitting and receiving reports, type of incidents reported, level of risk materialization, enforcement of reporting, anonymity of reporters, and post-reporting actions. We then examine nine case studies of incident reporting in safety-critical industries to extract design considerations for AI incident reporting in the United States. We discuss, among other factors, differences in systems operated by regulatory vs. non-regulatory government agencies, near miss reporting, the roles of mandatory reporting thresholds and voluntary reporting channels, how to enable safety learning after reporting, sharing incident information, and clarifying legal frameworks for reporting. Our aim is to inform researchers and policymakers about when particular design choices might be more or less appropriate for AI incident reporting.
\end{abstract}

\setcounter{figure}{0}
\renewcommand{\thefigure}{\arabic{figure}}
\setcounter{table}{0}
\renewcommand{\thetable}{\arabic{table}}

\section{Introduction} \label{sec:Intro}

General-purpose artificial intelligence (GPAI) systems, including large language models (LLMs), have recently contributed to a number of high-profile cases of harm. For instance, GPAI systems have helped perpetrate a \$25.6 million financial scam \cite{chen_finance_2024}, assisted in planning explosives attacks \cite{palmer_fbi_2025, winter_driver_2025}, created explicit deepfakes \cite{weatherbed_trolls_2024}, accidentally deleted all of a company’s code \cite{lee_replits_2025}, been demonstrated to be capable of blackmail and deception \cite{lynch_agentic_2025}, and spread election disinformation in the United States (US) and across the world \cite{seitz-wald_fake_2024, row_rest_2024}. Increased AI capabilities, more agentic AI systems \cite{chan_harms_2023}, and wider adoption of GPAI also suggest that GPAI will soon contribute to more, and more severe, safety incidents and rights incidents: events in which AI systems cause or nearly cause harm to people, property, the environment, legal or human rights, infrastructure, or the public interest. %

Normal accident theory implies that in complex systems such as GPAI, severe incidents are inevitable over time \cite{bianchi_viewpoint_2023, maas_regulating_2018}; the risk of system failures and accidents rises as GPAI systems are becoming increasingly complex \cite{cook_how_2000, maas_regulating_2018, amodei_concrete_2016, jatho_concrete_2023, zaharia_shift_2024}. Governance interventions occurring prior to model deployment such as audits also may not succeed in preventing all AI incidents \cite{obrien_deployment_2023}. For instance, emergent capabilities in LLMs may arise unexpectedly after deployment \cite{zoph_emergent_2022} and create unanticipated incident types \cite{casper_defending_2024, woodside_emergent_2024}. Moreover, because GPAI can be deployed across domains, GPAI-caused harms might involve a wide range of forms, hazards, settings, and affected parties \cite{jatho_concrete_2023}.

These dynamics increase the importance of post-deployment governance interventions---e.g., incident reporting, post-deployment evaluations, and other risk management practices---that can uncover new risks \cite{gailmard_known_2025} and enable greater visibility into GPAI deployments \cite{schuett_principles_2024, ec_code_2025}. In particular, incident reporting has been implemented in many safety-critical industries such as nuclear power, civilian aviation, healthcare, and pharmaceuticals. The experiences of these domains suggest that incident reporting could be an important mechanism for managing safety and rights risks from AI systems \cite{guha_ai_2023, leveson_engineering_2011} by enabling learning and promoting accountability \cite{WHO_2005}.

AI incident reporting initiatives have yet to be implemented on a wide scale (Section \ref{sec:Review}). Moreover, discussions around incident reporting in the AI governance literature are still developing, and little guidance exists concerning AI incident reporting systems' institutional design---i.e., the construction of organizations, rules, and norms, that shape the tasks and responsibilities between actors \cite{klijn_institutional_2006, koppenjan_institutional_2005}. 

We fill this gap in the literature by 1) introducing a framework for conceptualizing the institutional design of AI incident reporting systems, and 2) reviewing nine case studies of safety-critical industries to provide the first, systematic examination of design considerations for AI incident reporting in the US. Our scope is limited to institutional design, and we exclude, e.g., the design of AI systems themselves or the operational-level details of IR; our scope is also limited only to safety and rights incidents \textit{caused by} GPAI, not on security incidents in which GPAI systems could be compromised by external actors (though these can be related). We further exclude discussions of AI whistleblowing \citetext{see \citealt{bullock_protecting_2025, hilton_right_2024}}. We hope to inform US researchers and policymakers who are considering incident reporting as a mechanism for mitigating harms from GPAI. 

\section{The AI Incident Reporting Landscape} \label{sec:Review}

Incident reporting systems allow companies, users, victims of harm, and others with knowledge of incidents to convey such information to institutions with responsibility for or oversight over AI products. By providing visibility into on-the-ground impacts and allowing stakeholders to understand system-caused harms, incident reporting systems can enhance safety outcomes \cite{kornecki_aviation_2015, petschnig_critical_2017, gardener_safety_2003} and initiate safety learning to reduce reoccurrences of past harms \cite{turetsky_success_2020, swartz_incident_1993}. They provide empirical data that can be used to identify existing or novel dangers \cite{naiac_recommendation_2023, dunbar_hazard_2014, goekcimen_addressing_2023, dunn_incident_2003, alarcon_cdcniosh_2021, hardy_chapter_2012}, create early warning indicators for future risks \cite{oleary_early_1996, oien_remote_2013}, and improve safety training programs \cite{evans_safety_2023}. Incident reporting can also help hold actors accountable for harm by enabling investigations of deployed systems, evaluations of existing safety measures \cite{niosh_pesticide-related_2005, ntsb_investigative_nodate}, risk estimation for insurance policies \cite{kvist_underwriting_2025}, and real-world system feedback \cite{vifladt_culture_2015}.

For the purposes of this article, we adopt an effects-based working definition of ``incident:'' incidents are events that either resulted in real-world harm (harm events) or that could have but did not ultimately result in harm (near misses). As of July 2025, no consensus definition for AI ``incidents'' has emerged, and we do not suggest that our working definition is ideal or operationalizable as-is. Rather, our definition aims to capture most events that are of interest and is also consistent with the safety science literature.

Below, we survey the AI governance literature on incident reporting. We then turn to discussing governmental and non-governmental initiatives around incident reporting for general-purpose AI. Appendix \ref{sec:Appendix_Background} has additional background.

\subsubsection{Literature Review} 

The AI governance literature is generally supportive of incident reporting \cite{goodman_ai_2024, raji_outsider_2022, dixon_argument_2024, obrien_deployment_2023, dsit_emerging_2023, schuett_towards_2023}. Existing literature has analyzed incident databases or operational-level factors such as documentation \cite{mcgregor_preventing_2021, longpre_position_2025, cattell_coordinated_2024, oecd_towards_2025}; sources have also proposed a variety of incident reporting systems for GPAI with different or even conflicting design choices and policy goals (see Appendix \ref{sec:Appendix_Background}). To date, however, no comprehensive analysis exists in the literature concerning when particular institutional design choices may be more or less appropriate, which is the focus of this article.

\subsubsection{Governmental Initiatives for General-Purpose AI Incident Reporting}

As of November 2025, China, the European Union (EU), California, and South Korea\footnote{See \citetalias[Art. 32(1)(2)]{republic_of_korea_basic_2024}.} are the only jurisdictions that have adopted incident reporting mandates specific to general-purpose AI. In China, an August 2023 ``generative AI'' regulation requires that AI service providers create channels for users to report problems, as well as that companies report any ``illegal content'' to the government \cite{cac_translation_2023}.\footnote{A recent policy document from \citet{chinese_ministry_of_foreign_affairs_global_2025} titled the ``Global AI Governance Action Plan'' indicates support for the ``promot[ion of] the development of threat information sharing and emergency response mechanisms'' at the global level, though it is uncertain how this policy will be operationalized.} In the EU, the AI Act, once implemented, will require that AI service providers report ``serious incidents'' to national regulators \citetext{\citetalias[Art. 73]{AIA}; see Appendix \ref{sec:Appendix_Policy_Initiatives}}. And in California, SB-53 was enacted in September 2025 and requires frontier developers to report certain safety and security incidents to the state government, as well as creating a hotline for public reporting \cite{wiener_sb-53_2025}. 

Many other jurisdictions, including the United States, have seen proposals for incident reporting but have yet to enact and implement such proposals. A number of federal legislators and experts have endorsed incident reporting for AI \cite{blumenthal_bipartisan_2023, murdick_written_2023, blumenthal_oversight_2023, bengio_managing_2023, arnold_statement_2023},\footnote{President Trump's 2025 ``AI Action Plan'' also contains a recommendation to ``Promote Mature Federal Capacity for AI Incident Response'' \cite{trump_winning_2025}. However, this recommendation appears to be concerned mostly with security incidents, rather than safety or rights incidents.} the National AI Advisory Committee in the US has recommended ``adverse event reporting'' for AI \cite{naiac_recommendation_2023}, bills introduced in Congress have proposed establishing or exploring voluntary reporting systems within the National Institute of Standards and Technology (NIST) \cite{warner_s_2024, ross_hr_2024}, White House memoranda have required government contractors to report ``serious AI incidents'' to federal agencies \cite{young_memorandum_2024, young_memorandum_2024-1}, and state level bills in California and New York would require reporting of certain AI incidents \cite{bores_a6453a_2025, wiener_sb-53_2025}. Reporting mandates have also been proposed in Brazil \cite{pacheco_bill_2023} and Canada \cite{champagne_letter_2023}, while companies in Canada have voluntarily committed to retaining incident-related information \cite{ised_voluntary_2023}. Appendix \ref{sec:Appendix_Policy_Initiatives} contains a partial list of such initiatives.

Similarly, institutions at the international level have called for but have yet to enact or implement AI incident reporting. The UN’s AI Advisory Group, for instance, called for a global incident reporting system \cite{UN_AI_Report}. The OECD has also established an Expert Group on AI Incidents, which launched the ``AI Incident Monitor'' \cite{AIM} in 2023 to crawl the web for media reports of AI incidents and classify them.

\subsubsection{Non-Governmental Initiatives for General-Purpose AI Incident Reporting} 

Non-governmental private actors have also established systems for reporting AI incidents. The most prominent of these initiatives are the Artificial Intelligence Incident Database \cite{AIID, mcgregor_preventing_2021}; the AI, Algorithmic, and Automation Incidents and Controversies Repository \cite{AIAAIC}; and the AI Vulnerability Database \cite{AVID}. These systems all take the form of voluntary, crowdsourced incident databases that solicit public submissions of alleged incidents, filter those submissions, and make the curated databases publicly available.

Researchers have also built various catalogs and visualizations of AI harms \cite{buolamwini_community_nodate, badnessai_ai_2023, smith_mitigating_2020, walker_merging_2024}, AI security vulnerabilities \cite{robust_intelligence_ai_2023, balunovic_lve_2024, MITRE_ATLAS}, and AI hazards \cite{dao_awful_2022, eticas_foundation_oasi_nodate}. These initiatives are generally less well-known, less comprehensive, and/or not as well-maintained as the AIID, AIAAIC, and AVID.

\section{The Institutional Design of Incident Reporting Systems} \label{sec:Framework}

Because incident reporting systems are well-established in industries such as aviation and agriculture, it is possible to study lessons from other industries and assess whether those learnings can be applied to AI \cite{guha_ai_2023, west_lessons_2024, gailmard_known_2025}. Methodologically, we thus adopt a three-step case study approach that is adapted from \citet{raji_outsider_2022}, \citet{ayling_putting_2022}, and \citet{stein_public_2024} to identify design considerations for AI incident reporting. 

First, we select nine safety-critical industries with robust incident reporting regimes as case studies, identified via seed articles (see Appendix \ref{sec:Appendix_Methodology}). Then, through a background literature review of incident reporting in these industries and of the AI governance literature, we develop a framework for the institutional design of incident reporting systems that consists of seven dimensions (Table \ref{tab:Framework_Defs}). Finally, we purposively select incident reporting systems from our nine case study industries and categorize them according to our framework. By identifying best or common practices in these industries, we extract design considerations for AI incident reporting systems and discuss when particular institutional design choices may be appropriate (Section \ref{sec:Recs}).

\begin{figure}[!htb]
    \centering
    \includegraphics[width=0.96\linewidth]{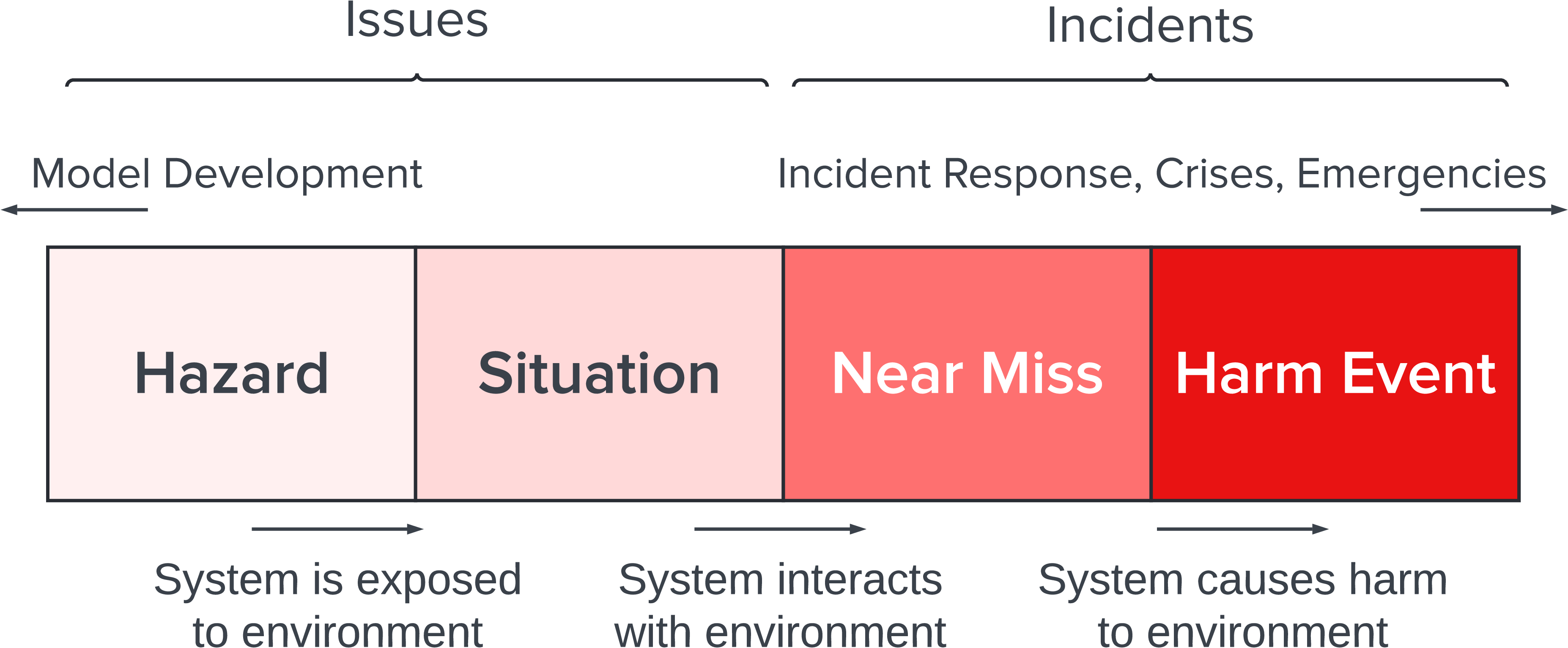}
    \caption{Visualization of the level of risk materialization dimension, i.e., the lifecycle of an (AI) incident}
    \label{fig:Risk_Materialization}
\end{figure}

\begin{figure*}[!hbt]
    \centering
    \includegraphics[width=0.96\textwidth]{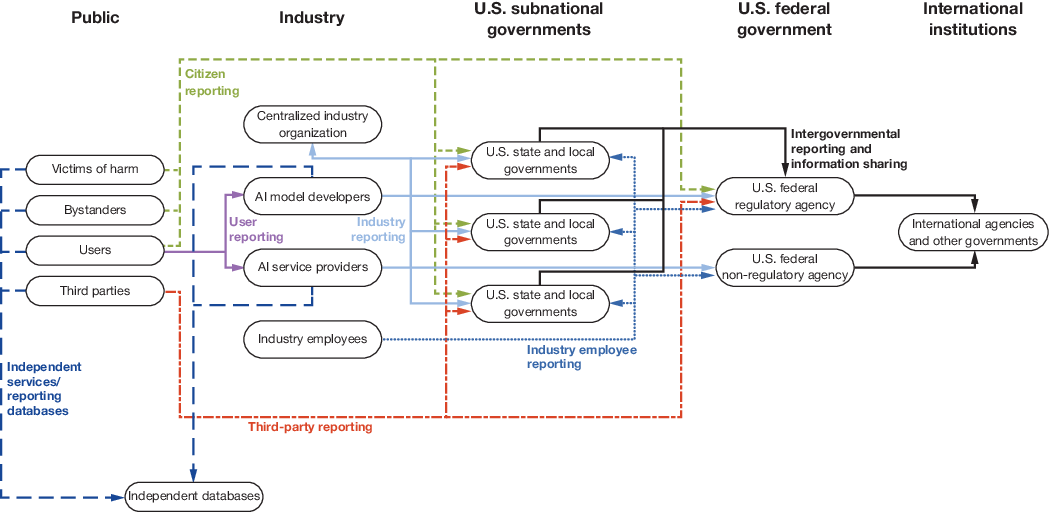}
    \caption[Visualization of the actors dimension, i.e., incident reporting systems involving different actors as defined in Table \ref{tab:Def_Actors}]{Visualization of the actors dimension, i.e., incident reporting systems involving different actors, as defined in Table \ref{tab:Def_Actors}. Note: each category of reporting system in Table \ref{tab:Def_Actors} is represented in the figure by a different color and dashed line (as labelled). Dashed lines are for accessibility only and do not represent additional distinctions beyond the colors and labels. US state and local governments are repeated to emphasize that there are many such governments, all of which may operate independent reporting systems in the same reporting regime in parallel (in contrast, with the notable exception of cybersecurity, systems in most incident reporting regimes are not normally run by multiple different agencies at the federal/international levels).}
    \label{fig:Framework_Actors}
\end{figure*}

Table \ref{tab:Framework_Defs} presents our framework for conceptualizing the institutional design of incident reporting systems. Because other literature is often unclear about the actors dimension and the level of risk materialization dimension, we highlight these two dimensions in this section. Possible sets of actors submitting/receiving reports are visualized in Figure \ref{fig:Framework_Actors}, while the the level of risk materialization dimension is defined in Table \ref{tab:Def_Scope_of_Risk} and visualized in Figure \ref{fig:Risk_Materialization}.

\begin{table}[!htb]
    \centering
    \small
    \begin{tabularx}{\linewidth}{ >{\hsize=0.30\hsize}X >{\hsize=0.70\hsize}X }
        \toprule
         \textbf{Dimension} & \textbf{Definition} \\
         \midrule
         
         Policy Goal & The policy aim that the incident reporting system attempts to achieve: safety \textit{learning} or \textit{accountability} for harm. \\ \midrule
         
         Actors Submitting \& Receiving Reports & Possible actors include users, victims of harm, third-party individuals or organizations, companies, industry employees, and governments at various levels. See Appendix \ref{subsec:Framework_Actors} for details. \\ \midrule
         
         Type of Incidents \newline Reported & The type of incident reported in the system: \textit{safety}, \textit{rights}, or \textit{security} incidents. \\ \midrule
         
         Level of Risk \newline Materialization & The level of risk materialization reported in the system: \textit{hazards}, \textit{situations}, \textit{near misses}, or \textit{harm events}. See Figure \ref{fig:Risk_Materialization}. \\ \midrule
         
         Enforcement of \newline Reporting & The procedures that incentivize actors to submit incident reports: \textit{voluntary} or \textit{mandatory} (by law). \\ \midrule
         
         Anonymity of \newline Reporters & The actors who have access to the reporter's identity: \textit{open}, \textit{confidential}, or \textit{anonymous}. \\ \midrule
         
         Post-Reporting \newline Actions & The actions taken by the party receiving incident reports, after reports are received: \textit{information sharing}, \textit{information disclosure}, \textit{audit}, or \textit{regulatory action}. \\
         
        \bottomrule
    \end{tabularx}
    \caption{Seven dimensions of the institutional design of incident reporting systems. Options/categories for each dimension are italicized.}
    \label{tab:Framework_Defs}
\end{table}

\begin{table}[!htb]
    \centering
    \small
    \begin{tabularx}{\linewidth}{ >{\hsize=0.30\hsize}X >{\hsize=0.70\hsize}X }
        \toprule
        \textbf{Category} & \textbf{Definition} 
        \\
        \midrule

        Hazard &
        A set of conditions or capabilities in a system that enables it to cause harm \cite{hendrycks_overview_2023, leveson_engineering_2011, birnbaum_research_2016, nasa_nasa_2004, ieee_ieee_1993, hutiri_not_2024}. An AI hazard could be, e.g., specific model capabilities and could be discovered through, e.g., red teaming, audits, or model evaluations.
        \\
        
        Situation (or hazardous situation) &
        A hazard that has been exposed to an environment in which it could cause harm. Situations are ``incidents waiting to happen.'' \citetext{See the discussion and definitions in \citet{leveson_engineering_2011, ieee_ieee_1993, hendrycks_x-risk_2022, hutiri_not_2024, iso_guide_2014}}. An AI situation could be, e.g., a publicly deployed AI model with dangerous capabilities (hazards).
        \\
        
        Near miss (or close call) &
        An event where a system's interactions with its environment could have caused harm but did not, generally due to some circumstances beyond the system's control. An AI near miss could, e.g., be an event in which an AI system generated a political deepfake for a user, but social media platforms removed such content before it was spread. See \citet{ec_code_2025}.
        \\
        
        Harm event &
        An event where a system did cause harm. An AI harm event could be, e.g., an event in which an AI system generated an offense cyberweapon that was then deployed by malicious actors in an attack.
        \\

        \bottomrule
    \end{tabularx}
    \caption{Definitions for the level of risk materialization dimension}
    \label{tab:Def_Scope_of_Risk}
\end{table}

We approach the level of risk materialization dimension as setting out the ``lifecycle'' of an AI incident. In particular, we make a distinction between AI issues (or AI flaws) and AI incidents. Issues are system \textit{conditions} (hazards) that once exposed to an external environment (situations) become prerequisites for incidents, while incidents are \textit{events} that could have caused harm (near misses) or did cause harm (harm event). Note that incidents can escalate into emergencies or crises \cite{gor_what_2025}, which we do not discuss in this paper. 

Full definitions for the options/categories in each dimension are in Appendix \ref{sec:Appendix_Framework_Definitions} (options meaning, e.g., ``learning'' and ``accountability'' for the Policy Goal dimension). 


\section{Design Considerations for AI Incident Reporting Systems} \label{sec:Recs}

Applying the framework in Section \ref{sec:Framework}, we review incident reporting systems from nine safety-critical industries in the US: nuclear power, aviation, pesticides, pharmaceuticals, cybersecurity, dams, rail, workplace safety, and healthcare. We discuss design considerations from these nine industries for AI incident reporting. Our discussion is organized by design dimension; Appendix \ref{sec:Appendix_Full_Results} contains full results of our review.

\begin{table*}[p]
    \centering
    \small
    \begin{tabularx}{\textwidth}{ m{0.25cm} s X s s s s s s s X s s s X } %
        \toprule

        \multicolumn{1}{c}{} &
        \multicolumn{1}{c}{\textbf{Goal}} &
        \multicolumn{1}{c}{\textbf{Actors}} &
        \multicolumn{3}{c}{\textbf{Incident Type}} &
        \multicolumn{4}{c}{\textbf{Risk Materialization}} &
        \multicolumn{1}{c}{\textbf{Enforcement}} &
        \multicolumn{3}{c}{\textbf{Anonymity}} &
        \multicolumn{1}{c}{\textbf{Post-Reporting}} 
        \\

        \multicolumn{1}{c}{} &
        \multicolumn{1}{c}{Learning?} &
        \multicolumn{1}{c}{} &
        \multicolumn{1}{c}{Saf} &
        \multicolumn{1}{c}{Rig} &
        \multicolumn{1}{c}{Sec} &
        \multicolumn{1}{c}{H} &
        \multicolumn{1}{c}{S} &
        \multicolumn{1}{c}{NM} &
        \multicolumn{1}{c}{HE} &
        \multicolumn{1}{c}{Mandatory?} &
        \multicolumn{1}{c}{O} &
        \multicolumn{1}{c}{C} &
        \multicolumn{1}{c}{A} &
        \multicolumn{1}{c}{} 
        \\

        \midrule

        \multirow{12}{*}{\rotatebox[origin=c]{90}{Nuclear}} &  
        \multicolumn{14}{l}{\textit{Nuclear Regulatory Commission (NRC), Safety Hotline}}
        \\ 

        & 
        &  
        \multicolumn{1}{c}{Citizen} &
        \multicolumn{1}{c}{\scalebox{1.25}{$\bullet$}} &
        \multicolumn{1}{c}{\scalebox{1.25}{$\bullet$}} &
        \multicolumn{1}{c}{\scalebox{1.25}{$\bullet$}} &
        \multicolumn{1}{c}{\scalebox{1.25}{$\bullet$}} &
        \multicolumn{1}{c}{\scalebox{1.25}{$\bullet$}} &
        \multicolumn{1}{c}{\scalebox{1.25}{$\bullet$}} &
        \multicolumn{1}{c}{\scalebox{1.25}{$\bullet$}} &
        &
        &
        \multicolumn{1}{c}{\scalebox{1.25}{$\bullet$}} &
        &
        \multicolumn{1}{c}{A; RA}
        \\

        &  
        \multicolumn{14}{l}{\textit{Nuclear Regulatory Commission (NRC), Statutory Reporting Requirements and Operations Center}}
        \\

        & 
        &
        \multicolumn{1}{c}{Company} &  
        \multicolumn{1}{c}{\scalebox{1.25}{$\bullet$}} &
        &
        \multicolumn{1}{c}{\scalebox{1.25}{$\bullet$}} &
        \multicolumn{1}{c}{\scalebox{1.25}{$\bullet$}} &  
        \multicolumn{1}{c}{\scalebox{1.25}{$\bullet$}} &
        \multicolumn{1}{c}{\scalebox{1.25}{$\bullet$}} &
        \multicolumn{1}{c}{\scalebox{1.25}{$\bullet$}} & 
        \multicolumn{1}{c}{\scalebox{1.25}{$\bullet$}} & 
        \multicolumn{1}{c}{\scalebox{1.25}{$\bullet$}} &
        &
        &
        \multicolumn{1}{c}{ID; RA}
        \\

        &  
        \multicolumn{14}{l}{\textit{International Atomic Energy Agency (IAEA), Incident Reporting System (IRS)}} 
        \\ 

        &
        \multicolumn{1}{c}{\scalebox{1.25}{$\bullet$}} &
        \multicolumn{1}{c}{Inter-Gov.} 
        &
        \multicolumn{1}{c}{\scalebox{1.25}{$\bullet$}} &
        &
        \multicolumn{1}{c}{\scalebox{1.25}{$\bullet$}} &
        \multicolumn{1}{c}{\scalebox{1.25}{$\bullet$}} &
        \multicolumn{1}{c}{\scalebox{1.25}{$\bullet$}} &
        \multicolumn{1}{c}{\scalebox{1.25}{$\bullet$}} &
        \multicolumn{1}{c}{\scalebox{1.25}{$\bullet$}} &
        &
        \multicolumn{1}{c}{\scalebox{1.25}{-}} &
        \multicolumn{1}{c}{\scalebox{1.25}{-}} &
        \multicolumn{1}{c}{\scalebox{1.25}{-}} &
        \multicolumn{1}{c}{IS}
        \\

        &  
        \multicolumn{14}{l}{\textit{International Atomic Energy Agency (IAEA), Fuel Incident Notification and Analysis System (FINAS)}}
        \\ 

        &
        \multicolumn{1}{c}{\scalebox{1.25}{$\bullet$}} &
        \multicolumn{1}{c}{Inter-Gov.} &
        \multicolumn{1}{c}{\scalebox{1.25}{$\bullet$}} &
        &
        \multicolumn{1}{c}{\scalebox{1.25}{$\bullet$}} &
        \multicolumn{1}{c}{\scalebox{1.25}{$\bullet$}} &
        \multicolumn{1}{c}{\scalebox{1.25}{$\bullet$}} &
        \multicolumn{1}{c}{\scalebox{1.25}{$\bullet$}} &
        \multicolumn{1}{c}{\scalebox{1.25}{$\bullet$}} &
        &
        \multicolumn{1}{c}{\scalebox{1.25}{-}} &
        \multicolumn{1}{c}{\scalebox{1.25}{-}} &
        \multicolumn{1}{c}{\scalebox{1.25}{-}} &
        \multicolumn{1}{c}{IS}
        \\

        &  
        \multicolumn{14}{l}{\textit{International Atomic Energy Agency (IAEA), Incident Reporting Systems for Research Reactors (IRSRR)}}
        \\ 

        &
        \multicolumn{1}{c}{\scalebox{1.25}{$\bullet$}} &
        \multicolumn{1}{c}{Inter-Gov.} &
        \multicolumn{1}{c}{\scalebox{1.25}{$\bullet$}} &
        &
        \multicolumn{1}{c}{\scalebox{1.25}{$\bullet$}} &
        \multicolumn{1}{c}{\scalebox{1.25}{$\bullet$}} &
        \multicolumn{1}{c}{\scalebox{1.25}{$\bullet$}} &
        \multicolumn{1}{c}{\scalebox{1.25}{$\bullet$}} &
        \multicolumn{1}{c}{\scalebox{1.25}{$\bullet$}} &
        &
        \multicolumn{1}{c}{\scalebox{1.25}{-}} &
        \multicolumn{1}{c}{\scalebox{1.25}{-}} &
        \multicolumn{1}{c}{\scalebox{1.25}{-}} &
        \multicolumn{1}{c}{IS}
        \\

        &  
        \multicolumn{14}{l}{\textit{James Martin Center for Nonproliferation Studies (CNS), Global Incidents and Trafficking Database}}
        \\ 

        &
        \multicolumn{1}{c}{\scalebox{1.25}{$\bullet$}} &
        \multicolumn{1}{c}{Indep. Db.} & 
        \multicolumn{1}{c}{\scalebox{1.25}{$\bullet$}} & 
        &
        & 
        & 
        \multicolumn{1}{c}{\scalebox{1.25}{$\bullet$}} & 
        \multicolumn{1}{c}{\scalebox{1.25}{$\bullet$}} & 
        \multicolumn{1}{c}{\scalebox{1.25}{$\bullet$}} & 
        & 
        &
        \multicolumn{1}{c}{\scalebox{1.25}{$\bullet$}} & &
        \multicolumn{1}{c}{ID} 
        \\

        \midrule

         \multirow{14}{*}{\rotatebox[origin=c]{90}{Civilian Aviation}} &  
        \multicolumn{14}{l}{\textit{National Transportation Safety Board (NTSB)}}
        \\

        &
        \multicolumn{1}{c}{\scalebox{1.25}{$\bullet$}} &
        \multicolumn{1}{c}{Company} & 
        \multicolumn{1}{c}{\scalebox{1.25}{$\bullet$}} & 
        &  
        &  
        &  
        & 
        & 
        \multicolumn{1}{c}{\scalebox{1.25}{$\bullet$}} & 
        \multicolumn{1}{c}{\scalebox{1.25}{$\bullet$}} & 
        \multicolumn{1}{c}{\scalebox{1.25}{$\bullet$}} &  
        &  
        &
        \multicolumn{1}{c}{ID; A}
        \\

        &  
        \multicolumn{14}{l}{\textit{Federal Aviation Administration (FAA), Hotline}}
        \\ 

        & 
        &
        \multicolumn{1}{c}{Citizen} &  
        \multicolumn{1}{c}{\scalebox{1.25}{$\bullet$}} &
        \multicolumn{1}{c}{\scalebox{1.25}{$\bullet$}} &
        &  
        \multicolumn{1}{c}{\scalebox{1.25}{$\bullet$}} &
        \multicolumn{1}{c}{\scalebox{1.25}{$\bullet$}} &
        \multicolumn{1}{c}{\scalebox{1.25}{$\bullet$}} &
        \multicolumn{1}{c}{\scalebox{1.25}{$\bullet$}} &
        &
        &
        \multicolumn{1}{c}{\scalebox{1.25}{$\bullet$}} &
        &
        \multicolumn{1}{c}{IS; A; RA}
        \\

        &  
        \multicolumn{14}{l}{\textit{Federal Aviation Administration (FAA), Voluntary Safety Reporting Program (VSRP)}} 
        \\ 

        & 
        \multicolumn{1}{c}{\scalebox{1.25}{$\bullet$}} &  
        \multicolumn{1}{c}{Employee} & 
        \multicolumn{1}{c}{\scalebox{1.25}{$\bullet$}} & 
        & 
        & 
        \multicolumn{1}{c}{\scalebox{1.25}{$\bullet$}}& 
        \multicolumn{1}{c}{\scalebox{1.25}{$\bullet$}} &
        &
        &
        &
        &
        \multicolumn{1}{c}{\scalebox{1.25}{$\bullet$}} &
        &
        \multicolumn{1}{c}{IS; A; RA}
        \\

        &  
        \multicolumn{14}{l}{\textit{Federal Aviation Administration (FAA), Aviation Safety Action Program (ASAP)}}
        \\ 

        & 
        \multicolumn{1}{c}{\scalebox{1.25}{$\bullet$}} &  
        \multicolumn{1}{c}{Employee} & 
        \multicolumn{1}{c}{\scalebox{1.25}{$\bullet$}} & 
        & 
        & 
        \multicolumn{1}{c}{\scalebox{1.25}{$\bullet$}}& 
        \multicolumn{1}{c}{\scalebox{1.25}{$\bullet$}} &
        &
        &
        &
        &
        \multicolumn{1}{c}{\scalebox{1.25}{$\bullet$}} &
        &
        \multicolumn{1}{c}{IS; A; RA}
        \\

        &  
        \multicolumn{14}{l}{\textit{Federal Aviation Administration (FAA), Voluntary Disclosure Reporting Program (VDRP)}}
        \\ 

        & 
        \multicolumn{1}{c}{\scalebox{1.25}{$\bullet$}} &  
        \multicolumn{1}{c}{Company} & 
        \multicolumn{1}{c}{\scalebox{1.25}{$\bullet$}} & 
        & 
        & 
        \multicolumn{1}{c}{\scalebox{1.25}{$\bullet$}} & 
        \multicolumn{1}{c}{\scalebox{1.25}{$\bullet$}} &
        \multicolumn{1}{c}{\scalebox{1.25}{$\bullet$}} &
        &
        &
        &
        \multicolumn{1}{c}{\scalebox{1.25}{$\bullet$}} &
        &
        \multicolumn{1}{c}{A}
        \\

        &  
        \multicolumn{14}{l}{\textit{National Aeronautics and Space Administration (NASA), Aviation Safety Reporting System (ASRS)}}
        \\ 

        & 
        \multicolumn{1}{c}{\scalebox{1.25}{$\bullet$}} &  
        \multicolumn{1}{c}{Employee} & 
        \multicolumn{1}{c}{\scalebox{1.25}{$\bullet$}} & 
        & 
        & 
        \multicolumn{1}{c}{\scalebox{1.25}{$\bullet$}}& 
        \multicolumn{1}{c}{\scalebox{1.25}{$\bullet$}} &
        \multicolumn{1}{c}{\scalebox{1.25}{$\bullet$}} &
        \multicolumn{1}{c}{\scalebox{1.25}{$\bullet$}} &
        &
        &
        &
        \multicolumn{1}{c}{\scalebox{1.25}{$\bullet$}} &
        \multicolumn{1}{c}{ID}
        \\

        &  
        \multicolumn{14}{l}{\textit{International Civil Aviation Organisation (ICAO), Annex 13 Information Sharing}}
        \\ 

        & 
        \multicolumn{1}{c}{\scalebox{1.25}{$\bullet$}} &  
        \multicolumn{1}{c}{Inter-Gov.} & 
        \multicolumn{1}{c}{\scalebox{1.25}{$\bullet$}} & 
        & 
        &
        &
        &
        \multicolumn{1}{c}{\scalebox{1.25}{$\bullet$}} &
        \multicolumn{1}{c}{\scalebox{1.25}{$\bullet$}} &
        \multicolumn{1}{c}{\scalebox{1.25}{$\bullet$}} &
        \multicolumn{1}{c}{\scalebox{1.25}{$\bullet$}} &
        &
        &
        \multicolumn{1}{c}{A}
        \\

        \midrule

        \multirow{8}{*}{\rotatebox[origin=c]{90}{Pesticides}} & 
        \multicolumn{14}{l}{\textit{California Environmental Protection Agency (CalEPA), Environmental Complaint System}} 
        \\ 

        &
        &
        \multicolumn{1}{c}{Citizen} &
        \multicolumn{1}{c}{\scalebox{1.25}{$\bullet$}} &
        &
        &
        &
        \multicolumn{1}{c}{\scalebox{1.25}{$\bullet$}} &
        \multicolumn{1}{c}{\scalebox{1.25}{$\bullet$}} &
        \multicolumn{1}{c}{\scalebox{1.25}{$\bullet$}} &
        &
        &
        &
        \multicolumn{1}{c}{\scalebox{1.25}{$\bullet$}} &
        \multicolumn{1}{c}{IS; A; RA}
        \\

        &  
        \multicolumn{14}{l}{\textit{California Environmental Protection Agency (CalEPA), California Pesticide Illness Surveillance Program}}
        \\

        & 
        &
        \multicolumn{1}{c}{Third Party} & 
        \multicolumn{1}{c}{\scalebox{1.25}{$\bullet$}} &
        &
        &
        &
        &
        &
        \multicolumn{1}{c}{\scalebox{1.25}{$\bullet$}} &
        \multicolumn{1}{c}{\scalebox{1.25}{$\bullet$}} &
        &
        \multicolumn{1}{c}{\scalebox{1.25}{$\bullet$}} &
        &
        \multicolumn{1}{c}{IS; A, RA}
        \\

        &  
        \multicolumn{14}{l}{\textit{Environmental Protection Agency (EPA), Enforcement and Compliance History Online (ECHO) Reporting System}}
        \\

        &
        &
        \multicolumn{1}{c}{Citizen} &
        \multicolumn{1}{c}{\scalebox{1.25}{$\bullet$}} &
        &
        &
        &
        \multicolumn{1}{c}{\scalebox{1.25}{$\bullet$}} &
        \multicolumn{1}{c}{\scalebox{1.25}{$\bullet$}} &
        \multicolumn{1}{c}{\scalebox{1.25}{$\bullet$}} &
        &
        &
        &
        \multicolumn{1}{c}{\scalebox{1.25}{$\bullet$}} &
        \multicolumn{1}{c}{IS; A; RA}
        \\

        &  
        \multicolumn{14}{l}{\textit{Environmental Protection Agency (EPA), Pesticide Manufacturer Reporting Requirements}}
        \\ 

        &
        &
        \multicolumn{1}{c}{Company} &
        \multicolumn{1}{c}{\scalebox{1.25}{$\bullet$}} &
        &
        &
        &
        &
        &
        \multicolumn{1}{c}{\scalebox{1.25}{$\bullet$}} &
        \multicolumn{1}{c}{\scalebox{1.25}{$\bullet$}} &
        &
        \multicolumn{1}{c}{\scalebox{1.25}{$\bullet$}} &
        &
        \multicolumn{1}{c}{ID}
        \\

        \bottomrule
    \end{tabularx}   
    \caption[Classification of incident reporting systems per our framework.]{Classification of incident reporting systems per our framework. \newline \newline Legend: incident types can include safety (``Saf''), rights (``Rig''), or security (``Sec''). Level of risk materialization can include hazards (``H''), situations (``S''), near misses (``NM''), or harm events (``HE''). Anonymity can be open (``O''), confidential (``C''), or anonymous (``A''). Post-reporting actions include information sharing (``IS''), information disclosure (``ID''), audit (``A''), or regulatory action (``RA''). Hyphens indicate where we found no publicly available information. References are available in Table \ref{tab:Results_References}.}
    \label{tab:Full_Results}
\end{table*}

\setcounter{table}{2}
\begin{table*}[!htbp]
    \centering
    \small
    \begin{tabularx}{\textwidth}{ m{0.25cm} s X s s s s s s s X s s s X } %
        \toprule

        \multicolumn{1}{c}{} &
        \multicolumn{1}{c}{\textbf{Goal}} &
        \multicolumn{1}{c}{\textbf{Actors}} &
        \multicolumn{3}{c}{\textbf{Incident Type}} &
        \multicolumn{4}{c}{\textbf{Risk Materialization}} &
        \multicolumn{1}{c}{\textbf{Enforcement}} &
        \multicolumn{3}{c}{\textbf{Anonymity}} &
        \multicolumn{1}{c}{\textbf{Post-Reporting}} 
        \\

        \multicolumn{1}{c}{} &
        \multicolumn{1}{c}{Learning?} &
        \multicolumn{1}{c}{} &
        \multicolumn{1}{c}{Saf} &
        \multicolumn{1}{c}{Rig} &
        \multicolumn{1}{c}{Sec} &
        \multicolumn{1}{c}{H} &
        \multicolumn{1}{c}{S} &
        \multicolumn{1}{c}{NM} &
        \multicolumn{1}{c}{HE} &
        \multicolumn{1}{c}{Mandatory?} &
        \multicolumn{1}{c}{O} &
        \multicolumn{1}{c}{C} &
        \multicolumn{1}{c}{A} &
        \multicolumn{1}{c}{} 
        \\

        \midrule

        \multirow{12}{*}{\rotatebox[origin=c]{90}{Pharmaceuticals}} & 
        \multicolumn{14}{l}{\textit{Food and Drug Administration (FDA), MedWatch (Patients)}} 
        \\

        & 
        &  
        \multicolumn{1}{c}{Citizen} &  
        \multicolumn{1}{c}{\scalebox{1.25}{$\bullet$}} & 
        & 
        & 
        & 
        \multicolumn{1}{c}{\scalebox{1.25}{$\bullet$}} & 
        \multicolumn{1}{c}{\scalebox{1.25}{$\bullet$}} & 
        \multicolumn{1}{c}{\scalebox{1.25}{$\bullet$}} & 
        & 
        \multicolumn{1}{c}{\scalebox{1.25}{$\bullet$}} & 
        & 
        & 
        \multicolumn{1}{c}{IS; A; RA} 
        \\

        &
        \multicolumn{14}{l}{\textit{Food and Drug Administration (FDA), MedWatch (Healthcare Providers)}} 
        \\

        & 
        &  
        \multicolumn{1}{c}{Third Party} &  
        \multicolumn{1}{c}{\scalebox{1.25}{$\bullet$}} & 
        & 
        & 
        & 
        \multicolumn{1}{c}{\scalebox{1.25}{$\bullet$}} & 
        \multicolumn{1}{c}{\scalebox{1.25}{$\bullet$}} & 
        \multicolumn{1}{c}{\scalebox{1.25}{$\bullet$}} & 
        & 
        \multicolumn{1}{c}{\scalebox{1.25}{$\bullet$}} & 
        & 
        & 
        \multicolumn{1}{c}{IS; A; RA} 
        \\

        &
        \multicolumn{14}{l}{\textit{Food and Drug Administration (FDA), Medical Device Reporting (Device Manufacturers)}} 
        \\
        
        & 
        &  
        \multicolumn{1}{c}{Company} &  
        \multicolumn{1}{c}{\scalebox{1.25}{$\bullet$}} & 
        &  
        & 
        & 
        & 
        \multicolumn{1}{c}{\scalebox{1.25}{$\bullet$}} & 
        \multicolumn{1}{c}{\scalebox{1.25}{$\bullet$}} & 
        \multicolumn{1}{c}{\scalebox{1.25}{$\bullet$}} & 
        \multicolumn{1}{c}{\scalebox{1.25}{$\bullet$}} & 
        & 
        &
        \multicolumn{1}{c}{ID; A; RA} 
        \\

        &
        \multicolumn{14}{l}{\textit{Food and Drug Administration (FDA), Medical Device Reporting (User Facilities)}} 
        \\
        
        & 
        &  
        \multicolumn{1}{c}{User} &  
        \multicolumn{1}{c}{\scalebox{1.25}{$\bullet$}} & 
        &  
        & 
        & 
        & 
        \multicolumn{1}{c}{\scalebox{1.25}{$\bullet$}} & 
        \multicolumn{1}{c}{\scalebox{1.25}{$\bullet$}} & 
        \multicolumn{1}{c}{\scalebox{1.25}{$\bullet$}} & 
        \multicolumn{1}{c}{\scalebox{1.25}{$\bullet$}} & 
        & 
        & 
        \multicolumn{1}{c}{ID; A; RA} 
        \\

        &  
        \multicolumn{14}{l}{\textit{Food and Drug Administration (FDA), Voluntary Malfunction Summary Reporting (VMSR)}} 
        \\
        
        &
        \multicolumn{1}{c}{\scalebox{1.25}{$\bullet$}} & 
        \multicolumn{1}{c}{Company} &  
        \multicolumn{1}{c}{\scalebox{1.25}{$\bullet$}} & 
        &  
        & 
        & 
        \multicolumn{1}{c}{\scalebox{1.25}{$\bullet$}} & 
        \multicolumn{1}{c}{\scalebox{1.25}{$\bullet$}} & 
        & 
        & 
        \multicolumn{1}{c}{\scalebox{1.25}{$\bullet$}} & 
        &
        &
        \multicolumn{1}{c}{ID} 
        \\

        &
        \multicolumn{14}{l}{\textit{Centers for Disease Control and Prevention (CDC), Vaccine Adverse Events Reporting System (VAERS) (Patients and Family)}} 
        \\
        
        & 
        &  
        \multicolumn{1}{c}{Citizen} &
        \multicolumn{1}{c}{\scalebox{1.25}{$\bullet$}} &
        &
        & 
        &
        &
        &
        \multicolumn{1}{c}{\scalebox{1.25}{$\bullet$}} &
        &
        &
        \multicolumn{1}{c}{\scalebox{1.25}{$\bullet$}} &
        & 
        \multicolumn{1}{c}{ID; A; RA} 
        \\

        &
        \multicolumn{14}{l}{\textit{Centers for Disease Control and Prevention (CDC), Vaccine Adverse Events Reporting System (VAERS) (Healthcare Providers)}} 
        \\
        
        & 
        &  
        \multicolumn{1}{c}{Third Party} &
        \multicolumn{1}{c}{\scalebox{1.25}{$\bullet$}} &
        &
        & 
        &
        &
        &
        \multicolumn{1}{c}{\scalebox{1.25}{$\bullet$}} &
        \multicolumn{1}{c}{\scalebox{1.25}{$\bullet$}} &
        \multicolumn{1}{c}{ - } &
        \multicolumn{1}{c}{ - } &
        \multicolumn{1}{c}{ - } &
        \multicolumn{1}{c}{ID; A; RA} 
        \\

        \midrule

        \multirow{2}{*}{\rotatebox[origin=c]{90}{Cyber}} & 
        \multicolumn{14}{l}{\textit{Information Security Analysis Centers / Organizations (ISACs / ISAOs)}} 
        \\ 

        &
        \multicolumn{1}{c}{\scalebox{1.25}{$\bullet$}} &
        \multicolumn{1}{c}{Indep. Db.} & 
        & 
        &
        \multicolumn{1}{c}{\scalebox{1.25}{$\bullet$}} & 
        & 
        \multicolumn{1}{c}{\scalebox{1.25}{$\bullet$}} & 
        \multicolumn{1}{c}{\scalebox{1.25}{$\bullet$}} & 
        \multicolumn{1}{c}{\scalebox{1.25}{$\bullet$}} & 
        & 
        &
        \multicolumn{1}{c}{\scalebox{1.25}{$\bullet$}} & 
        & 
        \multicolumn{1}{c}{IS} 
        \\

        \midrule

        \multirow{4}{*}{\rotatebox[origin=c]{90}{Dams}} & 
        \multicolumn{14}{l}{\textit{Association of State Dam Safety Officials (ASDSO), Dam Incident Database}} 
        \\ 

        &
        \multicolumn{1}{c}{\scalebox{1.25}{$\bullet$}} &
        \multicolumn{1}{c}{Indep. Db.} & 
        \multicolumn{1}{c}{\scalebox{1.25}{$\bullet$}} & 
        &
        & 
        & 
        \multicolumn{1}{c}{\scalebox{1.25}{$\bullet$}} & 
        \multicolumn{1}{c}{\scalebox{1.25}{$\bullet$}} & 
        \multicolumn{1}{c}{\scalebox{1.25}{$\bullet$}} & 
        & 
        & 
        \multicolumn{1}{c}{\scalebox{1.25}{$\bullet$}} & 
        & 
        \multicolumn{1}{c}{ID} 
        \\

        &
        \multicolumn{14}{l}{\textit{Stanford University National Performance on Dams Program (NPDP), Dam Directory and Incident Database}} 
        \\ 

        &
        \multicolumn{1}{c}{\scalebox{1.25}{$\bullet$}} &
        \multicolumn{1}{c}{Indep. Db.} & 
        \multicolumn{1}{c}{\scalebox{1.25}{$\bullet$}} & 
        &
        & 
        & 
        \multicolumn{1}{c}{\scalebox{1.25}{$\bullet$}} & 
        \multicolumn{1}{c}{\scalebox{1.25}{$\bullet$}} & 
        \multicolumn{1}{c}{\scalebox{1.25}{$\bullet$}} & 
        & 
        &
        \multicolumn{1}{c}{\scalebox{1.25}{$\bullet$}} & 
        & 
        \multicolumn{1}{c}{ID} 
        \\

        \midrule

        \multirow{2}{*}{\rotatebox[origin=c]{90}{Rail}} & 
        \multicolumn{14}{l}{\textit{National Aeronautics and Space Administration (NASA), Confidential Close Call Reporting System (C\textsuperscript{3}RS)}} 
        \\ 

        & 
        \multicolumn{1}{c}{\scalebox{1.25}{$\bullet$}} &  
        \multicolumn{1}{c}{Employee} & 
        \multicolumn{1}{c}{\scalebox{1.25}{$\bullet$}} & 
        & 
        & 
        \multicolumn{1}{c}{\scalebox{1.25}{$\bullet$}}& 
        \multicolumn{1}{c}{\scalebox{1.25}{$\bullet$}} &
        \multicolumn{1}{c}{\scalebox{1.25}{$\bullet$}} &
        \multicolumn{1}{c}{\scalebox{1.25}{$\bullet$}} &
        & 
        & 
        & 
        \multicolumn{1}{c}{\scalebox{1.25}{$\bullet$}} &
        \multicolumn{1}{c}{ID} 
        \\
        
        \bottomrule
    \end{tabularx}   
    \caption[]{Classification of incident reporting systems per our framework (cont'd). \newline \newline Legend: incident types can include safety (``Saf''), rights (``Rig''), or security (``Sec''). Level of risk materialization can include hazards (``H''), situations (``S''), near misses (``NM''), or harm events (``HE''). Anonymity can be open (``O''), confidential (``C''), or anonymous (``A''). Post-reporting actions include information sharing (``IS''), information disclosure (``ID''), audit (``A''), or regulatory action (``RA''). Hyphens indicate where we found no publicly available information. References are available in Table \ref{tab:Results_References}.}
\end{table*}

\subsection{Policy Goal of the Incident Reporting System} \label{subsec:Recs_Goal}

\subsubsection{Addressing Different Policy Goals of Incident Reporting} Policymakers and system operators may wish to achieve multiple policy goals with incident reporting systems, such as safety learning (helping stakeholders adapt from past incidents and improving processes to prevent future harms) or corporate accountability (holding actors responsible for past harms via corrective actions or penalties). Although systems can in theory be oriented toward both learning and accountability, dual-goal systems are rare in practice because these goals often point toward opposing design choices \cite{WHO_2005, mills_promise_2010}. For instance, a voluntary reporting system (for learning) that encouraged responsible actors to report directly to a regulatory agency would be ineffective if reporters were deterred by the prospects of fines (for accountability). Thus, pursuing multiple goals via incident reporting systems could require the creation of multiple single-goal systems---e.g., creating an FAA-style voluntary information sharing system, as in \citet{shrishak_how_2023} (for learning), as well as a separate citizen reporting system that would enable regulatory audits, as in \citet{raji_outsider_2022} (for accountability). 

\subsection{Actors Submitting \& Receiving Reports} \label{subsec:Recs_Actors}

\subsubsection{Shortcomings of Existing AI Incident Databases} Third-party incident databases such as the AIID, AIAAIC, or AVID are good first steps for creating visibility into the AI risk landscape \cite{mcgregor_indexing_2022, whittlestone_why_2021}. However, these databases lack stakeholder buy-in and often lack information necessary for safety learning \cite{richards_incidents_2025}. Moreover, the stated goal of these databases is generally to facilitate safety learning; it is uncertain that databases alone can advance accountability beyond raising public awareness of AI-caused harms \cite{rodrigues_when_2023, richards_incidents_2025}. Note also that incident investigation and analysis remain immature in AI, and it is unclear whether independent databases can facilitate these practices within AI developers or regulatory agencies.

Buy-in from both industry and government actors is needed for incident databases to contribute meaningfully to safety learning \cite{wolff_models_2014}. An example is the Dam Directory of the National Performance on Dams Program, a database of dam-related safety incidents that contains material drawn from multiple federal programs, dam engineers, professional organizations, and private collections; the program also maintains a real-time incident notification database with guidelines developed by state, federal, and industry dam safety engineers \cite{mccann_national_nodate}. Similar levels of buy-in have yet to be achieved by existing AI incident databases: for instance, most top contributors to the AIID are from civil society \cite{aiid_submissions_2024}.

Because of these shortcomings of independent databases, the AI governance literature has called for official or centralized incident reporting systems \cite{brundage_lessons_2022, shevlane_model_2023, schuett_risk_2023}. Eight of the nine safety-critical industries we examined have implemented reporting regimes that include other systems beyond independent incident databases (Table \ref{tab:Full_Results}).

\subsubsection{Facilitating Data Collection by Expanding Coverage} Reporting ``coverage'' refers to the audiences from whom incident reports are solicited and for whom an incident reporting system is designed. Increasing coverage means that more information is collected from different parties, thus generating a more complete picture of the incident landscape \cite{wolff_models_2014, mehran_post-market_2004}. Higher coverage is particularly critical when incident information is not concentrated but rather spread between different parties.

Different parties may have information about new, unreported incidents: one emergency room study found that although reports by healthcare providers most often helped identify new incidents, reports by other parties---patients, their families, and hospital risk management---nevertheless helped in incident identification \cite{reznek_patient_2015}. Many reporting systems also see skewed reporting distributions, where most reports are submitted by one class of actors but are supplemented by reports from other classes of actors \cite{norman_whether_2023, saurin_findings_2015, dolbeer_trends_2015}. In the context of general-purpose AI systems, information may be in the hands of many different actors \cite{ec_code_2025} and harms could be diverse \cite{slattery_ai_2025}, so expanding reporting coverage may be important to facilitate a broader understanding of AI issues and incidents.

Different parties may also corroborate or have different information about the same incidents that have already been reported by other actors \cite{wolff_models_2014}. A study of wildlife collisions with commercial aircraft estimated that 91\% of such incidents were being reported via the FAA’s incident reporting systems \cite{dolbeer_trends_2015}. Although most reports came from one central system, at least 10 other systems also contributed reports \cite{dolbeer_trends_2015, dolbeer_wildlife_2023}. In particular, reports from Air Traffic Organization employees comprised 7\% of all reports in the FAA’s database and provided supplementary information to another 7\% of reports that were duplicated from other sources \cite{dolbeer_trends_2015}. Given the complexity of the AI supply chain \cite{hopkins_ai_2025}, information about AI incidents may also be distributed.

An illustrative case study of an incident reporting regime that has significant coverage is that of agricultural pesticides, which can cause health or environmental harm both to immediate users and to downstream consumers \cite{damalas_pesticide_2011}. The complexity of pesticide distribution and the distributed nature of harm from pesticides has resulted in an enormous web of reporting and data sharing that involves, in various capacities: agriculture workers, poison control centers, doctors, medical labs, hospitals, most US state governments, the Department of Agriculture, the Environmental Protection Agency, the Bureau of Labor Statistics, the World Health Organization (WHO), the Intergovernmental Forum on Chemical Safety, and the UN Food and Agriculture Organization \cite{calvert_surveillance_2010}. Similarly, AI systems can cause harm in a variety of ways---importantly, in ways not anticipated by traditional safety science such as catastrophic misuse, psychological manipulation, mental health harms, or civil liberties violations \cite{bengio_international_2025, weidinger_taxonomy_2022}. Research into LLMs has found an impressive breadth of adoption \cite{mcelheran_ai_2024}, which could foreshadow risk models and distribution chains that are as complicated as, if not more so, than those in agriculture \citetext{see \citealt{hopkins_ai_2025}}. These factors may justify reporting systems for diverse segments of society, especially as lower-severity incidents become more common.

Coverage can be increased by establishing multiple incident reporting systems each targeted at different parties likely to have knowledge of incidents. Many design options can satisfy this criteria. One option could be to create multiple industry employee reporting systems for different classes of industry employees: The Federal Aviation Administration (FAA) maintains no fewer than eight distinct voluntary reporting programs for industry employees---dispatchers, air traffic controllers, pilots, flight attendants, technicians, airport employees, and FAA employees---to report aviation-related incidents \cite{FAA_Voluntary_Programs}. Another option is multiple industry reporting systems for actors at different parts of the supply chain or for service providers in different verticals---as in agriculture above. Nuclear energy also follows this model: the International Atomic Energy Agency (IAEA) has established three separate systems for nuclear power plants, fuel cycle facilities, and research reactors \cite{IAEA_IRS}. Yet another option is to create one primary reporting system with many reporting formats for different parties, e.g., the MedWatch system for pharmaceuticals. Since knowledge about AI incidents may be diffused along the supply chain \cite{ec_code_2025, hopkins_ai_2025}, and since incidents are likely to be diverse \cite{slattery_ai_2025}, AI incident reporting could take similar approaches to these industries.

With multiple incident reporting systems, information flows between systems also becomes important. Options for tackling this issue include ``federated'' regime consisting of a unified reporting standard implemented across government, industry, and non-profits \cite{dixon_argument_2024}, or a more centralized regime as in nuclear or aviation with a few government agencies operating parallel reporting systems. Another factor to take into account is operational simplicity for parties submitting reports; systems such as the US Coast Guard’s National Response Center (NRC) create a single, unified point of contact for reporting parties. After reports are submitted, the Center then re-routes to relevant US agencies a wide range of incidents, including ecological damage (routed to the Environmental Protection Agency (EPA) and other relevant actors), railroad accidents (routed to the Federal Railroad Administration (FRA)), and suspicious maritime activity \cite{neuman_privacy_2015, epa_national_2024, epa_national_2024-1, fra_accident_2024}. 

\subsubsection{Regulatory vs. Non-Regulatory Governmental System Operators} For government-run incident reporting systems, whether systems are operated by regulatory vs. non-regulatory agencies can affect the system’s policy goals and post-reporting actions. Throughout our case study industries, regulators generally oversee reporting by companies, industry employees, citizens, third parties, or product users (Table \ref{tab:Full_Results})---in particular, regulators usually oversee reporting when industry actors are mandated to report incidents to the government (for accountability). Additionally, systems that allow members of the public, third parties, or product users to report to regulators have been shown to improve product safety \cite{geier_review_2004}, especially if regulators can take enforcement actions as a result of those reports \cite{raji_outsider_2022}. Public reporting hotlines may also have secondary benefits such as providing consumer information or access to resources \cite{calvert_surveillance_2010}. Some industries such as aviation have multiple incident reporting systems, some run by regulators and others by non-regulators.
    
On the other hand, systems where incidents are reported to non-regulatory agencies are generally non-punitive and oriented toward learning---since accountability could be difficult to achieve without regulatory authority. Thus, non-regulatory agencies usually help coordinate reporting for companies or industry employees. Because these agencies cannot take punitive actions against reporters, systems administered by non-regulators could also engender more trust in reporters \cite{mills_role_2017}. Moreover, regulators may sometimes have perverse incentives to avoid solicitation of incident reports if they believe that such reports may reflect negatively on their reputations; non-regulators tend to be more insulated from these political pressures may thus avoid perverse incentives \cite{etienne_politics_2015, christensen_elements_2017}.

The gold standard for a non-regulator managed voluntary reporting system is the Aviation Safety Reporting System (ASRS), whose administration the FAA entirely outsourced to the National Aeronautics and Space Administration (NASA). Because NASA does not regulate airlines, its choice as the ASRS administrator helped promote trust and confidence in the system \cite{mills_promise_2010}---especially as it is able to guarantee reporters anonymity and some liability protections \cite{AC0046F} (further discussion of ASRS in Section \ref{subsec:Recs_Enforcement}). Similarly, the National Transportation Safety Board (NTSB) is a cross-cutting agency responsible for incident investigation; it has no regulatory authority, allows participation by outside experts and parties, and is insulated from the political process. These factors have all made the NTSB process more collaborative and made stakeholders more likely to participate to gain access to incident information \cite{fielding_national_2010}.

\subsubsection{User-to-Company Reporting Systems} AI model developers and service providers may benefit from establishing internal systems for accepting and investigating reports or complaints from product users \cite{mcgregor_err_2024}. Currently, not all model developers have harm reporting channels for users or members of the public, and existing forms do not appear to always capture information about what harm or incident has occurred. User reports can be a valuable source of safety information \cite{sarkar_consumer_2018}, so the absence of such reporting presents a gap that prevents companies from gaining important safety information and thus effectively learning from incidents.

In the context of consumer products, users most often report issues and incidents to organizations closest to them in the supply chain: according to a European Commission poll, consumers who had product issues reported them to retailers or service providers 81\% of the time, to product manufacturers 26\% of the time, but to government agencies only 10\% of the time \cite{dg_just_consumer_2023}. In the context of AI systems, the organizations closest to the user (and to AI-caused harm) are user-facing AI service providers. Although most model developers currently double as service providers, as AI systems transition away from single-model systems \cite{zaharia_shift_2024, wei_methodological_2025}, the market could see intermediary service providers that use a multitude of models from different developers. These trends suggest that model developers and service providers may both benefit from implementing user reporting systems. In fact, China has already mandated that these organizations develop user reporting systems \cite{cac_translation_2022,cac_translation_2023}.

Industry actors may also wish to standardize reporting formats to enable intra-industry information sharing, as has been suggested in the context of workplace safety reporting systems \cite{brazier_summary_1994} and in the context of AI \cite{longpre_position_2025, oecd_towards_2025}.

\subsection{Type of Incidents Reported} \label{subsec:Recs_Incident_Type}

\subsubsection{Distinguishing AI Safety, Security, and Rights Incidents} Operators of AI incident reporting systems may benefit from recognizing one possible categorization of incidents in terms of safety, security, and rights (defined in Appendix \ref{sec:Appendix_Framework_Definitions}); incidents in different categories may have different risk profiles and require different responses. Notably, safety and security are traditionally distinct fields. However, security, rights, and safety incidents also occur conjointly (e.g., a model jailbreak that results in property damage), and a single incident may involve harms of multiple types (e.g., a discriminatory output resulting in physical harm). Distinctions may thus be blurred in practice \cite{johnston_adversarial_2004, derock_convergence_2010}, and additional research is needed to understand the confluence between security, rights, and safety incidents, as well as how incident responses may differ between security, rights, and safety incidents. Reporting distinct types of incidents, for instance, could be done through separate systems, or reporting could occur within one reporting system that routes and responds to different incident types accordingly.

The literature draws sharp lines between safety, security, and rights incidents. Historically, incident reporting systems were first established for safety reporting and only later adapted into reporting systems for security incidents \cite{johnson_architectures_2015, johnson_failure_2003, johnson_tools_2014}; safety and security are traditionally considered distinct fields \cite{piggin_safety_2015, jore_conceptual_2019, qi_ai_2024}. Although there is recognition that software security and safety incidents can be intertwined \cite{piggin_safety_2015}, security incidents have generally been managed by different regulatory actors or even reported through distinct systems \cite{connell_cross-industry_2004}.

These distinctions may apply to general-purpose AI. The goal of safety mitigations is to protect external actors from AI systems, whereas the goal of security is to protect AI systems from external actors \cite{khlaaf_toward_2023, roumani_examining_2020, qi_ai_2024}. Sharing information about security vulnerabilities (and perhaps also AI misuse) carries risks \cite{albakri_risks_2018, shevlane_offense-defense_2020, grotto_vulnerability_2021}, whereas sharing information about safety hazards is critical to learning. Security incidents could be patched, whereas AI safety incidents are probabilistic and are not as easily addressed at a model level \cite{cattell_coordinated_2024, kim_jailbreaking_2024}. Existing AI security frameworks differ widely from safety frameworks \cite{grotto_vulnerability_2021, MITRE_ATLAS, kumar_failure_2019}. However, safety and security incidents in the context of AI can also occur conjointly \cite{qi_ai_2024}.

Rights harms can be distinct from both security and safety harms. Though both rights and safety harms can be caused by AI systems, the nature of the harm caused is different; most incident reporting systems tend not to support reporting both safety and rights incidents (Table \ref{tab:Full_Results}). This limitation may be partially due to the fact that historically, rights harms such as civil liberties or human rights violations were primarily caused by people, not technical systems---but as AI systems become increasingly autonomous \cite{chan_harms_2023}, they may cause new types of rights harms \cite{dobson_intangible_2023, hoffman_adding_2023} and magnify existing rights harms or inequities \cite{critch_tasra_2023, shelby_sociotechnical_2023}.

Some jurisdictions have recognized the distinctions between incident types. In the US, recent policy documents have trended toward differentiating safety and rights incidents. The US Executive Order 14110 on AI contained distinct provisions addressing ``safety'' and ``rights'' protections \cite{biden_executive_2023}, and agency guidance has delineated between ``safety-impacting AI'' and ``rights-impacting AI'' \cite{young_memorandum_2024, young_memorandum_2024-1}. A state-level bill in New York State that would require AI companies to report incidents to the state Attorney General is similarly limited to ``safety incidents'' \cite{bores_a6453a_2025}. Other governmental institutions, however, have tended to blur the lines between incident types \citetext{e.g., \citealt{perset_stocktaking_2023}}. Most notably, the EU AI Act establishes a single incident reporting requirement, with the definition of an incident including both rights and safety incidents \citepalias[Art. 3]{AIA}. Many open questions remain regarding reporting for different types of incidents, and additional research is needed.

\subsection{Level of Risk Materialization} \label{subsec:Recs_Scope_Risk}

\subsubsection{Reporting AI Near Misses} Near misses---events that could have but ultimately did not cause harm---are a valuable source of data for safety learning, especially since the vast majority of safety incidents are not harm events but rather near misses. For instance, one hospital study found that less than 1\% of reported safety incidents caused major harm, 18\% caused minor/temporary harm, and 82\% resulted in no harm \cite{mansouri_rates_2016}; another estimate pins near misses as occurring at up to 300 times as often as harm events \cite{shojania_making_2001}. Reducing the number of near misses can also reduce the number of harm events \cite{jones_importance_1999}. If AI issues and incidents occur at rates proportionate to those in healthcare or other industries, then AI near misses may similarly be valuable for safety learning. Note that near misses and harm events are frequently reported via the same reporting system \citetext{\citealt{cheng_international_2011}; \citealt{manheim_results_2021}; Table \ref{tab:Full_Results}}.

Governments and industry organizations can consider enabling AI near miss reporting by industry employees, users, third parties, and citizens---either through reporting systems specific to near misses or by permitting near misses to be reported along with harm events or other issues. Near miss reporting systems tend to be voluntary, but they could also be mandatory; at the very least, there is consensus in the literature that self-reporting near misses should be non-punitive given that no harm was caused \cite{coyle_designing_2005}. Many industries have implemented near miss reporting systems---including aviation, nuclear, energy, chemical, and construction \cite{gnoni_near_2022, macrae_close_2014}---and some of these may be appropriate as guides for the AI context \cite{shrishak_how_2023}.

\subsubsection{Reporting of AI Issues vs. AI Incidents} AI issues may also be useful sources of information for safety learning, and the AI governance literature has recently begun to address ``flaw disclosures'' \cite{longpre_position_2025, cattell_coordinated_2024}. Whether issues should be reported via the same reporting channels as incidents may depend on whether the reporting parties could have access to knowledge about either issues or incidents. Some incident reporting systems, such as those in nuclear power and civilian aerospace, appear to permit the reporting of both issues and incidents (Table \ref{tab:Full_Results}). With AI systems, however, some hazards or situations may be discoverable only by third parties like experts or red-teamers, which may necessitate reporting systems for AI issues that are different from those designed for users, the public, or other actors. System administrators will also need to ensure that issue reporting does not overwhelm reporting systems, especially if the reports are primarily about product complaints rather than safety or rights issues \cite{havinga_hazard_2021}.

\subsection{Enforcement of Reporting} \label{subsec:Recs_Enforcement}

\subsubsection{Mandatory Reporting Thresholds} In some safety-critical industries, government-mandated incident reporting has seen some success. The EU AI Act has already taken a step in this direction \citepalias[Art. 73]{AIA}, though its incident definition remains somewhat ambiguous. A phased reporting mandate---requiring that organizations report incidents shortly after discovery but permitting reports to be later amended with details and investigative results---may also be appropriate for high-severity AI incidents. Because LLM systems are complex, an initial report can notify government actors and determine if an official response is necessary while leaving organizations time to investigate (e.g., as in \citetalias[Art. 73(5)]{AIA}).

Reporting mandates generally require reports to be submitted to government agencies or other centralized actors, and mandates can be accountability-oriented since they may result in regulatory actions \cite{WHO_2005}. For instance, \citet{kesari_data_2023} finds that mandatory cybersecurity incident reporting to state Attorneys General reduced consumer complaints of identity theft by 10.1\% on average, possibly by deterring firms from engaging in unsafe practices.

Mandatory reporting generally applies to (high-severity) harm events (e.g., cases of death or serious injury): hospitals must report severe incidents to state agencies \cite{cdph_dph-11-023_2021}, airlines must report certain accidents and collisions to the FAA \cite{faa_aeronautical_2024}, and medical device manufacturers must report drug reactions and device malfunctions to the Food and Drug Administration \cite{fda_providing_2022}. One analysis estimates that 90\% of reports in the FDA’s central database, MedWatch, are submitted by device manufacturers under mandatory requirements \cite{rajan_medical_2015}. Thus, mandatory reports can be important mechanism for regulatory visibility.

\subsubsection{Defining Reporting Requirements and Thresholds} Incident reporting systems usually need to develop clear, well-scoped reporting thresholds and definitions to be practically useful. Ideally, reporting thresholds capture all or most new hazards and incidents helpful for safety learning, but they cannot be so low that systems become inundated with reports that may or may not be useful \cite{johnson_failure_2003}. Approaches to defining incidents include encouraging reporting for any possible issue or incident, providing lists of reportable incident types, or establishing thresholds based on particular incident outcomes, system behaviors, or procedural violations. Which approach is appropriate for AI incident reporting is unclear; we offer examples of incident definitions in Appendix \ref{sec:Appendix_Example_Defs}, but additional research is needed to operationalize AI incident definitions and taxonomies.

Vagueness in reporting thresholds hampers safety learning \cite{stavropoulou_how_2015} and accountability while allowing industry actors to dodge compliance. For instance, when the National Highway Traffic Safety Administration (NHTSA)’s requirements were unclear as to what types of safety defect-related documents automobile manufacturers were required to turn over, manufacturers employed a range of strategies designed to evade responsibility. During NHTSA investigations, manufacturers denied that defects existed, responded to NHTSA requests for information with misleading and confusing language, and denied that the issues identified were safety-related---despite, in one case, a manufacturer later issuing a voluntary recall for the exact issues identified by NHTSA \cite{pecht_role_2005}. Similarly, \citet{kesari_data_2023} also found that when state laws exempted breaches involving encrypted data from cybersecurity notification requirements, companies did not report breaches where data was stolen along with the encryption key. States that closed this loophole saw a 13.1\% decrease in the number of data breach consumer complaints received.

Near miss reporting systems similarly need clear definitions \cite{gnoni_near_2022} and to collect sufficient supplementary information to be useful for learning, e.g., information about the user, or interactions/communications between users or systems \cite{thoroman_system_2018}.

\subsubsection{Voluntary Reporting Systems} Mandatory reporting requirements alone may be insufficient to achieve safety learning---because high-severity harm events are rare, significant learning can occur from information about issues and near misses, and such information may not be centralized (Sections \ref{subsec:Recs_Actors}, \ref{subsec:Recs_Scope_Risk}). Voluntary systems can fill those gaps by allowing citizens, users, third-parties, companies, and industry employees to submit reports to centralized actors. Commentators have suggested that AI incident reporting adopt the model of the voluntary systems of the FDA \cite{naiac_recommendation_2023} or---more commonly---of the FAA \cite{shrishak_how_2023, croxton_how_2024}.

The rest of this subsection examines the FAA's voluntary reporting programs (primarily ASRS), which are often considered the gold standard of voluntary reporting systems. ASRS's success is attributed to multiple factors. First, reports to the ASRS (run by NASA) are de-identified and---with certain exceptions---cannot be used by the FAA in regulatory enforcement actions \cite{AC0046F}; these guarantees ensure that ASRS is not viewed as punitive while creating incentives for reporting \cite{cohen_private_2020}. Additionally, the FAA also intentionally aimed to achieve industry buy-in to ASRS by involving stakeholders early in its design process \cite{asrs_asrs_2001, mills_role_2017}.

Ultimately, the FAA’s systems have successfully enabled safety learning \cite{connell_cross-industry_2004, mills_secondary_2014}. Since its inception in 1975, ASRS has received over 2,000,000 reports \cite{marfise_nasa_2023}, regularly generates feedback on safety hazards \cite{mills_promise_2010}, and surfaced unique information unavailable via other sources \cite{connell_cross-industry_2004}. %

The ASRS model, however, has not succeeded elsewhere. Inspired by ASRS, the Federal Railroad Administration has established a similar program, the Confidential Close Call Reporting System (C\textsuperscript{3}RS), also administered by NASA \cite{ranney_confidential_2019}. But C\textsuperscript{3}RS failed to achieve industry buy-in: only 23 of 800 railroad companies in the US participated \cite{gao_federal_2022}, and some companies even withdrew participation because they perceived C\textsuperscript{3}RS to be ineffective \cite{jeffries_letter_2023}. 

It is unclear whether the FAA's voluntary systems are an appropriate model for AI incident reporting. As in rail, buy-in from AI developers may be difficult to achieve. Unlike aviation, current competitive dynamics in AI may inhibit voluntary information reporting \cite{beers_beyond_2025}. For instance, developers have been reluctant to share data about their models, training data, and other technical features \cite{bommasani_foundation_2024}, but such information may be critical to enable safety learning. Aviation incidents are also industry-specific whereas harms from general-purpose AI systems are likely to traverse different verticals and involve significantly more actors than in the FAA’s reporting scheme, which complicates reporting structures. ASRS aside, FAA's other voluntary programs like the Aviation Safety Action Program are administered in partnership with labor unions \cite{mills_promise_2010}, which are virtually non-existent in the AI industry. Finally, FAA is known to have established a highly collaborative relationship with industry via its voluntary reporting programs \cite{mills_promise_2010}, but such relationships in AI may raise concerns about regulatory capture \cite{wei_how_2024}.

\subsection{Anonymity of Reporters} \label{subsec:Recs_Anonymity}

\subsubsection{Anonymous or Confidential Reporting} Whether parties submitting incident reports should be offered anonymity or confidentiality depends heavily on the identity of the parties and their perceptions about the possibility of retaliation if they report. For individual reporters, fear of identification and subsequent reprisal or reputational risks is a primary reason that they do not submit reports \cite{van_der_schaaf_biases_2004, durant_ignorance_2020, beers_beyond_2025}. These fears persist even in anonymous reporting systems, in which only the reporters themselves are aware of their identity \cite{harper_identifying_2005}, and they also affect reporting systems internal to organizations when government mandates require non-anonymous reporting downstream \cite{weissman_error_2005, wolf_error_2008}. Beyond the effect on overall reporting rates, non-confidentiality may also skew reported the types of incidents/issues reported \cite{sieberichs_why_2021}.

Anonymity can also be warranted when the policy goal is accountability and when reporting parties include those who fall outside the chain of incident responsibility, such as citizens, third parties, industry employees (when applicable), and in independent databases. When industry employees fall within the chain of incident responsibility, learning systems can be confidential or open but would require additional protections or assurances to facilitate participation \cite{flott_enhancing_2018}; the same may be true in learning systems involving industry organizations who fear legal or competitive repercussions from reporting, which may be the case in AI. Withholding identifiable information before information sharing could assuage these concerns. Overall, some incident reporting systems may benefit from offering (optional) anonymity to AI incident reporters \cite{kolt_responsible_2024}.

On the other hand, not all reporting systems need to be anonymous. Although there is little literature on the anonymity of company reporting mandates, most mandatory systems in practice appear to be open or confidential when the reporting parties are industry organizations (Table \ref{tab:Full_Results}). Anonymous systems may also hinder accountability because they may be less reliable and make follow-up investigations difficult or impossible \cite{barach_reporting_2000}.

\subsection{Post-Reporting Actions} \label{subsec:Recs_Post-Reporting}

\subsubsection{Facilitating Safety Learning After Reporting} Incident reporting is the first, but not the only step, toward safety learning and accountability. Researchers have developed mature models of safety learning in which incident reporting leads to incident investigation, classification, and analysis \cite{adole_accident_2020, benn_feedback_2009, hewitt_incident_2013, leveson_engineering_2011, obrien_deployment_2023} followed by the development, implementation, and monitoring of interventions at the systems, individual, and organizational levels \cite{briggs_design_2017, webster_safety_2016, drupsteen_critical_2013, zwetsloot_thinking_2019}. Reporting is crucial to enabling subsequent analysis \cite{kolly_benefits_2012}, increased visibility into issues and incidents \cite{mangold_use_1995}, and subsequent corrective actions.

This learning lifecycle is not yet mature in AI. AI incident databases have inspired research into incident types and taxonomies \cite{mcgregor_indexing_2022, stanley_exploring_2023, tidjon_threat_2022, wei_ai_2022, nasim_artificial_2022, abercrombie_collaborative_2024}, and they can help us understand the risk landscape \cite{whittlestone_why_2021} and raise awareness of AI risks \cite{feffer_ai_2023}. However, centralized repositories are necessary for transparency, accountability, and analysis \cite{mandel_system_2014, lupo_risky_2023}. In addition, lack of transparency in many general-purpose AI systems \cite{bommasani_foundation_2024} may impede safety learning research, and the community has not yet developed consensus classification taxonomies, investigation methods, or evaluation/intervention processes \cite{paeth_lessons_2024}. Additional research is needed to adapt and operationalize the post-reporting learning lifecycle in the context of general-purpose AI systems.

\subsubsection{Information Sharing Between Incident Reporting Systems} After an incident report has been submitted, information must be aggregated and/or routed to relevant stakeholders to facilitate safety or accountability. Successful information sharing requires both identifying the correct actors and ensuring that information can be easily shared (e.g., via standardization and interoperability between systems).

Experience from other industries indicates that incident reporting systems at the national and international level are more concerned with incidents that are predictive of greater risks, high-severity incidents or incidents that could escalate into emergencies or crises, or incidents from which system- or industry-wide learning is possible \cite{barach_reporting_2000, iaea_iaea_2022, novak_iaea_1985, naiac_recommendation_2023}. Local or industry-specific systems will be better positioned to handle incidents that are limited in scope or generalizability, and user reporting systems or third-party reporting systems to industry organizations may want to set lower thresholds so that low-severity incidents may be captured for learning \cite{frey_does_2002, webster_safety_2016}.

Standardization and interoperability between systems are also important to facilitate the flow of information \cite{shane_ai_2024}. A counterexample is the U.S. cybersecurity incident reporting regime, in which 22 federal US agencies have implemented at least 45 sometimes-duplicative incident reporting requirements \cite{dhs_harmonization_2023, kosseff_positive_2016}. Such fragmentation makes data aggregation difficult and hinders learning \cite{wood_mandatory_2005}. On the other hand, reporting systems can consider closer integration to avoid increasing the burden of filing reports, which can disincentivize reporting \cite{lubomski_building_2004, guffey_incident_2014}. Reporting standards may also need to carefully consider privacy policies, which must accommodate information aggregation \cite{dixon_argument_2024}, ensure that relevant actors can access information \cite{kolt_responsible_2024}, and also protect against (perceptions of) identity disclosure and retaliation (Section \ref{subsec:Recs_Enforcement}). Nascent efforts are attempting to create standards for GPAI incident documentation \cite{longpre_position_2025, oecd_towards_2025, ezell_incident_2025}, but industry adoption may pose a challenge.

Existing systems offer examples of how information sharing can succeed. In pesticide reporting, reporting systems range across five different levels: systems at the local, state, and national level are interconnected and create reporting ``chains,'' through which information is conveyed up the chain of command \cite{calvert_surveillance_2010}. The IAEA Incident Reporting System---which collects incident information at an international level---generally requires that countries who are members of the system must have established national-level incident reporting systems before applying for membership in the IAEA IRS \cite{iaea_iaea_2022}. Other examples include incident sharing via the International Civil Aviation Organisation (ICAO) in civilian aerospace, coordinated vulnerability disclosure \cite{cattell_coordinated_2024} and information sharing and analysis centers (ISACs) in cybersecurity, or various government-to-government reporting requirements \cite{guffey_incident_2014}. These multi-tiered structures allow lower-level systems to filter and transmit only generally applicable information to higher-level systems \cite{iaea_iaea_2022, novak_iaea_1985}. They can also allow different fora to focus on different types of risks \cite{sabel_regulation_2018}.

\subsubsection{Legal Liability and Regulatory Frameworks} Individuals and companies commonly cite legal uncertainty and fear of liability as top reasons for deciling to report \cite{nagamatsu_healthcare_2009, ic_ig_joint_2023, fukuda_impact_2010, carlfjord_experiences_2018}. Policymakers who wish to facilitate incident reporting in the context of GPAI can consider clarifying legal frameworks for reporting up front, setting clear reporting incentives \cite{glendinning_employee_2001, vredenburgh_organizational_2002, briggs_design_2017}, and communicating these guidelines to reporting parties. Some issues for policymakers to consider include what types of liability attach to reporting, whether incident reports are discoverable in court, the precision of the scope of reporting requirements, the relevant standards to apply where security incidents are intertwined \cite{johnson_tools_2014}, and how anti-trust law interacts with sharing of safety information \cite{anthropic_comment_2023}. In addition, which parties have access to incident data is a perennial source of concern, and determining with whom to share information will require balancing legal and competitive concerns with report receivers’ interests in visibility and accountability \cite{ic_ig_joint_2023}.

A few types of incentives are commonly used in different systems. In mandatory reporting, penalties for failing to make reports may be effective \cite{grepperud_medical_2005, yew_penalty_2022}. In voluntary reporting, limited liability protections for reporters are also common and can incentivize reporting: in a survey of healthcare providers, for instance, 72\% of physicians indicated that they would be more likely to submit incident reports if reports were protected from legal discovery \cite{harper_identifying_2005}.

Finally, legal loopholes that allow companies to avoid reports or disclosures can hinder incident reporting. Product manufacturers, for example, have obtained broad protective orders or confidential settlements in court to avoid public disclosure of defects and product safety issues \cite{engstrom_secrecy_2024, egilman_confidentiality_2020, cohen_private_2020, saver_deciphering_2017}. Firms fearing that incident reports or documentation will be used in subsequent litigation may also intentionally keep less documentation, making incident reporting requirements less useful \cite{schwarcz_how_2023}. AI incident reporting regimes may wish to take note of these problems.

\section{Limitations} \label{sec:Limitations}

The scope of our work is limited. Importantly, we do not conduct a full cost-benefit analysis of whether incident reporting systems are desirable in the context of GPAI systems; it is possible that goals of learning or accountability could be better achieved by other governance practices. Our case studies are also US-centric, and some lessons may not be easily transferred to non-US jurisdictions. Moreover, AI security incidents may require different processes and frameworks than safety and rights incidents \cite{qi_ai_2024}, which are our primary focus here. %

Our scope is also restricted to the institutional design of AI incident reporting systems. We do not address, for instance, some implementation features external to the institutional design of incident reporting systems---e.g., political feasibility, technical methods for incident monitoring and detection, safety culture \cite{guldenmund_nature_2000, miller_i_2019, probst_accident_2010, probst_organizational_2008}, reporter trust and communication \cite{briggs_design_2017, kingston_attitudes_2004}, documentation practices \cite{turri_why_2023}, organizational processes \cite{shedden_organisational_2010}, user interaction \cite{goode_translating_2016}, and other socio-technical factors \cite{shedden_organisational_2010, maslen_preventing_2016, probst_pressure_2013}. Future research can consider conducting practitioner interviews, user experiments, and industry surveys to explore these dynamics in practice. Similarly, our framework is also intended to provide conceptual clarity around AI incidents and the institutional structures of incident reporting systems, but additional research will be needed to make these definitions operationalizable and interoperable.

\section{Conclusion} \label{sec:Conclusion}

Incident reporting systems can help create a safer AI ecosystem and hold organizations responsible for harms from AI. Incident reporting systems’ success is rooted in their institutional structures, and this paper provides the first systematic examination of how incident reporting systems may be designed in the context of general-purpose AI. Through nine case studies of safety-critical industries, we provide institutional design considerations for GPAI incident reporting systems, and we discuss when particular design choices may be more or less appropriate based on stakeholders’ goals and other factors. We hope to inform to US stakeholders interested in establishing incident reporting systems at a time when GPAI is seeing increased adoption across the economy.

\section*{Ethical Considerations Statement}

This project was determined not to be human subjects research after an initial review by the RAND Human Subjects Protection Committee. No further review was required for this work.

\section*{Acknowledgements}

KW was the primary writer of this paper, and LH was responsible for supervision as well as review and editing. The authors are grateful for conversations with and feedback from (in random order): Karson Elmgren, Tommy Shaffer Shane, Morgan Simpson, Alan Chan, Markus Anderljung, Casey Mahoney, Emma Bluemke, Ben Garfinkel, Shaun Ee, Leonie Kessler, Fabian Ulmer, Gaurav Sett, Jason Greenlowe, Noam Kolt, Lisa Soder, Mauricio Baker, Toni Lorente, Cristian Trout, Michael Aird, Alexis Carlier, Ren Bin Lee Dixon, Francis Rhys Ward, and four anonymous AAAI reviewers. This work was initiated while KW and LH were at the Centre for the Governance of AI and completed while KW and LH were at RAND. Funding for this work was provided by gifts from RAND supporters and by the Centre for the Governance of AI.

\clearpage

\bibliography{references_AAAI_IR}

\appendix

\section{Appendix: Additional Background Information} \label{sec:Appendix_Background}

This Appendix contains additional background information; it is an extension of Section \ref{sec:Review}. We begin with a note about the scope of our discussion and about terminology. We then review discussions of incident reporting in the AI governance literature.

First, a note on terminology and on the scope of our discussion. Our discussion is focused on systems for reporting of safety and rights incidents caused by general-purpose AI (``incident reporting systems for general-purpose AI''). The following definitions help scope our discussion:
\begin{itemize}
    \item ``Safety and rights'': We limit our discussion to systems concerning incidents that could or did harm (human) safety and rights. We largely exclude discussion of security incidents (AI security or cybersecurity); security incidents are normally events in which (AI) systems are compromised by external actors. See extended discussion in Sections \ref{subsec:Recs_Incident_Type} and \ref{subsec:Framework_Incident_Type}.
    \item ``Incident:'' for the purposes of this article, we adopt a broad, effects-based working definition of ``incident:'' incidents are events that either resulted in real-world harm (harm events) or that could have but did not ultimately result in harm (near misses). See discussion below.
    \item ``Reporting:'' our working definition is that reporting systems are formal systems in which information (about incidents) are unidirectionally conveyed from an actor with knowledge of the incident to an institution(s) with responsibility for oversight over AI.\footnote{For instance, we largely exclude discussion of ``information sharing'' systems in which companies share information with each other.} See terminology in Appendix \ref{subsec:Framework_Post-Reporting}).
    \item ``Caused by general-purpose AI:'' We are interested in incidents directly or indirectly resulting from the involvement of general-purpose AI systems, as incidents where AI systems played no role in the chain of events leading to the incident (whether or not AI systems were present).\footnote{For definitions of AI incidents vs. non-AI incidents, see the definitions in Appendix \ref{sec:Appendix_Example_Defs}.} 
\end{itemize}

Regarding the definition of ``incident'' in particular, no consensus yet emerged for how to define an AI ``incident'' in the AI governance literature as of the time of this writing in July 2025.\footnote{For an overview of definitions of ``incident'' in the AI governance literature, see Appendix \ref{subsec:Review_Literature}. Others have also noted the difficulties of defining an AI incident \cite{turri_why_2023}.} We do not suggest that our working definition is ideal or operationalizable as-is.\footnote{See various sample definitions used in AI and in other domains in Appendices \ref{sec:Appendix_Example_Defs}, \ref{subsec:Review_Literature}. Additional work will be needed to better taxonomize and operationalize AI incident definitions.} Our working definition is intentionally broad, and it is chosen with the following considerations in mind:

\begin{itemize}
    \item Our definition is intended to capture many incidents of interest, and it is thus agnostic as to safety and rights risk models since general-purpose AI systems present a wide variety of risk models \cite{slattery_ai_2025};
    \item Our definition is consistent with many (though not all) definitions of ``incident'' in the traditional safety science literature \citetext{see Appendix \ref{sec:Appendix_Example_Defs}};
    \item Our definition does not exclude events that result from intentional misuse \citetext{e.g., \citealp{arnold_ai_2021}}, unlike definitions based in terms of ``unintended'' harm (e.g., ``accidents'');\footnote{In some industries, the majority of incidents of interest are accidents (see Table \ref{tab:Example_Incident_Defs}), but with general-purpose AI systems, incidents resulting from intentional misuse are also of interest \cite{zwetsloot_thinking_2019}.} 
    \item Our definition does not exclude events that result from normal system behavior or from human errors (e.g., the definition in \citealp{nis_cooperation_group_cybersecurity_2018}), unlike definitions based on ``unexpected'' system behavior (e.g., some definitions of ``failures''); and
    \item Our definition is independent of the AI model lifecycle because not all incidents occur post-deployment. Safety incidents from autonomous AI behavior \cite{anthropic_anthropics_2023}, for instance, could occur even pre-deployment.
\end{itemize}

Note that our focus is also on incidents from general-purpose AI systems rather than narrow AI systems. This focus is motivated by the facts that:

\begin{itemize}
    \item General-purpose AI systems exhibit different behaviors from narrow AI systems, e.g., emergent capabilities \cite{casper_defending_2024, woodside_emergent_2024};
    \item Incidents from narrow AI systems may be covered by existing sectoral frameworks (e.g., \citealp{nhtsa_second_2023}); and
    \item Since general-purpose AI systems could be deployed across a variety of domains, it may necessary to examine many different domains to inform incident reporting for GPAI.
\end{itemize}

We excluded from our discussion and review the following systems and programs: 

\begin{itemize}
    \item Reporting requirements for narrow AI systems \citetext{e.g., \citealp{nhtsa_second_2023}}; 
    \item Domain-specific incident reporting requirements that could apply to specific AI use cases; 
    \item Whistleblowing systems, because unlike many incident reporting systems, they often exist to promote legal or regulatory compliance and collect largely non-incident information \cite{muto_compilation_2024}
    \item Emergency response or crisis response systems (see Appendix \ref{subsec:Methods_Industries}); and 
    \item Post-reporting incident investigation and analysis (see Appendix \ref{subsec:Recs_Post-Reporting}). 
\end{itemize} 

\subsection{The Literature on AI Incident Reporting} \label{subsec:Review_Literature}

At the beginning of this study, we conducted a scoping review of the academic and gray literature on incident reporting for general-purpose AI systems. Our review methodology is detailed in Appendix \ref{subsec:Methods_Typologies}, and results are presented below.

Overall, the AI governance literature is generally supportive of incident reporting \cite{goodman_ai_2024, dixon_argument_2024, obrien_deployment_2023, dsit_emerging_2023, schuett_risk_2023, naiac_recommendation_2023, croxton_how_2024, shrishak_how_2023}. Our review found no comprehensive analysis of the institutional design considerations of such systems in the general-purpose AI context---though note that our review was conducted in 2023--2024.\footnote{The first author believes that this statement still holds as of July 2025.}

Specific findings follow.

\subsubsection{\enquote{Incident} does not have a consensus definition in the AI governance literature}

Many different definitions of an AI ``incident'' are in use in the AI governance literature (see, e.g., Table \ref{sec:Appendix_Example_Defs}; \citealp{shevlane_model_2023, raji_change_2023, lupo_risky_2023}). Often, no explicit definition is provided \citetext{e.g., \citealp{macrae_learning_2022, durso_analyzing_2022, manheim_building_2023}}. Two common approaches to incident definition are the effects-based and the cause-based definitions. 

Effects-based definitions scope incidents according to their resulting effects, i.e., by defining incidents as events that directly or indirectly contributed to the occurrence of some negative impact. A typical example of this type of definition is given by \citet{brundage_toward_2020}, which defines incidents as ``cases of undesired or unexpected behavior by an AI system that causes or could cause harm.'' This definition has given rise to the use of terminology such as ``harm events,'' ``harms,'' or ``harm incidents'' \citetext{e.g., \citealp{mcgregor_data-centric_2023, hoffman_adding_2023, raji_change_2023, costanza-chock_who_2022}}. We also adopt an effects-based approach in this article, as explained at the beginning of Appendix \ref{sec:Appendix_Background}.

Cause-based approaches define incidents based on the underlying factors that resulted in particular system behaviors. For instance, some articles use terms such as ``failure''\footnote{Earlier literature tended to use the ``failure'' or ``accident'' terminology more frequently. For instance, \citet{amodei_concrete_2016} (not included in our review) provides the following definition: ``Very broadly, an accident can be described as a situation where a human designer had in mind
a certain (perhaps informally specified) objective or task, but the system that was designed and
deployed for that task produced harmful and unexpected results.''} \cite{macrae_learning_2022, falco_governing_2021, turri_why_2023}, ``vulnerability'' \cite{avid_lifecycle_2024}, ``error'', or ``accident''\footnote{These may all be conceptually distinct from ``incidents,'' but the literature sometimes uses these terms interchangeably.} \cite{maas_regulating_2018}. For example, \citet{turri_why_2023} notes: ``no widespread definitions are yet agreed upon . . . we use the term AI incident interchangeably with AI failure to refer to instances in which an AI system results in unintentional negative impacts on humans.''

Other definitions are also present in the literature. Articles have limited ``incidents'' to ``concerning or otherwise noteworthy [model] evaluation results'' \cite{shevlane_model_2023} or to events occurring post-deployment of AI models \cite{schuett_towards_2023, avin_filling_2021, anderljung_frontier_2023}. Some definitions exclude near misses, while other definitions include them \cite{mcgregor_preventing_2021}. One definition even defines ``incident'' to include events where AI has ``introduced a benefit'' \cite{lefcourt_ai_2023}. 

Overall, our review suggests that to date, there is no consensus definition of an ``incident'' in the AI governance literature.\footnote{For additional context on different types of incidents and incident terminology, KW suggests referring to recent work from \citet{hoffman_mechanisms_2025}.}

\subsubsection{\enquote{Reporting} is used broadly in the literature}

Our review indicates that the AI governance literature uses the terms ``reporting,'' ``sharing,'' or ``disclosure'' interchangeably. Systems that have been categorized as ``incident reporting'' systems include ``a structured process for developers to share [certain information] with other developers, third parties, or regulators'' \cite{shevlane_model_2023}, ``a national incident reporting system to enable third parties to file complaints about algorithmic systems'' \cite{raji_outsider_2022}, and AI incident databases \cite{schuett_towards_2023}. Incident ``sharing'' has also been used to refer to submitting incident information to a database accessible to other industry actors \cite{avin_filling_2021, brundage_toward_2020} as well as information sharing agreements between countries \cite{shoker_confidence-building_2023}. 

We propose some distinctions between these terms in Appendix \ref{subsec:Framework_Post-Reporting}.

\subsubsection{Many policy proposals we reviewed are conflicting, unclear, or lack implementation details.}

Sources have proposed a variety of incident reporting systems, often with different design choices and policy goals. These include proposals for: systems for AI developers to submit voluntary reports to governments or third parties \cite{brundage_toward_2020, muehlhauser_12_2023, shrishak_how_2023}; user reporting systems operated by the UN or a new (inter)national body \cite{raji_change_2023, mulgan_case_2023}; a private system for AI developers to share dangerous model evaluation results \cite{shevlane_model_2023}; and international incident information-sharing \cite{shoker_confidence-building_2023}. Proposals vary in whether they call for reports to be publicly disclosed, confidential, or anonymous.

The stated goal of reporting also differs significantly. Sources have proposed that incident reporting systems should be designed to improve safety \cite{brundage_toward_2020, shevlane_model_2023}, monitor hazards from AI \cite{mcgregor_preventing_2021, muehlhauser_12_2023}, facilitate corporate accountability \cite{ajl_excoded_nodate}, increase media or public awareness of risks from AI \cite{raji_change_2023}, build public trust in AI systems \cite{falco_governing_2021}, assist regulatory enforcement \cite{raji_outsider_2022, costanza-chock_who_2022}, or facilitate research into AI risks \cite{mcgregor_preventing_2021}. %

Moreover, a number of articles we reviewed lacked key implementation details. Of the final 54 items included in our review, 30 were academic or gray literature articles (the rest included documentation from existing AI reporting systems or incident databases, legislation, voluntary frameworks, etc.). After we developer our framework for the institutional design of incident reporting systems, we returned to these 30 articles and coded them according to our framework. Of the 30 articles from the academic or gray literature, only three discussed all seven axes of our framework. Articles were primarily silent about the reporter anonymity (21 of 30 articles did not discuss), whether the system would be mandatory (18), and what post-incident reporting actions would be taken (14). In 13 articles, even one or both of the involved in submitting and receiving reports was not specified. %

\clearpage

\section{Appendix: Methodology} \label{sec:Appendix_Methodology}

\begin{figure*}[!htbp]
    \centering
    \includegraphics[width=\textwidth]{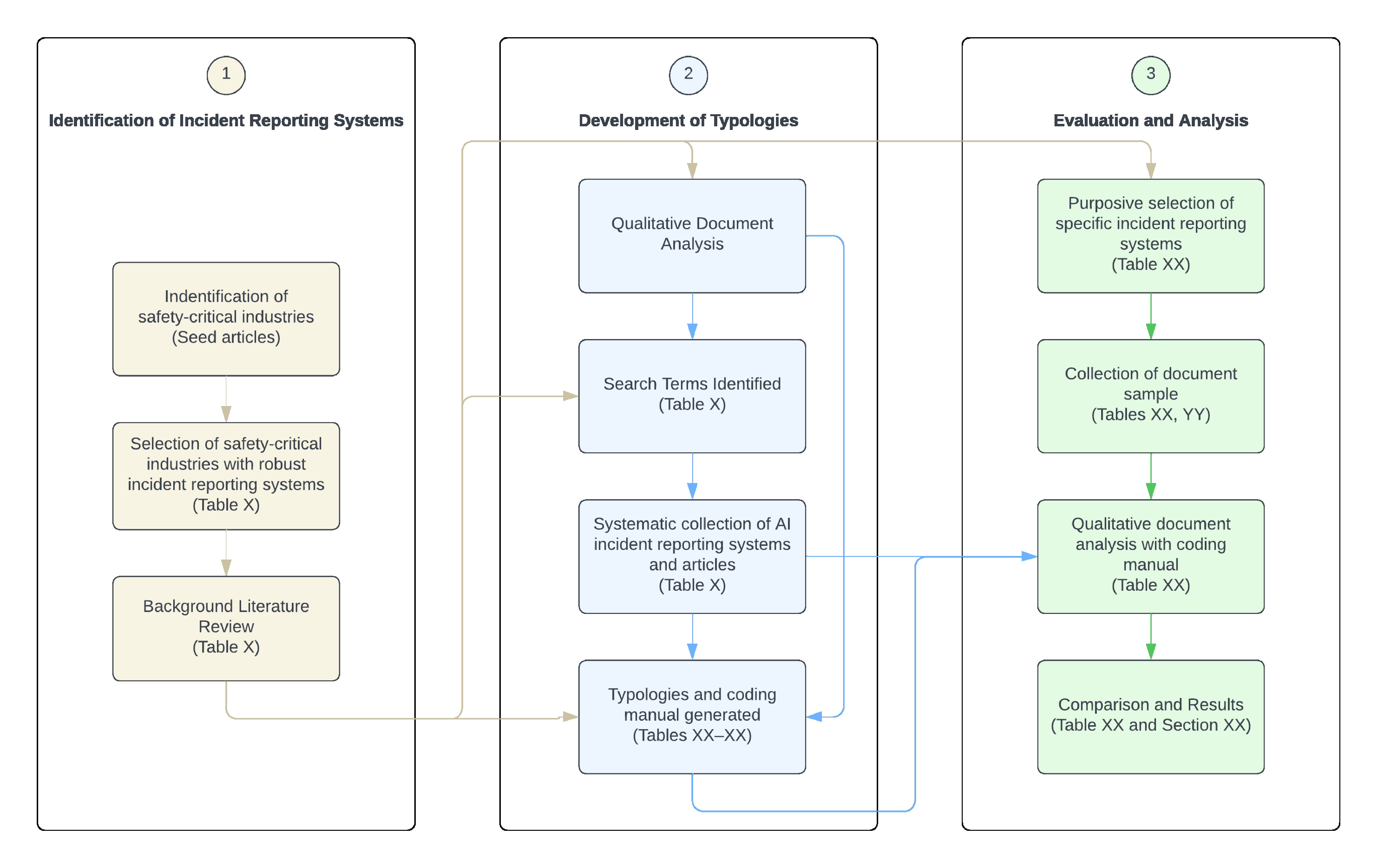}
    \caption{Diagram of methodology}
    \label{fig:Methods_Chart}
\end{figure*}

This Appendix contains details about the methodology used to identify our nine case studies, develop a framework for the institutional design of incident reporting systems, and review incident reporting systems in our case study industries. Our methodology is outlined in Figure \ref{fig:Methods_Chart} and is based on the approaches taken in \citet{raji_outsider_2022}, \citet{ayling_putting_2022}, and \citet{stein_public_2024}. 

We proceed in three parts:
\begin{enumerate}
    \item Identification of case studies (Appendix \ref{subsec:Methods_Industries}); 
    \item Development of a framework for the institutional design of incident reporting systems Appendix \ref{subsec:Methods_Typologies}); and
    \item Application of that framework to our case studies to generate insights for AI incident reporting Appendix \ref{subsec:Methods_Analysis}).
\end{enumerate}

\subsection{Identification of Case Studies} \label{subsec:Methods_Industries}

First, we identify case studies of safety-critical industries with existing and robust incident reporting frameworks. A case study approach is appropriate because incident reporting systems of various forms are mature and well-established in other industries such as aviation, and so it is possible to study lessons learned in those industries and assess whether those lessons can be applied to AI \cite{guha_ai_2023, west_lessons_2024}. Similar case study approaches have also been applied generally in the AI governance literature \citetext{e.g., \citealt{raji_outsider_2022, stein_public_2024}}. Moreover, general-purpose AI systems are general-purpose technologies and present a wide array of risks \cite{slattery_ai_2025, weidinger_taxonomy_2022, critch_tasra_2023, anderljung_frontier_2023}, so it may be useful to examine many different industries embodying different risk models in order to understand AI incidents.\footnote{Various industries have more well-scoped and narrowly-defined incident definitions (see Appendix \ref{sec:Appendix_Example_Defs}) than have been proposed for AI currently \citetext{e.g., \citealp{orssich_defining_2024}}.}

We begin by identifying 22 candidate safety-critical industries using the 11 seed articles outlined in Table \ref{tab:Seed_Articles}, which were tracked as a list in a Google Sheet. The seed articles were identified using the search terms ``safety critical'' and ``incident reporting'' in Google Scholar, backwards snowballing from those results, then purposively excluding articles that did not discuss incident reporting systems or that did not contain a list of safety-critical industries. We conducted a background literature review on Google Scholar using keywords that consisted of the industry names in Table \ref{tab:Industries_List} and the term ``incident reporting.'' 

\begin{table}[!hbt]
    \centering
    \begin{tabularx}{\linewidth}{ X }
        \toprule

        \multirow{2}{*}{}{\textbf{Seed articles}}
        \\
        
        \midrule

        \citet{barach_reporting_2000} 
        \\
        
        \citet{cisa_critical_nodate}
        \\
        
        \citet{koessler_risk_2023}
        \\
        
        \citet{knight_safety_2002} 
        \\
        
        \citet{johnson_software_2002}
        \\
        
        \citet{laplante_software_2017}
        \\
        
        \citet{martins_requirements_2017}
        \\
        
        \citet{raji_outsider_2022}
        \\
        
        \citet{saunders_conceptualising_2015}
        \\
        
        \citet{singh_reliability_2021}
        \\
        
        \citet{turri_why_2023}
        \\

        \bottomrule
    \end{tabularx}
    \caption{Seed articles to identify safety-critical industries}
    \label{tab:Seed_Articles}
\end{table}

Using the inclusion criteria in Table \ref{tab:Industries_Inclusion}, we then filtered the 22 candidate industries down to the nine ultimately selected for inclusion. Once the list of nine case study industries was finalized (Table \ref{tab:Industries_List}), we then updated our background literature review to focus only on those articles relevant to the nine industries selected for inclusion. 

\begin{table}[!ht]
    \centering
    \begin{tabularx}{\linewidth}{ X }
        \toprule
        \textbf{Inclusion criteria}
        \\
        
        \midrule

        Publicly Available: Information about incident reporting systems is publicly available.
        \\

        Well Documented: Incident reporting systems are documented via academic studies, technical documentation, or reports commissioned by organizations responsible for those systems.
        \\

        Common Practice: Industry has incident reporting systems that involve at least two different types of reporting parties.
        \\

        Safety Systems Incident reporting systems do not primarily exist for emergency or crisis response.
        \\

        \bottomrule
    \end{tabularx}
    \caption{Inclusion criteria for safety-critical industries}
    \label{tab:Industries_Inclusion}
\end{table}

\begin{table}[!ht]
    \centering
    \begin{tabularx}{\linewidth}{ X X }
        \toprule
        \textbf{Included} &
        \textbf{Excluded}
        \\

        \midrule
        
        Nuclear energy &
        Financial services
        \\
        
        Civilian aviation &
        Biosecurity
        \\
        
        Pesticides (agriculture) &
        Defense
        \\
        
        Pharmaceuticals &
        Maritime
        \\
        
        Cybersecurity &
        Automotive
        \\
        
        Dams &
        Space
        \\
        
        Rail &
        Oil or (petro)chemical
        \\
        
        Workplace safety &
        Manufacturing
        \\
        
        Medical or healthcare &
        Telecommunications
        \\
        
        &
        (Waste)water systems
        \\
        
        &
        Weapons
        \\
        
        &
        Emergency services
        \\
        
        &
        Energy or electricity
        \\

        \bottomrule
    \end{tabularx}
    \caption{Included and excluded safety-critical industries}
    \label{tab:Industries_List}
\end{table}

\subsection{Development of a Framework for the Institutional Design of Incident Reporting Systems} \label{subsec:Methods_Typologies}

\subsubsection{Framework Development} After identifying the relevant industry case studies, we develop a framework that encompasses key features in the institutional design of AI incident reporting systems. Our framework consists of seven typologies (or dimensions): typologies allow us to generate a common language for discussing a set of observations, identify common characteristics in different cases, synthesize concepts across multiple fields, and perform systematic comparisons \cite{collier_putting_2012, marradi_classification_1990, smith_typologies_2002, wallace_beyond_2006, bailey_typologies_1994}. Methodologically, typologies are also a common and well-accepted method in both AI \citetext{e.g., \citealp{schiff_whats_2020}} and in public policy research \cite{smith_public_2009, smith_typologies_2002}.

To develop our typologies, we adopted a qualitative content analysis methodology similar to that of \citet{ayling_putting_2022}. ``Content analysis is a research technique for making replicable and valid inferences from texts . . . to the contexts of their use'' \cite{krippendorff_content_2019}, and it can help unearth important background information \cite{yanow_qualitative-interpretive_2007} as well as allow for cross-case comparisons \cite{sadovnik_qualitative_2007}. In particular, content analysis is particularly suited to investigating ``[i]nstitutional realities'' \cite{krippendorff_content_2019} and for complementing other methods (here, typologies) to iterate on the scope of data collection \cite{bowen_document_2009}. 

The first step in building our typologies was to examine the documents in our background literature review to identify keywords and themes appearing in the incident reporting literature in other industries (references were stored in the Zotero reference manager, and relevant themes were listed in a Google Sheet), iteratively refining these themes to create the dimensions listed in Table \ref{tab:Framework_Defs}. The criterion for inclusion of each typology was \textit{exhaustiveness} \cite{marradi_classification_1990}, i.e., whether the dimension on which the typology categorizes incident reporting systems is complete (in the sense that it would not be possible to specify an incident reporting system without falling somewhere on the typological dimension).

After identifying the seven typologies, we developed the categories within each typology. A content analysis approach was again adopted, identifying keywords and themes iteratively from the background literature, i.e., from incident reporting systems in other industries. At this stage in the process, we queried both our background literature, as well as the literature on AI incident reporting for validation. There is a rich academic literature studying incident reporting systems in our case study industries, and studying this body of work allowed us to identify key categories and concepts already in use in other contexts.

\subsubsection{Scoping Literature Review} At the start of this project, we also conducted a scoping review of existing and proposed AI incident reporting systems. This review was used to give us an overview of the literature and to help contextualize the framework in Section \ref{sec:Framework} to general-purpose AI.

In September 2023 and again in January 2024, we queried the ACM Full-Text Collection, IEEE Xplore, and arXiv for articles containing, in either full text or any metadata field, both one of the AI keywords and a combination of any two words from different sets of the incident reporting keywords presented in Table \ref{tab:Review_Keywords}. Queries were limited to articles published between January 2010 and September 2023, which yielded a total of 537 articles (54, 358, 125 from each database, respectively). 

The first author then analyzed paper titles and, where necessary, paper abstracts to include/exclude articles based on the criteria in Tables \ref{tab:Review_Inclusion} and \ref{tab:Review_Exclusion}.\footnote{For efficiency, we began by screening by title only. Our search term was broad enough to catch a number of articles that very obviously fell outside our exclusion criteria---in particular many articles that used AI technologies to conduct incident analysis. The title alone was usually sufficient to exclude these articles. We then screened the remaining articles by title and abstract.} After this step, the first author added in articles from the academic and the gray literatures through limited backwards snowballing, and manual searches via Google Search using combinations of the keywords. This procedure yielded a total of $n=81$ articles, and the first author then screened the full text of each article for inclusion based on the rules in Tables \ref{tab:Review_Inclusion} and \ref{tab:Review_Exclusion}, resulting in a final $n=54$ articles.

\begin{table}[!htb]
    \centering
    \small
    \begin{tabularx}{\linewidth}{ l X X }
        \toprule
        \textbf{\makecell{AI \\ Keywords}} & 
        \textbf{\makecell{Incident \\ Keywords}} & 
        \textbf{\makecell{Reporting \\ Keywords}} 
        \\
        
        \midrule
        
        ``artificial intelligence'' & 
        ``incident'' & 
        ``report'' 
        \\
        
        ``AI ethics'' & 
        ``accident'' & 
        ``sharing'' 
        \\
        
        ``AI governance'' & 
        ``hazard'' & 
        ``disclosure'' 
        \\
        
        ``AI policy'' & 
        ``failure'' & 
        \\
        
        ``ethics of AI'' & 
        ``harm'' & 
        \\
        
        ``governance of AI'' & 
        & 
        \\
        
        \bottomrule
    \end{tabularx}
    \caption{Search Terms for Scoping Review}
    \label{tab:Review_Keywords}
\end{table}

\begin{table}[!htbp]
    \centering
    \small
    \begin{tabularx}{\linewidth}{ X }
        \toprule
        
        \textbf{Inclusion Criteria} 
        \\ 
        
        \midrule
        
        Discusses incidents from general-purpose AI systems (at the review stage, we considered any negative event resulting from AI systems to be in-scope)
        \\ 
        
        \midrule
        
        Discusses reporting of such incidents
        \\

        \midrule

        Discusses an existing or proposed system for such reporting
        \\
        
        \bottomrule
    \end{tabularx}
\caption{Inclusion Criteria for Scoping Review}
\label{tab:Review_Inclusion}
\end{table}

\begin{table}[!htb]
    \centering
    \small
    \begin{tabularx}{\linewidth}{ X }
        \toprule
        
        \textbf{Exclusion Criteria} 
        \\ 
        
        \midrule
        
        Applies AI technologies or uses ML methods to answer other research questions (e.g., using ML to facilitate incident reporting systems in a different industries)
        \\ 
        
        \midrule
        
        Uses data from AI incident reports to answer other research questions (e.g., creating a taxonomy of AI incident types)
        \\

        \midrule

        Does not discuss systems for reporting but only discusses the reports themselves
        \\ 
        
        \midrule

        Does not contain a proposal for an AI incident reporting system or discuss an existing AI incident reporting system
        \\ 
        
        \midrule

        Discusses only AI security incidents or AI security vulnerabilities, and does not discuss safety or rights incidents
        \\ 
        
        \bottomrule
    \end{tabularx}
\caption{Exclusion Criteria for Scoping Review}
\label{tab:Review_Exclusion}
\end{table}

Table \ref{tab:Review_Result_Articles} contains a list of included/excluded articles.

\begin{table*}[!htbp]
    \centering
    \small
    \begin{tabularx}{\textwidth}{s X}
        \toprule

        &
        \textbf{Articles}
        \\

        \midrule

        Included ($n = 54$) &
        \citet{AIAAIC, AIID, ajao_safety_2023, ajl_excoded_nodate, anderljung_frontier_2023, anthropic_anthropics_2023, arnold_ai_2021, avid_lifecycle_2024, avin_filling_2021, biden_executive_2023, blumenthal_bipartisan_2023, boine_general_2023, brundage_toward_2020, brundage_lessons_2022, brynjolfsson_big_2023, buolamwini_community_nodate, cac_translation_2023, costanza-chock_who_2022, dsit_emerging_2023, durso_analyzing_2022}; \citetalias{AIA}; \citet{ falco_governing_2021, fli_policymaking_2023, google_comment_2023, govai_consultation_2020, ised_voluntary_2023, kenway_bug_2022, lefcourt_ai_2023, lupo_risky_2023, macrae_close_2014, manheim_building_2023, mcgregor_preventing_2021, mcgregor_data-centric_2023, meta_overview_2023, muehlhauser_12_2023, mulgan_case_2023, nist_ai_2023, nist_rmf_playbook, obrien_deployment_2023, oecd_oecd_2023, openai_model_nodate, plonk_developing_2022, raji_outsider_2022, schmidt_final_2021, schuett_risk_2023, schuett_towards_2023, shevlane_model_2023, shneiderman_human-centered_2022, shoker_confidence-building_2023, shrishak_how_2023, tc_260_translation_2023, the_white_house_ensuring_2023, turri_why_2023}
        \\

        \midrule

        Excluded ($n = 27$) &
        \citet{anthropic_comment_2023, anthropic_reporting_2023, aws_uk_and_ireland_ai_2023, barrett_identifying_2023, bengio_managing_2023, blauth_artificial_2022, cabrera_discovering_2021, coeckelbergh_-responsible_2020, executive_office_of_the_president_preparing_2016, google_deepmind_ai_2023, hoffman_adding_2023, kassab_investigating_2022, li_trustworthy_2023, maas_regulating_2018, mcgregor_indexing_2022, microsoft_microsofts_2023, nedzhvetskaya_role_2022, ogrady_trust_2022, openai_frontier_2023, petkovic_it_2023, raji_fallacy_2022, schuett_defining_2022, schwartz_towards_2022, wei_ai_2022, winfield_robot_2020, yew_penalty_2022}
        \\

        \bottomrule
    \end{tabularx}
    \caption{A complete list of the 81 articles remaining after the initial screen in our scoping review, including the final 54 articles included in our review.}
    \label{tab:Review_Result_Articles}
\end{table*}

\subsection{Review of Case Studies} \label{subsec:Methods_Analysis}

Following the methodology outlined in \citet{raji_outsider_2022}, we apply our framework to review real-world incident reporting systems in the nine selected case study industries, and we generate lessons from these other safety-critical industries for the institutional design of AI incident reporting systems. From our background literature review, we purposively identify a number of real-world incident reporting systems for each case study industry listed in Table \ref{tab:Industries_List} (see Table \ref{tab:Full_Results} for a list of incident reporting systems). We collect a sample of documents that describe the institutional design of each selected incident reporting system; these documents were gathered using Google search and using the search functions on the websites of the organizations that operate each incident reporting system. Keywords used in these searches and other inclusion criteria for this sample were generated from the methodology in \citet{ayling_putting_2022} and are presented in Table \ref{tab:Methods_Sample_Selection}, and the final sample of documents is in Table \ref{tab:Results_References}.

Using this sample of documents, we categorize each of the incident reporting systems identified along the seven dimensions of our framework. The results of this categorization are presented in Appendix \ref{sec:Appendix_Full_Results}. Finally, we examine patterns in these systems and synthesize these results and the background literature to present lessons for the institutional design of AI incident reporting. These lessons are generated by identifying best practices in or common practices between our case study industries, as evidenced by the academic literature and by our categorization, and the methodology for generating lessons was informed by \citet{byman_writing_2024}. Generally, we identified best practices or points of consensus in institutional design for incident reporting systems across our case study industries under the assumption that the goals of such systems are learning and accountability (see Table \ref{tab:Def_Goals}), and areas where design choices differed are indicated clearly in Sections \ref{sec:Recs} and \ref{sec:Recs}.

\begin{table*}[!bhtp]
    \centering
    \begin{tabularx}{\textwidth}{ X X X }
        \toprule
        
        \textbf{Filter Type} &
        \textbf{Inclusion Criteria} &
        \textbf{Exclusion Criteria}
        \\

        \midrule

        Document Type &
        Technical documentation, federal/state regulations, official or unofficial agency guidance, legislation or compiled US code, webpages &
        News reports, op-eds, audio-visual materials
        \\

        Keywords &
        ``Documentation'', ``operating manual'', ``guidance'', ``report'', ``circular'', ``advisory'' [name of incident reporting system], ``site:[website of operating organization]'' (for Google search) &
        \\

        Author &
        Organizations operating the system (e.g., government agency, agency staff/officials, or commissioned experts), academics &
        Non-academic commentators, industry organizations
        \\

        Language &
        English &
        \\

        Availability &
        Public, online
        \\

        \bottomrule
    \end{tabularx}
    \caption{Inclusion criteria for sample document selection}
    \label{tab:Methods_Sample_Selection}
\end{table*}

\clearpage

\section{Appendix: Full Definitions for Incident Reporting Framework} \label{sec:Appendix_Framework_Definitions}

This Appendix contains full definitions for all dimensions and categories in our framework for the institutional design of incident reporting systems. Our framework contains seven typologies (dimensions) for understanding the institutional structure of incident reporting systems; each dimension consists of a number of discrete categories.

Table \ref{tab:Framework_Defs} (reproduced below from Section \ref{sec:Framework}) presents an overview of the seven typologies in our framework. Additional definitions for each typology are provided in each subsection below. \footnote{Note that with the exception of the anonymity axis, the categories in most axes of our framework are not mutually exclusive. Although we treat learning and accountability as mutually exclusive in the goal axis, they may not necessarily be so in every system in practice.}

\setcounter{table}{0}
\begin{table}[!thb]
    \centering
    \small
    \begin{tabularx}{\linewidth}{ >{\hsize=0.30\hsize}X >{\hsize=0.70\hsize}X }
        \toprule
         \textbf{Dimension} & \textbf{Definition} \\
         \midrule
         
         Policy Goal & The policy aim that the incident reporting system attempts to achieve: safety \textit{learning} or \textit{accountability} for harm. \\ \midrule
         
         Actors Submitting \& Receiving Reports & Possible actors include users, victims of harm, third-party individuals or organizations, companies, industry employees, and governments at various levels. See Appendix \ref{subsec:Framework_Actors} for details. \\ \midrule
         
         Type of Incidents \newline Reported & The type of incident reported in the system: \textit{safety}, \textit{rights}, or \textit{security} incidents. \\ \midrule
         
         Level of Risk \newline Materialization & The level of risk materialization reported in the system: \textit{hazards}, \textit{situations}, \textit{near misses}, or \textit{harm events}. See Figure \ref{fig:Risk_Materialization}. \\ \midrule
         
         Enforcement of \newline Reporting & The procedures that incentivize actors to submit incident reports: \textit{voluntary} or \textit{mandatory} (by law). \\ \midrule
         
         Anonymity of \newline Reporters & The actors who have access to the reporter's identity: \textit{open}, \textit{confidential}, or \textit{anonymous}. \\ \midrule
         
         Post-Reporting \newline Actions & The actions taken by the party receiving incident reports, after reports are received: \textit{information sharing}, \textit{information disclosure}, \textit{audit}, or \textit{regulatory action}. \\
         
        \bottomrule
    \end{tabularx}
    \caption[]{Seven dimensions of the institutional design of incident reporting systems. Options for each dimension are italicized. Reproduced from Section \ref{sec:Framework}.}
\end{table}
\setcounter{table}{11}

\clearpage

\subsection{Policy Goal} \label{subsec:Framework_Goal}

System objectives can determine design choices. The \textit{goal} of an incident reporting system is the policy aim the system attempts to achieve. At the highest level of generality, systems generally have one of two goals \cite{WHO_2005, mills_promise_2010}: learning or accountability. These are defined in Table \ref{tab:Def_Goals}.

Reporting systems can naturally serve other goals as well. For instance, nearly every reporting system by definition aims to achieve transparency or visibility. Systems may also aim at facilitating (legal) compliance, deterrence, or incident/emergency response. We primarily discuss learning and accountability in this article because these goals are well-studied in the general incident reporting literature and because systems aiming primarily to achieve other goals (e.g., emergency response) entail sharply different design considerations. 

\begin{table}[!htb]
    \centering
    \small
    \begin{tabularx}{\linewidth}{ l X }
        \toprule
         \textbf{Category} & \textbf{Definition} 
         \\
         \midrule

         Learning & 
         The system aims to reduce or prevent future harms from occurring by ensuring that stakeholders adapt from past incidents and improve products, systems, and processes.
         \\

         Accountability &
         The system aims to hold actors responsible for past harms by imposing corrective actions or penalties, and/or to reduce or prevent future harms by deterring unsafe or negligent behavior.
         \\
         
        \bottomrule
    \end{tabularx}
    \caption{Definitions for the policy goals dimension}
    \label{tab:Def_Goals}
\end{table}

\clearpage

\subsection{Actors Submitting \& Receiving Reports} \label{subsec:Framework_Actors}

\setcounter{figure}{1}
\begin{figure*}[!hbt]
    \centering
    \includegraphics[width=0.96\textwidth]{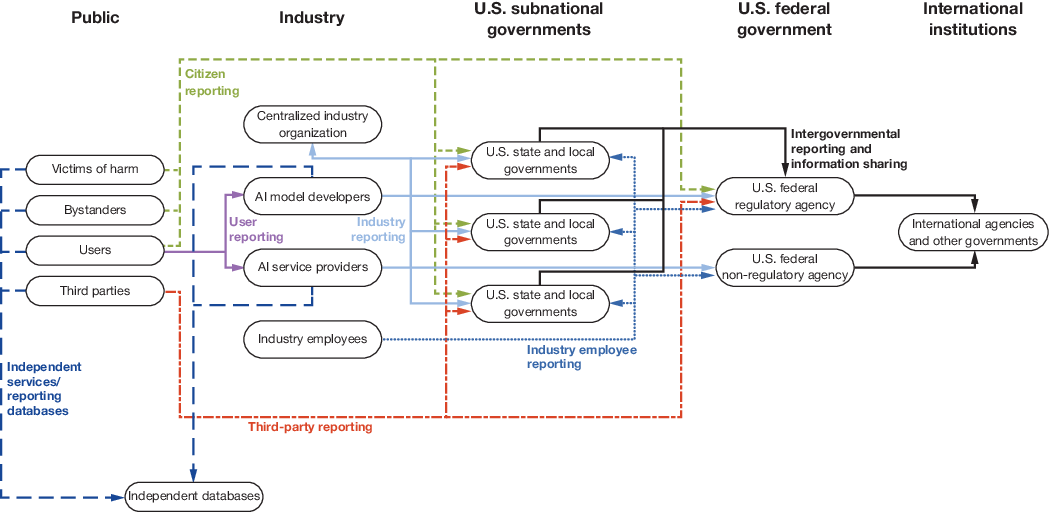}
    \caption[]{Visualization of the actors dimension, i.e., incident reporting systems involving different actors as defined in Table \ref{tab:Def_Actors}. Note: each category of reporting system in Table \ref{tab:Def_Actors} is represented in the figure by a different color and dashed line (as labelled). Dashed lines are for accessibility only and do not represent additional distinctions beyond the colors and labels. US state and local governments are repeated to emphasize that there are many such governments, all of which may operate independent reporting systems in the same reporting regime in parallel (in contrast, with the notable exception of cybersecurity, systems in most incident reporting regimes are not normally run by multiple different agencies at the federal/international levels). Reproduced from Section \ref{sec:Framework}.}
\end{figure*}

Multiple parties may possess varying types of incident information at different levels of granularity, and they may have different incentives to report or keep silent after an issue or incident surfaces. The \textit{actors} in a reporting system refers to the parties that are submitting and receiving incident reports. While there are many possible configurations of actors in theory, we identify seven general categories of systems that exist in practice, as defined in Table \ref{tab:Def_Actors} and visualized in Figure \ref{fig:Framework_Actors} (each category consists of a reporting party and a receiving party).\footnote{Some configurations of actors are nonsensical in the context of reporting, e.g., a government reporting to an individual. But these same configurations may be valid in other contexts such as information sharing or information disclosure (e.g., a government disclosing information to the public).}

One further consideration for any systems involving government actors is whether the agency receiving reports has \textit{regulatory authority} over the relevant industry; Section \ref{subsec:Recs_Actors} discusses the importance of this factor to incident reporting systems.

\begin{table*}[!htp]
    \centering
    \small
    \begin{tabularx}{\linewidth}{ X X }
        \toprule
         \textbf{Category} & \textbf{Definition} 
         \\
         \midrule

         Independent Service or Independent Database &
         A private organization operates a database or service where reports are either submitted by the public or actively collected by the organization.
         \\

         Citizen Reporting System &
         Members of the public submit reports to a government agency \cite{frase_ai_2023}
         \\

         User Reporting System &
         System users submit reports to a service provider or product developer.
         \\

         Third-Party Reporting System &
         Third parties submit reports to a government agency. Third parties are individuals or organizations who are neither the product user nor the victim of harm but gains knowledge of the incident through other means (e.g., third-party testers/developers, red-teamers or model evaluators, treating physicians, etc.). The government agency receiving reports could be regulatory or non-regulatory.
         \\

         Industry Employee Reporting System &
         Employees in or adjacent to the industry submit reports to a government agency or (more rarely) to a professional organization.
         \\

         Industry Reporting System &
         Industry organizations submit reports to a government agency. More rarely, reports may be submitted to a centralized industry group; in the AI context, we usually refer to government-related systems. 
         \\

         Inter-Governmental Reporting System &
         A governmental actor submits reports to or shares information with another governmental actor (generally from a lower to a higher jurisdiction; e.g., state to federal or federal to international).
         \\
         
        \bottomrule
    \end{tabularx}
    \caption{Definitions for the actors dimension}
    \label{tab:Def_Actors}
\end{table*}

\clearpage

\subsection{Type of Incidents Reported} \label{subsec:Framework_Incident_Type}

The proper response to an issue or incident may be highly dependent on the victims and the forms of (potential) harms. To capture these differences, we delineate three general \textit{incident types}, i.e., broad types of harms. These distinctions are presented in Table \ref{tab:Def_Type_of_Incident} and have been recognized in policy documents such as \citep{young_memorandum_2024, biden_executive_2023}.

\begin{table}[!htb]
    \centering
    \small
    \begin{tabularx}{\linewidth}{ l X }
        \toprule
         \textbf{Category} & \textbf{Definition} 
         \\
         \midrule

         Safety &
         Issues or incidents in which a system could or did cause harm to a person, property, or environment external to the system.
         \\

         Rights &
         Issues or incidents in which a system could or did cause a violation of individual legal, civil, or human rights or liberties.
         \\

         Security &
         Issues or incidents in which an individual external to the system could or did exploit a system vulnerability to cause harm to, failure of, or misbehavior of a system (see \citet{nist_vulnerability_nodate}).
         \\
         
        \bottomrule
    \end{tabularx}
    \caption{Definitions for the type of incident dimension}
    \label{tab:Def_Type_of_Incident}
\end{table}

This article is primarily concerned with designing systems for reporting AI safety and rights incidents as opposed to AI security incidents (see Section \ref{subsec:Recs_Incident_Type}). Security incidents are included in this typology for completeness and because some reporting systems we reviewed are type-agnostic. 

\clearpage

\subsection{Level of Risk Materialization} \label{subsec:Scope_Risk}

\setcounter{figure}{0}
\begin{figure*}[!hbtp]
    \centering
    \includegraphics[width=0.75\textwidth]{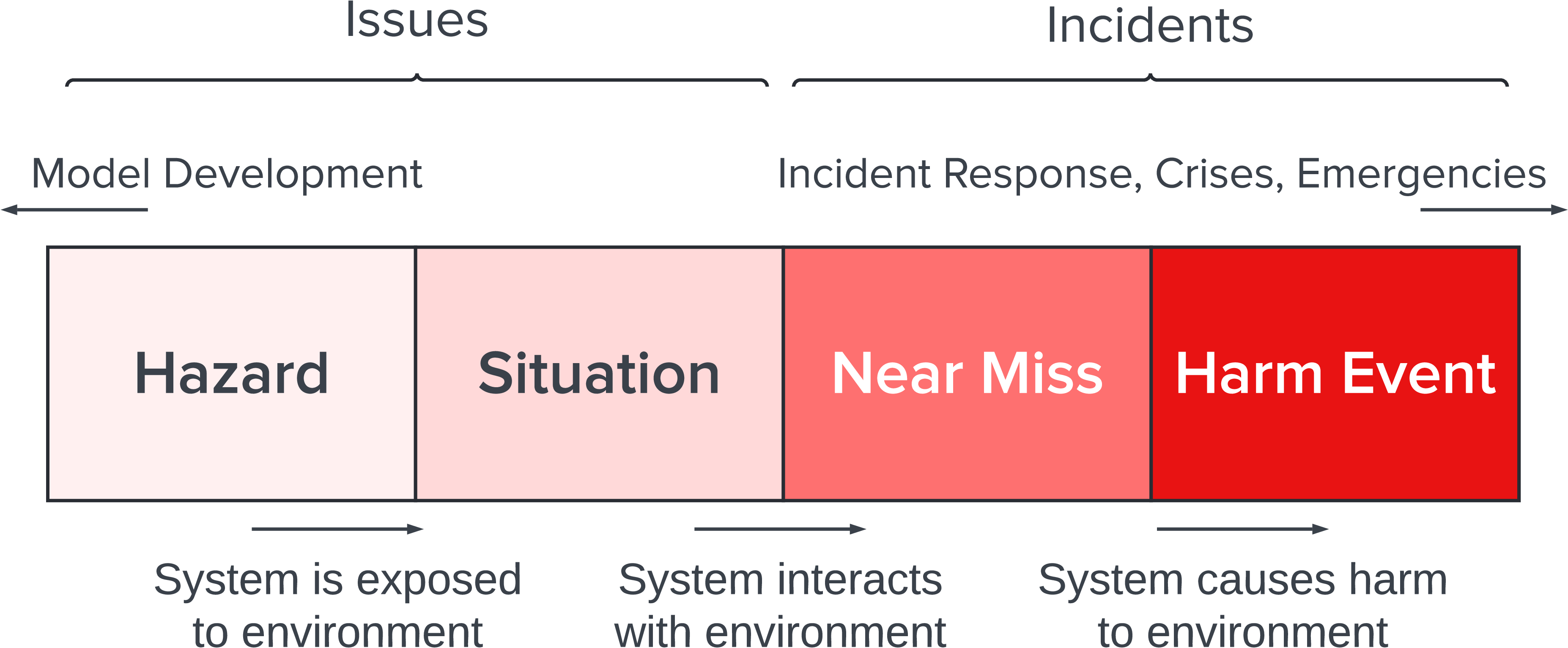}
    \caption[]{Visualization of the level of risk materialization dimension, i.e., the lifecycle of an (AI) incident as defined in Table \ref{tab:Def_Scope_of_Risk}. Reproduced from Section \ref{sec:Framework}.}
\end{figure*}

Whether or not harm has been caused---and how close harm is to occurring---can also affect issue or incident response.\footnote{Incident response consists of the actions taken by a system operator to mitigate the effects or harms of an incident \cite{nist_incident_nodate, uk_hpa_chemical_2009}.} A hazard (defined in Table \ref{tab:Def_Scope_of_Risk}) in an AI model may only require a developer to take internal steps to re-train or fine-tune the model, whereas a harm event (defined in Table \ref{tab:Def_Scope_of_Risk}) could involve external parties and audits. \textit{Level of risk materialization} is a spectrum that measures how close a reported event is to resulting in harm. Based on a review of the safety science literature, we conceptualize level of risk materialization categories as progressive stages in the incident lifecycle, from risk inception to harm occurrence.\footnote{Note that incident response occurs after an incident. In addition, an incident could escalate into a crisis or emergency; see \citealt{gor_what_2025})}. Figure \ref{fig:Risk_Materialization} (reproduced from Section \ref{sec:Framework}) illustrates this lifecycle, and key definitions are presented in Table \ref{tab:Def_Scope_of_Risk}.\footnote{We avoid use of the term ``vulnerabilities'' when referring to AI hazards or AI situations. In cybersecurity, a vulnerability traditionally refers to conditions that can be exploited by (malicious) outside actors \cite{nist_vulnerability_nodate}---these are \textit{security} hazards or security risks, not necessarily \textit{safety} hazards or safety risks \cite{qi_ai_2024}. Moreover, AI safety or rights incidents do not necessarily occur as a result of external exploitation of an AI system. The term ``flaw'' has recently also been used to refer to hazards and situations (collectively, AI issues).}

\setcounter{table}{1}
\begin{table*}[!htb]
    \centering
    \small
    \begin{tabularx}{\linewidth}{ l X }
        \toprule
        \textbf{Category} & \textbf{Definition} 
        \\
        \midrule

        Hazard &
        A set of conditions or capabilities in a system that enables it to cause harm \cite{hendrycks_overview_2023, leveson_engineering_2011, birnbaum_research_2016, nasa_nasa_2004, ieee_ieee_1993, hutiri_not_2024}. An AI hazard could be, e.g., specific model capabilities and could be discovered through, e.g., red teaming, audits, or model evaluations.
        \\
        
        Situation (or hazardous situation) &
        A hazard that has been exposed to an environment in which it could cause harm. Situations are ``incidents waiting to happen.'' \citetext{See the discussion and definitions in \citet{leveson_engineering_2011, ieee_ieee_1993, hendrycks_x-risk_2022, hutiri_not_2024, iso_guide_2014}}. An AI situation could be, e.g., a publicly deployed AI model with dangerous capabilities (hazards).
        \\
        
        Near miss (or close call) &
        An event where a system's interactions with its environment could have caused harm but did not, generally due to some circumstances beyond the system's control. An AI near miss could, e.g., be an event in which an AI system generated a political deepfake for a user, but social media platforms removed such content before it was spread. See \citet{ec_code_2025}.
        \\
        
        Harm event &
        An event where a system did cause harm. An AI harm event could be, e.g., an event in which an AI system generated an offense cyberweapon that was then deployed by malicious actors in an attack.
        \\
        
        \bottomrule
    \end{tabularx}
    \caption[]{Definitions for the level of risk materialization dimension. Reproduced from Section \ref{sec:Framework}.}
\end{table*}
\setcounter{table}{14}

These working definitions are consistent with the safety science literature\footnote{For instance, see the discussions in \citet{kornecki_aviation_2015} and \citet{leveson_engineering_2011}.} and extend models developed by others in the context of AI \cite{hoffman_adding_2023, hendrycks_overview_2023, oecd_expert_group_on_ai_incidents_oecd_2023, agarwal_structural_2007, mcgregor_indexing_2022}.\footnote{Our working definitions are also largely independent of the AI model lifecycle. While some sources seem to consider safety incidents to be events that occur only after deployment of a model, not all safety incidents arise post-deployment. Some risks could exist just by mere possession of the model, e.g., autonomous replication \cite{anthropic_anthropics_2023}; these risks could materialize at any point in a training or testing phase of the model. Incidents could also occur where an AI model causes harm to a company employee during training or testing, even when the model has not been deployed to the broader public.} They also represent a modification and extension of other work in the AI governance literature \cite{hoffman_adding_2023, hendrycks_overview_2023, oecd_expert_group_on_ai_incidents_oecd_2023, orssich_defining_2024}. 

Our typology here has two significant departures from other AI governance literature. \textbf{In particular, our definitions \textit{are not} consistent with those proposed by the OECD in \citet{orssich_defining_2024}, nor are they consistent with some AI governance literature such as \citet{gor_what_2025}.} Specifically, we delineate between:

\begin{enumerate}
    \item \textit{Hazards} vs. \textit{Situations}: this distinction has been recognized in traditional safety science \cite{iso_guide_2014} but has been under-theorized in AI governance. In the context of AI, this distinction follows from the distinction made in \citet{weidinger_sociotechnical_2023} between \textit{model capabilities} and \textit{human interaction}: risks only materialize once the system has interacted with an external actor or environment. Note that with our working definitions, the interaction need not necessarily be with a \textit{human}.
    \item \textit{Issues} vs. \textit{Incidents}: We follow \citet{hoffman_adding_2023} in referring to hazards and situations as \textit{issues},\footnote{Hazards and situations are not events, and thus, our working definition does not consider them to be incidents even if some literature does categorize them as such. However, the term ``incident reporting'' is frequently used to refer to systems that permit reporting of hazards or situations \cite{havinga_hazard_2021}. It may sometimes be useful to distinguish between these types of information (see Section \ref{subsec:Recs_Scope_Risk}).} and to near misses and harm events as \textit{incidents}. The distinction is that issues are \textit{conditions} in systems or environments that are prerequisites to harm occurring, while incidents are \textit{events} in which harm could have or did occur. Note that recent work has also introduced the terminology of AI ``flaws'' \citetext{e.g., \citealt{longpre_position_2025}}, which appear to be synonymous with AI issues and which seem to cover issues of all types (safety, rights, or security).\footnote{\citet{longpre_position_2025} writes: ``we refer to AI \textit{flaws}, broadly referring to conditions in a system that lead to undesirable effects or policy violations. We intentionally define AI flaws more broadly than traditional software security vulnerabilities to reflect the range of potential sociotechnical risks with GPAI systems.'' This definition seems to include both hazards and situations.}
\end{enumerate}

The issues/incidents distinction has one major practical implication: the methods of detection for incidents and issues are distinct. Issues that exist in AI systems and result in incidents can be detected by their resulting (near) harm, but issues that have \textit{not} yet resulted in (near) harm can only detected through other methods---e.g., red-teaming, model evaluations, etc. On the other hand, incidents may \textit{only} be detected after a system with an issue has interacted with its operating environment in such a way as to (nearly) result in harm (but note that detection of the existence of an incident is distinct from the attribution of that incident to AI systems generally or to a specific AI system). Issues and incidents may be thus more easily detected by different types of actors (see Section \ref{subsec:Recs_Scope_Risk}). 

Finally, note that ``harm'' should be understood broadly: victims of harm could be users, people, the natural environment, property, or systems \cite{hoffman_adding_2023, orssich_defining_2024}. The types of harms that general-purpose AI systems may cause are also broader than in the traditional safety science literature and include (but are not limited to) physical, emotional, psychological harms as well as civil rights or human rights violations \cite{hoffman_adding_2023}.\footnote{See also the discussion in \citet{diberardino_algorithmic_2024}.} 

\clearpage

\subsection{Enforcement of Reporting} \label{subsec:Framework_Enforcement}

Incident reporting systems are effective only with buy-in from the parties that submit reports. \textit{Enforcement} refers to the procedures for encouraging actors to submit reports to an incident reporting system \citetext{see also \citealt{kolt_responsible_2024}}); our typology for enforcement is presented in Table \ref{tab:Def_Enforcement}.

\begin{table}[!htb]
    \centering
    \small
    \begin{tabularx}{\linewidth}{ l X }
        \toprule
         \textbf{Category} & \textbf{Definition} 
         \\
         \midrule

         Voluntary &
         Participants generally face no formal consequences after reporting or failing to report an incident. Participants may receive positive incentives to report, e.g., access to information, technical assistance, or financial incentives.
         \\
         
        Mandatory &
        A government or other authority requires participants to report incidents, and failure to report may result in civil or criminal penalties.
        \\
         
        \bottomrule
    \end{tabularx}
    \caption{Definitions for the enforcement dimension}
    \label{tab:Def_Enforcement}
\end{table}

\clearpage

\subsection{Anonymity of Reporters} \label{subsec:Framework_Anonymity}

Actors may be reluctant to report incidents if they believe that they or their affiliates may suffer  (reputational, financial, legal, or regulatory) repercussions as a result of filing a report. \textit{Anonymity} refers to which other actors have knowledge of reporters' identities \cite{johnson_failure_2003}, and our typology for levels of anonymity is presented in Table \ref{tab:Def_Anonymity}.

\begin{table}[!htb]
    \centering
    \small
    \begin{tabularx}{\linewidth}{ l X }
        \toprule
        \textbf{Category} & \textbf{Definition} 
        \\
        \midrule

        Open &
        The identity of the reporting party is known to all parties to whom incident information is made available.
        \\
        
        Confidential &
        The identity of the reporting party is known only to some trusted actors.
        \\
        
        Anonymous (or de-identified) &
        The identity of the reporting party is known only to the reported party  themselves.
        \\
         
        \bottomrule
    \end{tabularx}
    \caption{Definitions for the anonymity dimension}
    \label{tab:Def_Anonymity}
\end{table}

\clearpage

\subsection{Post-Reporting Actions} \label{subsec:Framework_Post-Reporting}

In order for incident reporting systems to support learning or accountability, stakeholders must take action after incidents have been reported.  \textit{Post-reporting actions} refers to actions taken by the actors receiving reports, and our typology for post-reporting actions is presented in Table \ref{tab:Def_Post-Reporting}.

\begin{table}[!htb]
    \centering
    \small
    \begin{tabularx}{\linewidth}{ l X }
        \toprule
        \textbf{Category}
        \textbf{Definition} 
        \\
        \midrule

        Information Sharing &
        Information in the incident report is made accessible to other relevant, trusted actors.
        \\
        
        Information Disclosure &
        Information in the incident report is released to the public or to affected parties.
        \\
        
        Audit (or investigation) &
        An analysis is conducted to verify incident information, identify the causes of incidents, or examine industry actors' practices or legal compliance. Audits and investigations could be punitive or non-punitive, and they could be conducted both by industry actors themselves or by external bodies.
        \\
        
        Regulatory Action &
        A system operator or developer could suffer negative consequences as a direct result of the report (e.g., fines, criminal penalties, etc.).
        \\
         
        \bottomrule
    \end{tabularx}
    \caption{Definitions for the post-reporting actions dimension}
    \label{tab:Def_Post-Reporting}
\end{table}

We use the terms ``reporting,'' ``sharing,'' and ``disclosure'' to refer to systems that are conceptually distinct, though they are often used interchangeably in the literature (see Appendix \ref{subsec:Review_Literature}). Although the boundaries between these concepts is somewhat fuzzy in practice, Table \ref{tab:Reporting_Def} outlines some typical distinguishing features of each type of system;\footnote{Table \ref{tab:Reporting_Def} is adapted from \citet{norton_providing_1996}. The features are adapted from uses of the terms ``reporting,'' ``sharing,'' and ``disclosure'' in the safety science and financial regulation literatures \cite{wolf_error_2008, iedema_anatomy_2012, dumay_critical_2016, boot_many_2001, greiling_accountability_2010}.} these features provide general guidelines for when to use the ``reporting,'' ``sharing,'' and ``disclosure'' terminology. Table \ref{tab:Full_Results} includes primarily reporting systems as defined in Table \ref{tab:Reporting_Def}, as well as some sharing systems (as marked) for completeness.

\begin{table}[!hbt]
    \centering
    \begin{tabularx}{\linewidth}{ X X X } 
        \toprule
        
        \textbf{Sharing} & 
        \textbf{Reporting} & 
        \textbf{Disclosure} 
        \\ 
        \midrule 
        
        Bi- or Multi-directional & 
        Uni-directional & 
        Uni-directional
        \\ 
        \midrule
         
        Many-to-many & 
        One-to-one &
        One-to-many
        \\  
        \midrule
        
        Non-hierarchical &  
        Hierarchical & 
        Hierarchical
        \\  
        \midrule
        
        Structured or Unstructured & 
        Structured & 
        Structured\\ 
        \midrule
        
        Formal or \newline Informal &
        Formal & 
        Formal
        \\ 
        
        \bottomrule
    \end{tabularx}
    \caption{Reporting vs. Sharing vs. Disclosure}
    \label{tab:Reporting_Def}
\end{table}

\clearpage

\section{Appendix: Full Results from Review of Case Studies} \label{sec:Appendix_Full_Results}

This Appendix presents full results from our review of incident systems in case study industries, which are contained in Table \ref{tab:Full_Results}. Citations for the results in Table \ref{tab:Full_Results} are provided in Table \ref{tab:Results_References}.

Note that Table \ref{tab:Full_Results} contains reviews of specific systems from only seven of our nine case study industries. Workplace safety and medical/healthcare (as distinct from pharmaceuticals) are not included below because the incident reporting regimes for these industries are highly decentralized and sometimes irregular (because much reporting is done by individual employers, hospitals, etc.). These case studies are nevertheless discussed in text where appropriate, and we included literature on incident reporting from these both of these industries in our literature review.

\setcounter{table}{2}
\begin{table*}[p]
    \centering
    \small
    \begin{tabularx}{\textwidth}{ m{0.25cm} s X s s s s s s s X s s s X } %
        \toprule

        \multicolumn{1}{c}{} &
        \multicolumn{1}{c}{\textbf{Goal}} &
        \multicolumn{1}{c}{\textbf{Actors}} &
        \multicolumn{3}{c}{\textbf{Incident Type}} &
        \multicolumn{4}{c}{\textbf{Risk Materialization}} &
        \multicolumn{1}{c}{\textbf{Enforcement}} &
        \multicolumn{3}{c}{\textbf{Anonymity}} &
        \multicolumn{1}{c}{\textbf{Post-Reporting}} 
        \\

        \multicolumn{1}{c}{} &
        \multicolumn{1}{c}{Learning?} &
        \multicolumn{1}{c}{} &
        \multicolumn{1}{c}{Saf} &
        \multicolumn{1}{c}{Rig} &
        \multicolumn{1}{c}{Sec} &
        \multicolumn{1}{c}{H} &
        \multicolumn{1}{c}{S} &
        \multicolumn{1}{c}{NM} &
        \multicolumn{1}{c}{HE} &
        \multicolumn{1}{c}{Mandatory?} &
        \multicolumn{1}{c}{O} &
        \multicolumn{1}{c}{C} &
        \multicolumn{1}{c}{A} &
        \multicolumn{1}{c}{} 
        \\

        \midrule

        \multirow{12}{*}{\rotatebox[origin=c]{90}{Nuclear}} &  
        \multicolumn{14}{l}{\textit{Nuclear Regulatory Commission (NRC), Safety Hotline}}
        \\ 

        & 
        &  
        \multicolumn{1}{c}{Citizen} &
        \multicolumn{1}{c}{\scalebox{1.25}{$\bullet$}} &
        \multicolumn{1}{c}{\scalebox{1.25}{$\bullet$}} &
        \multicolumn{1}{c}{\scalebox{1.25}{$\bullet$}} &
        \multicolumn{1}{c}{\scalebox{1.25}{$\bullet$}} &
        \multicolumn{1}{c}{\scalebox{1.25}{$\bullet$}} &
        \multicolumn{1}{c}{\scalebox{1.25}{$\bullet$}} &
        \multicolumn{1}{c}{\scalebox{1.25}{$\bullet$}} &
        &
        &
        \multicolumn{1}{c}{\scalebox{1.25}{$\bullet$}} &
        &
        \multicolumn{1}{c}{A; RA}
        \\

        &  
        \multicolumn{14}{l}{\textit{Nuclear Regulatory Commission (NRC), Statutory Reporting Requirements and Operations Center}}
        \\

        & 
        &
        \multicolumn{1}{c}{Company} &  
        \multicolumn{1}{c}{\scalebox{1.25}{$\bullet$}} &
        &
        \multicolumn{1}{c}{\scalebox{1.25}{$\bullet$}} &
        \multicolumn{1}{c}{\scalebox{1.25}{$\bullet$}} &  
        \multicolumn{1}{c}{\scalebox{1.25}{$\bullet$}} &
        \multicolumn{1}{c}{\scalebox{1.25}{$\bullet$}} &
        \multicolumn{1}{c}{\scalebox{1.25}{$\bullet$}} & 
        \multicolumn{1}{c}{\scalebox{1.25}{$\bullet$}} & 
        \multicolumn{1}{c}{\scalebox{1.25}{$\bullet$}} &
        &
        &
        \multicolumn{1}{c}{ID; RA}
        \\

        &  
        \multicolumn{14}{l}{\textit{International Atomic Energy Agency (IAEA), Incident Reporting System (IRS)}} 
        \\ 

        &
        \multicolumn{1}{c}{\scalebox{1.25}{$\bullet$}} &
        \multicolumn{1}{c}{Inter-Gov.} 
        &
        \multicolumn{1}{c}{\scalebox{1.25}{$\bullet$}} &
        &
        \multicolumn{1}{c}{\scalebox{1.25}{$\bullet$}} &
        \multicolumn{1}{c}{\scalebox{1.25}{$\bullet$}} &
        \multicolumn{1}{c}{\scalebox{1.25}{$\bullet$}} &
        \multicolumn{1}{c}{\scalebox{1.25}{$\bullet$}} &
        \multicolumn{1}{c}{\scalebox{1.25}{$\bullet$}} &
        &
        \multicolumn{1}{c}{\scalebox{1.25}{-}} &
        \multicolumn{1}{c}{\scalebox{1.25}{-}} &
        \multicolumn{1}{c}{\scalebox{1.25}{-}} &
        \multicolumn{1}{c}{IS}
        \\

        &  
        \multicolumn{14}{l}{\textit{International Atomic Energy Agency (IAEA), Fuel Incident Notification and Analysis System (FINAS)}}
        \\ 

        &
        \multicolumn{1}{c}{\scalebox{1.25}{$\bullet$}} &
        \multicolumn{1}{c}{Inter-Gov.} &
        \multicolumn{1}{c}{\scalebox{1.25}{$\bullet$}} &
        &
        \multicolumn{1}{c}{\scalebox{1.25}{$\bullet$}} &
        \multicolumn{1}{c}{\scalebox{1.25}{$\bullet$}} &
        \multicolumn{1}{c}{\scalebox{1.25}{$\bullet$}} &
        \multicolumn{1}{c}{\scalebox{1.25}{$\bullet$}} &
        \multicolumn{1}{c}{\scalebox{1.25}{$\bullet$}} &
        &
        \multicolumn{1}{c}{\scalebox{1.25}{-}} &
        \multicolumn{1}{c}{\scalebox{1.25}{-}} &
        \multicolumn{1}{c}{\scalebox{1.25}{-}} &
        \multicolumn{1}{c}{IS}
        \\

        &  
        \multicolumn{14}{l}{\textit{International Atomic Energy Agency (IAEA), Incident Reporting Systems for Research Reactors (IRSRR)}}
        \\ 

        &
        \multicolumn{1}{c}{\scalebox{1.25}{$\bullet$}} &
        \multicolumn{1}{c}{Inter-Gov.} &
        \multicolumn{1}{c}{\scalebox{1.25}{$\bullet$}} &
        &
        \multicolumn{1}{c}{\scalebox{1.25}{$\bullet$}} &
        \multicolumn{1}{c}{\scalebox{1.25}{$\bullet$}} &
        \multicolumn{1}{c}{\scalebox{1.25}{$\bullet$}} &
        \multicolumn{1}{c}{\scalebox{1.25}{$\bullet$}} &
        \multicolumn{1}{c}{\scalebox{1.25}{$\bullet$}} &
        &
        \multicolumn{1}{c}{\scalebox{1.25}{-}} &
        \multicolumn{1}{c}{\scalebox{1.25}{-}} &
        \multicolumn{1}{c}{\scalebox{1.25}{-}} &
        \multicolumn{1}{c}{IS}
        \\

        &  
        \multicolumn{14}{l}{\textit{James Martin Center for Nonproliferation Studies (CNS), Global Incidents and Trafficking Database}}
        \\ 

        &
        \multicolumn{1}{c}{\scalebox{1.25}{$\bullet$}} &
        \multicolumn{1}{c}{Indep. Db.} & 
        \multicolumn{1}{c}{\scalebox{1.25}{$\bullet$}} & 
        &
        & 
        & 
        \multicolumn{1}{c}{\scalebox{1.25}{$\bullet$}} & 
        \multicolumn{1}{c}{\scalebox{1.25}{$\bullet$}} & 
        \multicolumn{1}{c}{\scalebox{1.25}{$\bullet$}} & 
        & 
        &
        \multicolumn{1}{c}{\scalebox{1.25}{$\bullet$}} & &
        \multicolumn{1}{c}{ID} 
        \\

        \midrule

         \multirow{14}{*}{\rotatebox[origin=c]{90}{Civilian Aviation}} &  
        \multicolumn{14}{l}{\textit{National Transportation Safety Board (NTSB)}}
        \\

        &
        \multicolumn{1}{c}{\scalebox{1.25}{$\bullet$}} &
        \multicolumn{1}{c}{Company} & 
        \multicolumn{1}{c}{\scalebox{1.25}{$\bullet$}} & 
        &  
        &  
        &  
        & 
        & 
        \multicolumn{1}{c}{\scalebox{1.25}{$\bullet$}} & 
        \multicolumn{1}{c}{\scalebox{1.25}{$\bullet$}} & 
        \multicolumn{1}{c}{\scalebox{1.25}{$\bullet$}} &  
        &  
        &
        \multicolumn{1}{c}{ID; A}
        \\

        &  
        \multicolumn{14}{l}{\textit{Federal Aviation Administration (FAA), Hotline}}
        \\ 

        & 
        &
        \multicolumn{1}{c}{Citizen} &  
        \multicolumn{1}{c}{\scalebox{1.25}{$\bullet$}} &
        \multicolumn{1}{c}{\scalebox{1.25}{$\bullet$}} &
        &  
        \multicolumn{1}{c}{\scalebox{1.25}{$\bullet$}} &
        \multicolumn{1}{c}{\scalebox{1.25}{$\bullet$}} &
        \multicolumn{1}{c}{\scalebox{1.25}{$\bullet$}} &
        \multicolumn{1}{c}{\scalebox{1.25}{$\bullet$}} &
        &
        &
        \multicolumn{1}{c}{\scalebox{1.25}{$\bullet$}} &
        &
        \multicolumn{1}{c}{IS; A; RA}
        \\

        &  
        \multicolumn{14}{l}{\textit{Federal Aviation Administration (FAA), Voluntary Safety Reporting Program (VSRP)}} 
        \\ 

        & 
        \multicolumn{1}{c}{\scalebox{1.25}{$\bullet$}} &  
        \multicolumn{1}{c}{Employee} & 
        \multicolumn{1}{c}{\scalebox{1.25}{$\bullet$}} & 
        & 
        & 
        \multicolumn{1}{c}{\scalebox{1.25}{$\bullet$}}& 
        \multicolumn{1}{c}{\scalebox{1.25}{$\bullet$}} &
        &
        &
        &
        &
        \multicolumn{1}{c}{\scalebox{1.25}{$\bullet$}} &
        &
        \multicolumn{1}{c}{IS; A; RA}
        \\

        &  
        \multicolumn{14}{l}{\textit{Federal Aviation Administration (FAA), Aviation Safety Action Program (ASAP)}}
        \\ 

        & 
        \multicolumn{1}{c}{\scalebox{1.25}{$\bullet$}} &  
        \multicolumn{1}{c}{Employee} & 
        \multicolumn{1}{c}{\scalebox{1.25}{$\bullet$}} & 
        & 
        & 
        \multicolumn{1}{c}{\scalebox{1.25}{$\bullet$}}& 
        \multicolumn{1}{c}{\scalebox{1.25}{$\bullet$}} &
        &
        &
        &
        &
        \multicolumn{1}{c}{\scalebox{1.25}{$\bullet$}} &
        &
        \multicolumn{1}{c}{IS; A; RA}
        \\

        &  
        \multicolumn{14}{l}{\textit{Federal Aviation Administration (FAA), Voluntary Disclosure Reporting Program (VDRP)}}
        \\ 

        & 
        \multicolumn{1}{c}{\scalebox{1.25}{$\bullet$}} &  
        \multicolumn{1}{c}{Company} & 
        \multicolumn{1}{c}{\scalebox{1.25}{$\bullet$}} & 
        & 
        & 
        \multicolumn{1}{c}{\scalebox{1.25}{$\bullet$}} & 
        \multicolumn{1}{c}{\scalebox{1.25}{$\bullet$}} &
        \multicolumn{1}{c}{\scalebox{1.25}{$\bullet$}} &
        &
        &
        &
        \multicolumn{1}{c}{\scalebox{1.25}{$\bullet$}} &
        &
        \multicolumn{1}{c}{A}
        \\

        &  
        \multicolumn{14}{l}{\textit{National Aeronautics and Space Administration (NASA), Aviation Safety Reporting System (ASRS)}}
        \\ 

        & 
        \multicolumn{1}{c}{\scalebox{1.25}{$\bullet$}} &  
        \multicolumn{1}{c}{Employee} & 
        \multicolumn{1}{c}{\scalebox{1.25}{$\bullet$}} & 
        & 
        & 
        \multicolumn{1}{c}{\scalebox{1.25}{$\bullet$}}& 
        \multicolumn{1}{c}{\scalebox{1.25}{$\bullet$}} &
        \multicolumn{1}{c}{\scalebox{1.25}{$\bullet$}} &
        \multicolumn{1}{c}{\scalebox{1.25}{$\bullet$}} &
        &
        &
        &
        \multicolumn{1}{c}{\scalebox{1.25}{$\bullet$}} &
        \multicolumn{1}{c}{ID}
        \\

        &  
        \multicolumn{14}{l}{\textit{International Civil Aviation Organisation (ICAO), Annex 13 Information Sharing}}
        \\ 

        & 
        \multicolumn{1}{c}{\scalebox{1.25}{$\bullet$}} &  
        \multicolumn{1}{c}{Inter-Gov.} & 
        \multicolumn{1}{c}{\scalebox{1.25}{$\bullet$}} & 
        & 
        &
        &
        &
        \multicolumn{1}{c}{\scalebox{1.25}{$\bullet$}} &
        \multicolumn{1}{c}{\scalebox{1.25}{$\bullet$}} &
        \multicolumn{1}{c}{\scalebox{1.25}{$\bullet$}} &
        \multicolumn{1}{c}{\scalebox{1.25}{$\bullet$}} &
        &
        &
        \multicolumn{1}{c}{A}
        \\

        \midrule

        \multirow{8}{*}{\rotatebox[origin=c]{90}{Pesticides}} & 
        \multicolumn{14}{l}{\textit{California Environmental Protection Agency (CalEPA), Environmental Complaint System}} 
        \\ 

        &
        &
        \multicolumn{1}{c}{Citizen} &
        \multicolumn{1}{c}{\scalebox{1.25}{$\bullet$}} &
        &
        &
        &
        \multicolumn{1}{c}{\scalebox{1.25}{$\bullet$}} &
        \multicolumn{1}{c}{\scalebox{1.25}{$\bullet$}} &
        \multicolumn{1}{c}{\scalebox{1.25}{$\bullet$}} &
        &
        &
        &
        \multicolumn{1}{c}{\scalebox{1.25}{$\bullet$}} &
        \multicolumn{1}{c}{IS; A; RA}
        \\

        &  
        \multicolumn{14}{l}{\textit{California Environmental Protection Agency (CalEPA), California Pesticide Illness Surveillance Program}}
        \\

        & 
        &
        \multicolumn{1}{c}{Third Party} & 
        \multicolumn{1}{c}{\scalebox{1.25}{$\bullet$}} &
        &
        &
        &
        &
        &
        \multicolumn{1}{c}{\scalebox{1.25}{$\bullet$}} &
        \multicolumn{1}{c}{\scalebox{1.25}{$\bullet$}} &
        &
        \multicolumn{1}{c}{\scalebox{1.25}{$\bullet$}} &
        &
        \multicolumn{1}{c}{IS; A, RA}
        \\

        &  
        \multicolumn{14}{l}{\textit{Environmental Protection Agency (EPA), Enforcement and Compliance History Online (ECHO) Reporting System}}
        \\

        &
        &
        \multicolumn{1}{c}{Citizen} &
        \multicolumn{1}{c}{\scalebox{1.25}{$\bullet$}} &
        &
        &
        &
        \multicolumn{1}{c}{\scalebox{1.25}{$\bullet$}} &
        \multicolumn{1}{c}{\scalebox{1.25}{$\bullet$}} &
        \multicolumn{1}{c}{\scalebox{1.25}{$\bullet$}} &
        &
        &
        &
        \multicolumn{1}{c}{\scalebox{1.25}{$\bullet$}} &
        \multicolumn{1}{c}{IS; A; RA}
        \\

        &  
        \multicolumn{14}{l}{\textit{Environmental Protection Agency (EPA), Pesticide Manufacturer Reporting Requirements}}
        \\ 

        &
        &
        \multicolumn{1}{c}{Company} &
        \multicolumn{1}{c}{\scalebox{1.25}{$\bullet$}} &
        &
        &
        &
        &
        &
        \multicolumn{1}{c}{\scalebox{1.25}{$\bullet$}} &
        \multicolumn{1}{c}{\scalebox{1.25}{$\bullet$}} &
        &
        \multicolumn{1}{c}{\scalebox{1.25}{$\bullet$}} &
        &
        \multicolumn{1}{c}{ID}
        \\

        \bottomrule
    \end{tabularx}   
    \caption[]{Classification of incident reporting systems per our framework. \newline \newline Legend: incident types can include safety (``Saf''), rights (``Rig''), or security (``Sec''). Level of risk materialization can include hazards (``H''), situations (``S''), near misses (``NM''), or harm events (``HE''). Anonymity can be open (``O''), confidential (``C''), or anonymous (``A''). Post-reporting actions include information sharing (``IS''), information disclosure (``ID''), audit (``A''), or regulatory action (``RA''). Hyphens indicate where we found no publicly available information. References are available in Table \ref{tab:Results_References}. Reproduced from Section \ref{sec:Recs}.}
\end{table*}

\setcounter{table}{2}
\begin{table*}[!htbp]
    \centering
    \small
    \begin{tabularx}{\textwidth}{ m{0.25cm} s X s s s s s s s X s s s X } %
        \toprule

        \multicolumn{1}{c}{} &
        \multicolumn{1}{c}{\textbf{Goal}} &
        \multicolumn{1}{c}{\textbf{Actors}} &
        \multicolumn{3}{c}{\textbf{Incident Type}} &
        \multicolumn{4}{c}{\textbf{Risk Materialization}} &
        \multicolumn{1}{c}{\textbf{Enforcement}} &
        \multicolumn{3}{c}{\textbf{Anonymity}} &
        \multicolumn{1}{c}{\textbf{Post-Reporting}} 
        \\

        \multicolumn{1}{c}{} &
        \multicolumn{1}{c}{Learning?} &
        \multicolumn{1}{c}{} &
        \multicolumn{1}{c}{Saf} &
        \multicolumn{1}{c}{Rig} &
        \multicolumn{1}{c}{Sec} &
        \multicolumn{1}{c}{H} &
        \multicolumn{1}{c}{S} &
        \multicolumn{1}{c}{NM} &
        \multicolumn{1}{c}{HE} &
        \multicolumn{1}{c}{Mandatory?} &
        \multicolumn{1}{c}{O} &
        \multicolumn{1}{c}{C} &
        \multicolumn{1}{c}{A} &
        \multicolumn{1}{c}{} 
        \\

        \midrule

        \multirow{12}{*}{\rotatebox[origin=c]{90}{Pharmaceuticals}} & 
        \multicolumn{14}{l}{\textit{Food and Drug Administration (FDA), MedWatch (Patients)}} 
        \\

        & 
        &  
        \multicolumn{1}{c}{Citizen} &  
        \multicolumn{1}{c}{\scalebox{1.25}{$\bullet$}} & 
        & 
        & 
        & 
        \multicolumn{1}{c}{\scalebox{1.25}{$\bullet$}} & 
        \multicolumn{1}{c}{\scalebox{1.25}{$\bullet$}} & 
        \multicolumn{1}{c}{\scalebox{1.25}{$\bullet$}} & 
        & 
        \multicolumn{1}{c}{\scalebox{1.25}{$\bullet$}} & 
        & 
        & 
        \multicolumn{1}{c}{IS; A; RA} 
        \\

        &
        \multicolumn{14}{l}{\textit{Food and Drug Administration (FDA), MedWatch (Healthcare Providers)}} 
        \\

        & 
        &  
        \multicolumn{1}{c}{Third Party} &  
        \multicolumn{1}{c}{\scalebox{1.25}{$\bullet$}} & 
        & 
        & 
        & 
        \multicolumn{1}{c}{\scalebox{1.25}{$\bullet$}} & 
        \multicolumn{1}{c}{\scalebox{1.25}{$\bullet$}} & 
        \multicolumn{1}{c}{\scalebox{1.25}{$\bullet$}} & 
        & 
        \multicolumn{1}{c}{\scalebox{1.25}{$\bullet$}} & 
        & 
        & 
        \multicolumn{1}{c}{IS; A; RA} 
        \\

        &
        \multicolumn{14}{l}{\textit{Food and Drug Administration (FDA), Medical Device Reporting (Device Manufacturers)}} 
        \\
        
        & 
        &  
        \multicolumn{1}{c}{Company} &  
        \multicolumn{1}{c}{\scalebox{1.25}{$\bullet$}} & 
        &  
        & 
        & 
        & 
        \multicolumn{1}{c}{\scalebox{1.25}{$\bullet$}} & 
        \multicolumn{1}{c}{\scalebox{1.25}{$\bullet$}} & 
        \multicolumn{1}{c}{\scalebox{1.25}{$\bullet$}} & 
        \multicolumn{1}{c}{\scalebox{1.25}{$\bullet$}} & 
        & 
        &
        \multicolumn{1}{c}{ID; A; RA} 
        \\

        &
        \multicolumn{14}{l}{\textit{Food and Drug Administration (FDA), Medical Device Reporting (User Facilities)}} 
        \\
        
        & 
        &  
        \multicolumn{1}{c}{User} &  
        \multicolumn{1}{c}{\scalebox{1.25}{$\bullet$}} & 
        &  
        & 
        & 
        & 
        \multicolumn{1}{c}{\scalebox{1.25}{$\bullet$}} & 
        \multicolumn{1}{c}{\scalebox{1.25}{$\bullet$}} & 
        \multicolumn{1}{c}{\scalebox{1.25}{$\bullet$}} & 
        \multicolumn{1}{c}{\scalebox{1.25}{$\bullet$}} & 
        & 
        & 
        \multicolumn{1}{c}{ID; A; RA} 
        \\

        &  
        \multicolumn{14}{l}{\textit{Food and Drug Administration (FDA), Voluntary Malfunction Summary Reporting (VMSR)}} 
        \\
        
        &
        \multicolumn{1}{c}{\scalebox{1.25}{$\bullet$}} & 
        \multicolumn{1}{c}{Company} &  
        \multicolumn{1}{c}{\scalebox{1.25}{$\bullet$}} & 
        &  
        & 
        & 
        \multicolumn{1}{c}{\scalebox{1.25}{$\bullet$}} & 
        \multicolumn{1}{c}{\scalebox{1.25}{$\bullet$}} & 
        & 
        & 
        \multicolumn{1}{c}{\scalebox{1.25}{$\bullet$}} & 
        &
        &
        \multicolumn{1}{c}{ID} 
        \\

        &
        \multicolumn{14}{l}{\textit{Centers for Disease Control and Prevention (CDC), Vaccine Adverse Events Reporting System (VAERS) (Patients and Family)}} 
        \\
        
        & 
        &  
        \multicolumn{1}{c}{Citizen} &
        \multicolumn{1}{c}{\scalebox{1.25}{$\bullet$}} &
        &
        & 
        &
        &
        &
        \multicolumn{1}{c}{\scalebox{1.25}{$\bullet$}} &
        &
        &
        \multicolumn{1}{c}{\scalebox{1.25}{$\bullet$}} &
        & 
        \multicolumn{1}{c}{ID; A; RA} 
        \\

        &
        \multicolumn{14}{l}{\textit{Centers for Disease Control and Prevention (CDC), Vaccine Adverse Events Reporting System (VAERS) (Healthcare Providers)}} 
        \\
        
        & 
        &  
        \multicolumn{1}{c}{Third Party} &
        \multicolumn{1}{c}{\scalebox{1.25}{$\bullet$}} &
        &
        & 
        &
        &
        &
        \multicolumn{1}{c}{\scalebox{1.25}{$\bullet$}} &
        \multicolumn{1}{c}{\scalebox{1.25}{$\bullet$}} &
        \multicolumn{1}{c}{ - } &
        \multicolumn{1}{c}{ - } &
        \multicolumn{1}{c}{ - } &
        \multicolumn{1}{c}{ID; A; RA} 
        \\

        \midrule

        \multirow{2}{*}{\rotatebox[origin=c]{90}{Cyber}} & 
        \multicolumn{14}{l}{\textit{Information Security Analysis Centers / Organizations (ISACs / ISAOs)}} 
        \\ 

        &
        \multicolumn{1}{c}{\scalebox{1.25}{$\bullet$}} &
        \multicolumn{1}{c}{Indep. Db.} & 
        & 
        &
        \multicolumn{1}{c}{\scalebox{1.25}{$\bullet$}} & 
        & 
        \multicolumn{1}{c}{\scalebox{1.25}{$\bullet$}} & 
        \multicolumn{1}{c}{\scalebox{1.25}{$\bullet$}} & 
        \multicolumn{1}{c}{\scalebox{1.25}{$\bullet$}} & 
        & 
        &
        \multicolumn{1}{c}{\scalebox{1.25}{$\bullet$}} & 
        & 
        \multicolumn{1}{c}{IS} 
        \\

        \midrule

        \multirow{4}{*}{\rotatebox[origin=c]{90}{Dams}} & 
        \multicolumn{14}{l}{\textit{Association of State Dam Safety Officials (ASDSO), Dam Incident Database}} 
        \\ 

        &
        \multicolumn{1}{c}{\scalebox{1.25}{$\bullet$}} &
        \multicolumn{1}{c}{Indep. Db.} & 
        \multicolumn{1}{c}{\scalebox{1.25}{$\bullet$}} & 
        &
        & 
        & 
        \multicolumn{1}{c}{\scalebox{1.25}{$\bullet$}} & 
        \multicolumn{1}{c}{\scalebox{1.25}{$\bullet$}} & 
        \multicolumn{1}{c}{\scalebox{1.25}{$\bullet$}} & 
        & 
        & 
        \multicolumn{1}{c}{\scalebox{1.25}{$\bullet$}} & 
        & 
        \multicolumn{1}{c}{ID} 
        \\

        &
        \multicolumn{14}{l}{\textit{Stanford University National Performance on Dams Program (NPDP), Dam Directory and Incident Database}} 
        \\ 

        &
        \multicolumn{1}{c}{\scalebox{1.25}{$\bullet$}} &
        \multicolumn{1}{c}{Indep. Db.} & 
        \multicolumn{1}{c}{\scalebox{1.25}{$\bullet$}} & 
        &
        & 
        & 
        \multicolumn{1}{c}{\scalebox{1.25}{$\bullet$}} & 
        \multicolumn{1}{c}{\scalebox{1.25}{$\bullet$}} & 
        \multicolumn{1}{c}{\scalebox{1.25}{$\bullet$}} & 
        & 
        &
        \multicolumn{1}{c}{\scalebox{1.25}{$\bullet$}} & 
        & 
        \multicolumn{1}{c}{ID} 
        \\

        \midrule

        \multirow{2}{*}{\rotatebox[origin=c]{90}{Rail}} & 
        \multicolumn{14}{l}{\textit{National Aeronautics and Space Administration (NASA), Confidential Close Call Reporting System (C\textsuperscript{3}RS)}} 
        \\ 

        & 
        \multicolumn{1}{c}{\scalebox{1.25}{$\bullet$}} &  
        \multicolumn{1}{c}{Employee} & 
        \multicolumn{1}{c}{\scalebox{1.25}{$\bullet$}} & 
        & 
        & 
        \multicolumn{1}{c}{\scalebox{1.25}{$\bullet$}}& 
        \multicolumn{1}{c}{\scalebox{1.25}{$\bullet$}} &
        \multicolumn{1}{c}{\scalebox{1.25}{$\bullet$}} &
        \multicolumn{1}{c}{\scalebox{1.25}{$\bullet$}} &
        & 
        & 
        & 
        \multicolumn{1}{c}{\scalebox{1.25}{$\bullet$}} &
        \multicolumn{1}{c}{ID} 
        \\
        
        \bottomrule
    \end{tabularx}   
    \caption[]{Classification of incident reporting systems per our framework (cont'd). \newline \newline Legend: incident types can include safety (``Saf''), rights (``Rig''), or security (``Sec''). Level of risk materialization can include hazards (``H''), situations (``S''), near misses (``NM''), or harm events (``HE''). Anonymity can be open (``O''), confidential (``C''), or anonymous (``A''). Post-reporting actions include information sharing (``IS''), information disclosure (``ID''), audit (``A''), or regulatory action (``RA''). Hyphens indicate where we found no publicly available information. References are available in Table \ref{tab:Results_References}. Reproduced from Section \ref{sec:Recs}.}
\end{table*}
\setcounter{table}{18}

\begin{table*}[!hbtp]
    \centering
    \small
    \begin{tabularx}{\textwidth}{ m{0.25cm} X X } %
        \toprule

        \multicolumn{2}{c}{System} &
        \multicolumn{1}{c}{References}
        \\
        
        \midrule

        \multirow{13}{*}{\rotatebox[origin=c]{90}{Nuclear}} &
        \textit{Nuclear Regulatory Commission (NRC), Safety Hotline} &
        \citet{nrc_allegation_2016, nrc_reporting_2017, nrc_backgrounder_2023}
        \\ 

        &
        \textit{Nuclear Regulatory Commission (NRC), Statutory Reporting Requirements and Operations Center} &
        \citetalias{10_CFR_20.2201, 10_CFR_20.2203, 10_CFR_20.2205, 10_CFR_30.34, 10_CFR_36.83, 10_CFR_50.72, 10_CFR_50.73, 10_CFR_70.32, 10_CFR_70.50, 10_CFR_72.74, 10_CFR_72.75, 10_CFR_73.1200, 10_CFR_74.11}; 
        \citet{nrc_about_2023, nrc_reports_2022}
        \\

        &  
        \textit{International Atomic Energy Agency (IAEA), Incident Reporting System (IRS)} &
        \citet{iaea_nuclear_2020, iaea_irs_2022, novak_iaea_1985}
        \\ 

        &  
        \textit{International Atomic Energy Agency (IAEA), Fuel Incident Notification and Analysis System (FINAS)} &
        \citet{iaea_iaeanea_2006, iaea_operating_2020}
        \\ 

        &  
        \textit{International Atomic Energy Agency (IAEA), Incident Reporting Systems for Research Reactors (IRSRR)} &
        \citet{hardesty_iaea_2022, iaea_guide_2011, iaea_operating_2015}
        \\ 

        &  
        \textit{James Martin Center for Nonproliferation Studies (CNS), Global Incidents and Trafficking Database} &
        \citet{nti_overview_2023}
        \\ 

        \midrule
        
        \multirow{13}{*}{\rotatebox[origin=c]{90}{Civilian Aviation}} &
        \textit{National Transportation Safety Board (NTSB)} &
        \citetalias{49_CFR_Part_830, 49_USC_1132, 49_USC_1155}; \citet{ntsb_ntsb_nodate}; \citet{ntsb_report_nodate}
        \\
        
        &  
        \textit{Federal Aviation Administration (FAA), Hotline} &
        \citet{65_FR_19476, 87_FR_61649, faa_order_2016, faa_order_2022, faa_order_2023}
        \\ 

        &  
        \textit{Federal Aviation Administration (FAA), Voluntary Safety Reporting Program (VSRP)} &
        \citet{faa_order_2021}
        \\ 

        &  
        \textit{Federal Aviation Administration (FAA), Aviation Safety Action Program (ASAP)} &
        \citet{faa_ac_2020, faa_aviation_2020, faa_notice_2023, faa_order_2023-1, faa_aviation_2024}
        \\ 

        &  
        \textit{Federal Aviation Administration (FAA), Voluntary Disclosure Reporting Program (VDRP)} &
        \citet{faa_vdrp_nodate-1}; \citet{faa_vdrp_nodate, faa_ac_2009, faa_order_2022-1, faa_ac_2023, faa_notice_2023, faa_voluntary_2024}
        \\ 

        &  
        \textit{National Aeronautics and Space Administration (NASA), Aviation Safety Reporting System (ASRS)} &
        \citet{asrs_asrs_2001, AC0046F, marfise_nasa_2023, nasa_asrs_2023, nasa_asrs_2024}
        \\ 

        &
        \textit{International Civil Aviation Organisation (ICAO), Annex 13 Information Sharing} &
        \citet{icao_annex_2020, icao_aircraft_2022, ICAO_Reporting_Taxonomy, ntsb_ntsbs_nodate}
        \\ 

        \midrule

        \multirow{8}{*}{\rotatebox[origin=c]{90}{Pesticides}} &
        \textit{California Environmental Protection Agency (CalEPA), Environmental Complaint System} &
        \citet{calepa_calepa_nodate, calepa_dpr_report_nodate}
        \\
        
        &
        \textit{California Environmental Protection Agency (CalEPA), California Pesticide Illness Surveillance Program} &
        \citet{ali_summary_2020, calepa_oehha_pesticide_2015, calepa_oehha_confidential_nodate, calepa_dpr_county_nodate, durant_ignorance_2020, state_of_california_california_1996}
        \\

        &  
        \textit{Environmental Protection Agency (EPA), Enforcement and Compliance History Online (ECHO) Reporting System} &
        \citet{EPA_Reporting_Instructions, epa_report_2013, EPA_Report_Form, epa_epa_2023, epa_report_2024}
        \\
        
        &  
        \textit{Environmental Protection Agency (EPA), Pesticide Manufacturer Reporting Requirements} &
        \citetalias{7_USC_136d, 7_USC_136j, 7_USC_136l}; \citet{epa_industrys_1998, epa_incident_2015, epa_about_2023, epa_epa_2023, johnson_pesticide_1998, mulkey_pesticide_1998}        
        \\ 
        
        \midrule

        \multirow{13}{*}{\rotatebox[origin=c]{90}{Pharmaceuticals}} &
        \textit{Food and Drug Administration (FDA), MedWatch (Patients)} & 
        \citet{fda_medwatch_nodate, fda_medwatch_2022, fda_medwatch_2024}
        \\ 

        &
        \textit{Food and Drug Administration (FDA), MedWatch (Healthcare Providers)} & 
        \citet{fda_medwatch_nodate, fda_medwatch_2022-1, fda_instructions_2022, fda_medwatch_2024}
        \\ 

        &  
        \textit{Food and Drug Administration (FDA), Medical Device Reporting (Device Manufacturers)} &
        \citetalias{79_FR_8832}; 
        \citet{fda_mandatory_2020, fda_medwatch_2022-2, fda_general_2022, fda_medical_2023}
        \\ 

        &  
        \textit{Food and Drug Administration (FDA), Medical Device Reporting (User Facilities)} &
        \citetalias{79_FR_8832}; 
        \citet{fda_mandatory_2020, fda_medwatch_2022-2, fda_general_2022, fda_medical_2023}
        \\ 
        
        &  
        \textit{Food and Drug Administration (FDA), Voluntary Malfunction Summary Reporting (VMSR)} &
        \citetalias{79_FR_8832, 87_FR_75634}; \citet{fda_medical_2023}
        \\ 

        &
        \textit{Centers for Disease Control and Prevention (CDC), Vaccine Adverse Events Reporting System (VAERS) (Patients and Family)} &
        \citet{vaers_vaers_nodate}
        \\ 

        &
        \textit{Centers for Disease Control and Prevention (CDC), Vaccine Adverse Events Reporting System (VAERS) (Healthcare Providers)} &
        \citet{vaers_vaers_nodate}
        \\ 

        \midrule

        \multirow{2}{*}{\rotatebox[origin=c]{90}{Cyber}} &
        \textit{Information Security Analysis Centers / Organizations (ISACs / ISAOs)} &
        \citet{cisa_critical_nodate}; \citet{cisa_frequently_nodate, dhs_critical_2016, dhs_guidance_2020, e-isac_guide_2019, isao_so_isao_2016, isao_so_isao_2016-1}
        \\ 

        \midrule

        \multirow{4}{*}{\rotatebox[origin=c]{90}{Dams}} &
        \textit{Association of State Dam Safety Officials (ASDSO), Dam Incident Database} &
        \citet{asdso_asdso_nodate}; \citet{asdso_dam_nodate}
        \\ 
        
        &
        \textit{Stanford University National Performance on Dams Program (NPDP), Dam Directory and Incident Database} &
        \citet{mccann_national_nodate, npdp_national_nodate}
        \\ 

        \midrule

        \multirow{2}{*}{\rotatebox[origin=c]{90}{Rail}} & 
        \textit{National Aeronautics and Space Administration (NASA), Confidential Close Call Reporting System (C\textsuperscript{3}RS}) &
        \citet{gao_federal_2022, ranney_confidential_2019}
        \\ 
        
        \bottomrule
    \end{tabularx}   
    \caption{List of references for Table \ref{tab:Full_Results}}
    \label{tab:Results_References}
\end{table*}

\clearpage

\section{Appendix: Example Real-World Incident Definitions} \label{sec:Appendix_Example_Defs}

This Appendix provides example definitions of incident- and issue-related terminology from various industries, including both AI and elsewhere (Table \ref{tab:Example_Incident_Defs}). 

Note that this list of definitions was compiled in May 2024 and is not meant to be up-to-date, complete, or exhaustive; rather, we intend to provide a sense of the diversity of terms and definitions used in different reporting systems (both in the context of AI and also in other industries). The definitions in Table \ref{tab:Example_Incident_Defs}, for instance, are of varying levels of generality and scopes; specific terms may be used for events resulting in different levels of severity of harm, events at different stages of materialization of risk, etc.

\renewcommand{\arraystretch}{1.5}
\begin{table*}[p]
    \centering
    \small
    \begin{tabularx}{\textwidth}{ s s s X }
        \toprule
        
        \textbf{Field} & 
        \textbf{Source} & 
        \textbf{Term} & 
        \textbf{Definition} 
        \\ 
        
        \midrule 

        AI & 
        AI, Algorithmic, and Automation Incidents and Controversies (AIAAIC) &
        ``Incident'' &
        ``A sudden known or unknown event (or ‘trigger’) that becomes public and which takes the form of a disruption, loss, emergency, or crisis. Examples: AI system or robot malfunction; Actual or perceived inappropriate or unethical behaviour by a system operator or developer; Data privacy or info confidentiality leak exposes system vulnerability.'' \cite{aiaaic_classifications_2024} 
        \\ 

        & 
        &
        ``Issue'' &
        ``Concerns publicly raised about the nature and/or potential impacts of a System, but without evidence of recognised harms. Examples: A public debate/controversy about an unlaunched system, technology, or patent; Research indicating unwarranted carbon emissions or water consumption by a system.'' \cite{aiaaic_classifications_2024} 
        \\ 
        
        &

        AI Incident Database (AIID) &

        ``AI incident'' &

        ``[A]n alleged harm or near harm event to people, property, or the environment where an AI system is implicated.'' \cite{aiid_editors_nodate} 
        \\ 
        
        &
        &

        ``AI issue'' &

        ``[A]n alleged harm or near harm by an AI system that has yet to occur or be detected.'' \cite{aiid_editors_nodate} 
        \\ 
        
        &
        European Union (EU) &
        ``Serious incident'' & 
        ``\,`[S]erious incident’ means an incident or malfunctioning of an AI system that directly or indirectly leads to any of the following:

        \begin{enumerate}
            \item [(a)] the death of a person, or serious harm to a person’s health;
            \item [(b)] a serious and irreversible disruption of the management or operation of critical infrastructure.
            \item [(c)] the infringement of obligations under Union law intended to protect fundamental rights;
            \item [(d)] serious harm to property or the environment[.]'' \citepalias[Art. 3]{AIA}
        \end{enumerate} 
        \\ 

        &
        Organisation for Economic Co-operation and Development (OECD) &
        ``AI Incident'' &
        ``An AI incident is an event, circumstance or series of events where the development, use or malfunction of one or more AI systems directly or indirectly leads to any of the following harms:

        \begin{enumerate}
            \item [(a)] injury or harm to the health of a person or groups of people;
            \item [(b)] disruption of the management and operation of critical infrastructure;
            \item [(c)] violations of human rights or a breach of obligations under the applicable law intended to protect fundamental, labour and intellectual property rights;
            \item [(d)] harm to property, communities or the environment.''
        \end{enumerate}
        \cite{orssich_defining_2024}
        \\

        &
        &
        ``AI hazard'' &
        ``An AI hazard is an event, circumstance or series of events where the development, use or malfunction of one or more AI systems could plausibly lead to an AI incident, i.e., any of the following harms:

        \begin{enumerate}
            \item [(a)] injury or harm to the health of a person or groups of people;
            \item [(b)] disruption of the management and operation of critical infrastructure;
            \item [(c)] violations to human rights or a breach of obligations under the applicable law intended to protect fundamental, labour and intellectual property rights;
            \item [(d)] harm to property, communities or the environment''
        \end{enumerate}
        \cite{orssich_defining_2024}
        \\
        
        \midrule
    \end{tabularx}
    \caption{Examples of Different Issue and Incident Definitions from Various Industries}
    \label{tab:Example_Incident_Defs}
\end{table*}

\setcounter{table}{19}
\begin{table*}[p]
    \centering
    \small
    \begin{tabularx}{\textwidth}{ s s s X }
        \toprule
        
        \textbf{Field} & 
        \textbf{Source} & 
        \textbf{Term} & 
        \textbf{Definition} 
        \\ 
        
        \midrule

        &
        Secure A.I. Act of 2024 (Proposed) &
        ``Artificial intelligence safety incident'' &
        ``(1) ARTIFICIAL INTELLIGENCE SAFETY INCIDENT.--The term `artificial intelligence safety incident' means an event that increases the risk that operation of an artificial intelligence system will--

        \,
        
        (A) result in physical or psychological harm; or

        \,
        
        (B) lead to a state in which human life, health, property, or the environment is endangered.'' \cite[Sec. 2]{warner_s_2024}
        \\

        &
        &
        ``Artificial intelligence security incident'' &
        ``(2) ARTIFICIAL INTELLIGENCE SECURITY INCIDENT.--The term `artificial intelligence security incident' means an event that increases--

        \,
        
        (A) the risk that operation of an artificial intelligence system occurs in a way that enables the extraction of information about the behavior or characteristics of an artificial intelligence system by a third party; or

        \,

        (B) the ability of a third party to manipulate an artificial intelligence system in order to subvert the confidentiality, integrity, or availability of an artificial intelligence system or adjacent system.'' \cite[Sec. 2]{warner_s_2024}
        \\

        &
        &
        ``Artificial intelligence security vulnerability'' &
        ``(3) ARTIFICIAL INTELLIGENCE SECURITY VULNERABILITY.--The term `artificial intelligence security vulnerability' means a weakness in an artificial intelligence system that could be exploited by a third party to subvert, without authorization, the confidentiality, integrity, or availability of an artificial intelligence system, including through techniques such as--

        \,

         \begin{enumerate}
            \item [(A)] data poisoning;
            \item [(B)] evasion attacks;
            \item [(C)] privacy-based attacks; and
            \item [(D)] abuse attacks.''
        \end{enumerate}
        \cite{warner_s_2024}
        \\

        &
        &
        ``Counter artificial intelligence'' &
        ``(4) COUNTER-ARTIFICIAL INTELLIGENCE.---The term `counter-artificial intelligence' means techniques or procedures to extract information about the behavior or characteristics of an artificial intelligence system, or to learn how to manipulate an artificial intelligence system, in order to subvert the confidentiality, integrity, or availability of an artificial intelligence system or adjacent system.'' \cite[Sec. 2]{warner_s_2024}
        \\
        
        \midrule
    \end{tabularx}
    \caption[]{Examples of Different Issue and Incident Definitions from Various Industries (cont'd)}
\end{table*}

\setcounter{table}{19}
\begin{table*}[p]
    \centering
    \small
    \begin{tabularx}{\textwidth}{ s s s X }
        \toprule
        \textbf{Field} & 
        \textbf{Source} & 
        \textbf{Term} & 
        \textbf{Definition} 
        \\ 
        
        \midrule 

        Aviation &
        Federal Aviation Administration (FAA) & 
        ``Aircraft Accident'' &
        ``An occurrence associated with the operation of an aircraft which takes place between the time any person boards the aircraft with the intention of flight and all such persons have disembarked, and in which any person suffers death or serious injury, or in which the aircraft receives substantial damage.'' \cite[p. 5]{faa_order_2018} 
        \\ 
        
        &
        & 
        ``Incident'' &
        ``An incident is defined as an occurrence other than an accident, associated with the operation of an aircraft, which affects or could affect the safety of operations. This definition covers a broad range of events and may include the following: 
        
        \begin{enumerate}
            \item [a.] Damage to an aircraft (other than an accident); 
            \item [b.] Runway incursion or excursion; 
            \item [c.] Pilot deviations (PD); 
            \item [d.] Near midair collision (NMAC); and 
            \item [e.] Unmanned Aircraft Systems (UAS) Reports.''
        \end{enumerate} 
        \cite[p. 54]{faa_order_2018} 
        \\ 

        &
        International Civil Aerospace Organization (ICAO) & 
        ``Accident'' &
        ``An occurrence associated with the operation of an aircraft which, in the case of a manned aircraft, takes place between the time any person boards the aircraft with the intention of flight until such time as all such persons have disembarked, or in the case of an unmanned aircraft, takes place between the time the aircraft is ready to move with the purpose of flight until such time as it comes to rest at the end of the flight and the primary propulsion system is shut down, in which:
        
        \begin{itemize}
            \item [a)] a person is fatally or seriously injured as a result of: 

            \begin{enumerate}
                \item [-] being in the aircraft, or 
                \item [-] direct contact with any part of the aircraft, including parts which have become detached from the aircraft, or 
                \item [-] direct exposure to jet blast,''
            \end{enumerate}
        \end{itemize} 

        \textit{except} when the injuries are from natural causes, self-inflicted or inflicted by other persons, or when the injuries are to stowaways hiding outside the areas normally available to the passengers and crew; or
        
        \begin{itemize}
            \item [b)] the aircraft sustains damage or structural failure which:

            \begin{itemize}
                \item [-] adversely affects the structural strength, performance or flight characteristics of the aircraft, and 
                \item [-] would normally require major repair or replacement of the affected component,'' (cont'd)
            \end{itemize}
            
            \textit{except} for engine failure or damage, when the damage is limited to a single engine (including its cowlings or accessories), to propellers, wing tips, antennas, probes, vanes, tires, brakes, wheels, fairings, panels, landing gear doors, windscreens, the aircraft skin (such as small dents or puncture holes), or for minor damages to main rotor blades, tail rotor blades, landing gear, and those resulting from hail or bird strike (including holes in the radome); or
            
            \item [c)] the aircraft is missing or is completely inaccessible.'' 
        \end{itemize} 
        \cite[p. 1-1]{icao_annex_2020} 
        \\ 
        
        &
        & 
        ``Incident'' &
        ``An occurrence, other than an accident, associated with the operation of an aircraft which affects or could affect the safety of operation.'' \cite[p. 1-2]{icao_annex_2020} 
        \\ 
        
        \midrule
    \end{tabularx}
    \caption[]{Examples of Different Issue and Incident Definitions from Various Industries (cont'd)}
\end{table*}
        
\setcounter{table}{19}
\begin{table*}[p]
    \centering
    \small
    \begin{tabularx}{\textwidth}{ s s s X }
        \toprule
        
        \textbf{Field} & 
        \textbf{Source} & 
        \textbf{Term} & 
        \textbf{Definition} 
        \\ 
        
        \midrule 

                Chemical &
        Federal Emergency Management Agency (FEMA) & 
        ``Chemical incident'' &
        ``A chemical incident includes a wide scope of events and refers to the release, or potential release, of a chemical substance

        \begin{enumerate}
            \item that harms people, animals and/or the environment, regardless of accidental or deliberate cause, and/or
            \item for which response needs have the potential to overwhelm state and local resources (both governmental and private sector), and/or
            \item for which the Environmental Protection Agency (EPA) and/or the United States Coast Guard (USCG), co-Chairs of the National Response Team (NRT) for Oil and Chemical Spills under the National Oil and Hazardous Substances Pollution Contingency Plan (NCP), deems that support is or will be required.''
        \end{enumerate}
        \cite[p. 2]{fema_key_2022}
        \\ 

        &
        Organisation for Economic Co-operation and Development (OECD) & 
        ``Chemical accident'' &
        ``Any unintentional event, such as a release, fire or explosion at a hazardous installation, involving hazardous substances that has the potential to cause harm to human health, the environment or property. This also covers chemical accidents triggered by the effects of natural hazards.'' \cite[p. 171]{oecd_oecd_2023} 
        \\ 

        &
        &
        ``Hazard'' &
        ``An inherent property of a substance, agent, source of energy or situation having the potential to cause undesirable consequences.'' \cite[p. 171]{oecd_oecd_2023} 
        \\ 

        &
        &
        ``Incident'' &
        ``Incident includes: 
        
        \begin{itemize}
            \item A chemical accident that has a significant impact on human health, company operations, the environment, the community or society at large.
            \item A near miss that does not have a serious impact but might have had under different circumstances, that is if the sequence of events had not been interrupted by a planned control measure or by chance.''
        \end{itemize} 
        \cite[p. 171]{oecd_oecd_2023} 
        \\ 

        &
        World Health Organization (WHO) & 
        ``Chemical incident'' &
        ``A chemical incident is the unexpected release of a substance that is (potentially) hazardous either to humans, other animals or the environment.'' \cite{who_chemical_2017} 
        \\ 

        \midrule
        
        Generic &
        National Safety Council (NSC) & 
        ``Accident'' &
        ``The occurrence in a sequence of events that produces unintended injury, death, or property damage. Accident refers to the event, not to the result of the event (see preventable injury). The term 'accident' has largely been replaced in the public health community with the term 'incident.'\,'' \cite{nsc_glossary_2024} 
        \\ 
        
        &
        &
        ``Incident'' &
        ``The preferred term for 'accident' in the public health community. It refers to the occurrence in a sequence of events that produces unintended injury, death or property damage.'' \cite{nsc_glossary_2024} 
        \\ 
        
        &
        Leveson & 
        ``Accident'' &
        ``An undesired and unplanned event that results in a loss (including loss of human life or injury, property damage, environmental pollution, and so on).'' \cite[p. 467]{leveson_engineering_2011} 
        \\ 

        &
        & 
        ``Hazard'' &
        ``A system state or set of conditions that, together with a particular set of worst-case environment conditions, will lead to an accident (loss).'' \cite[p. 467]{leveson_engineering_2011} 
        \\ 

        &
        &
        ``Risk Level'' &
        ``A function of the hazard level combined with (1) the likelihood of the hazard leading to an accident and (2) hazard exposure or duration.'' \cite[p. 467]{leveson_engineering_2011} 
        \\ 
        
        \midrule
    \end{tabularx}
    \caption[]{Examples of Different Issue and Incident Definitions from Various Industries (cont'd)}
\end{table*}

\setcounter{table}{19}
\begin{table*}[p]
    \centering
    \small
    \begin{tabularx}{\textwidth}{ s s s X }
        \toprule
        
        \textbf{Field} & 
        \textbf{Source} & 
        \textbf{Term} & 
        \textbf{Definition} 
        \\ 
        
        \midrule  

        Healthcare \& Pharmaceuticals &
        Department of Health and Human Services (HHS) & 
        ``Adverse event'' &
        ``An `adverse event' indicates that the harm prolonged the hospital stay; resulted in an elevation of care or transfer to another facility; caused permanent harm; required life-saving intervention; or contributed to death. '' \cite[p. 4]{maxwell_adverse_2023} 
        \\ 

        &
        &
        ``Patient harm'' &
        ``Patient harm refers collectively to adverse events and temporary harm events.'' \cite[p. 4]{maxwell_adverse_2023} 
        \\ 

        &
        &
        ``Temporary harm event'' &
        ``A `temporary harm event' is an event that resulted in patient harm and required medical intervention but did not prolong the patient stay, cause permanent harm, or require life-sustaining intervention.'' \cite[p. 4]{maxwell_adverse_2023} 
        \\

        &
        Food and Drug Administration (FDA) &
        ``Adverse event'' &
        ``Adverse event means any untoward medical occurrence associated with the use of a drug in humans, whether or not considered drug related.'' \cite{21_CFR_312.32} 
        \\

        &
        &
        ``Life-threatening adverse event or life-threatening suspected adverse reaction'' &
        ``An adverse event or suspected adverse reaction is considered `life-threatening' if, in the view of either the investigator or sponsor, its occurrence places the patient or subject at immediate risk of death. It does not include an adverse event or suspected adverse reaction that, had it occurred in a more severe form, might have caused death.'' \cite{21_CFR_312.32} 
        \\ 
        
        &
        &
        ``Serious adverse event or serious suspected adverse reaction'' &
        ``An adverse event or suspected adverse reaction is considered `serious' if, in the view of either the investigator or sponsor, it results in any of the following outcomes: Death, a life-threatening adverse event, inpatient hospitalization or prolongation of existing hospitalization, a persistent or significant incapacity or substantial disruption of the ability to conduct normal life functions, or a congenital anomaly/birth defect. Important medical events that may not result in death, be life-threatening, or require hospitalization may be considered serious when, based upon appropriate medical judgment, they may jeopardize the patient or subject and may require medical or surgical intervention to prevent one of the outcomes listed in this definition. '' \cite{21_CFR_312.32} 
        \\

        &
        &
        ``Suspected adverse reaction'' &
        ``Suspected adverse reaction means any adverse event for which there is a reasonable possibility that the drug caused the adverse event.'' \cite{21_CFR_312.32} 
        \\

        &
        &
        ``Unexpected adverse event or unexpected suspected adverse reaction'' &
        ``An adverse event or suspected adverse reaction is considered `unexpected' if it is not listed in the investigator brochure or is not listed at the specificity or severity that has been observed; or, if an investigator brochure is not required or available, is not consistent with the risk information described in the general investigational plan or elsewhere in the current application, as amended.'' \cite{21_CFR_312.32} 
        \\

        &
        National Cancer Institute & 
        ``Adverse event'' &
        ``An undesired effect of a drug or other type of treatment, such as surgery. Adverse events can range from mild to severe and can be life-threatening. Also called adverse effect and adverse reaction.'' \cite{nci_adverse_2011} 
        \\ 

        Nuclear &
        International Atomic Energy Agency (IAEA) & 
        ``Accident'' &
        ``Any unintended \textit{event}, including operating errors, equipment \textit{failures} and other mishaps, the consequences or potential consequences of which are not negligible from the point of view of \textit{protection and safety}.'' \cite[p. 10]{iaea_iaea_2022} 
        \\ 

        &
        &
        ``Event'' &
        ``In the context of the reporting and \textit{analysis} of \textit{events}, an event is any occurrence unintended by the \textit{operator}, including operating error, equipment \textit{failure} or other mishap, and deliberate action on the part of others, the consequences or potential consequences of which are not negligible from the point of view of \textit{protection and safety}.'' \cite[p. 77]{iaea_iaea_2022} 
        \\ 
        
        \midrule       
    \end{tabularx}
    \caption[]{Examples of Different Issue and Incident Definitions from Various Industries (cont'd)}
\end{table*}

\setcounter{table}{19}
\begin{table*}[p]
    \centering
    \small
    \begin{tabularx}{\textwidth}{ s s s X }
        \toprule
        
        \textbf{Field} & 
        \textbf{Source} & 
        \textbf{Term} & 
        \textbf{Definition} 
        \\ 
        \midrule 

        &
        &
        ``Failure (technical)'' &
        ``Loss of the ability of a \textit{structure}, \textit{system or component} to function within \textit{acceptance criteria}.'' \cite[p. 86]{iaea_iaea_2022} 
        \\ 
        
        &
        &
        ``Hazard'' &
        ``The potential for harm or other detriment, especially for \textit{radiation risks}; a factor or condition that might operate against \textit{safety}.'' \cite[p. 96]{iaea_iaea_2022} 
        \\ 

        &
        &
        ``Incident'' &
        ``Any unintended \textit{event}, including operating errors, equipment \textit{failures}, \textit{initiating events}, \textit{accident precursors}, \textit{near misses} or other mishaps, or unauthorized act, malicious or non-malicious, the consequences or potential consequences of which are not negligible from the point of view of protection and safety.  . . . The word \textit{incident} is sometimes used, for example in the INES 2008 Manual, to describe \textit{events} that are, in effect, minor \textit{accidents}, i.e. that are distinguished from \textit{accidents} only in terms of being less severe. . . . This is a distinction with little basis in general usage, in which an \textit{incident} can be minor or major, just as an \textit{accident} can; however, unlike an \textit{accident}, an \textit{incident} can be caused intentionally.'' \cite[p. 101]{iaea_iaea_2022} 
        \\
        
        &
        &
        ``Near miss'' &
        ``A potential significant \textit{event} that could have occurred as the consequence of a sequence of actual occurrences but did not occur owing to the conditions prevailing at the time.'' \cite[p. 132]{iaea_iaea_2022} 
        \\ 

        &
        &
        ``Threat'' &
        ``A person or group of persons with motivation, intention and capability to commit a \textit{malicious act}.'' \cite[p. 210]{iaea_iaea_2022} 
        \\ 

        &
        &
        ``Vulnerability'' &
        ``A physical feature or operational attribute that renders an entity, asset, system, network, facility, activity or geographic area open to exploitation or susceptible to a given \textit{threat}.'' \cite[p. 224]{iaea_iaea_2022} 
        \\ 
        
        &
        US statute & 
        ``Nuclear incident'' &
        ``The term `nuclear incident' means any occurrence within the United States causing, within or outside the United States, bodily injury, sickness, disease, or death, or loss of or damage to property, or loss of use of property, arising out of or resulting from the radioactive, toxic, explosive, or other hazardous properties of source, special nuclear, or byproduct material . . . .'' \cite{42_USC_2014} 
        \\ 

        \midrule

        Occupational Safety &
        International Labour Organization (ILO) & 
        ``Dangerous occurrence'' &
        ``A readily identifiable event, as defined under national laws and regulations, with the potential to cause an injury or disease to persons at work or to the public, e.g. a crane toppling over that results only in damage to property.'' \cite[p. iv]{ilo_investigation_2015} 
        \\ 

        &
        &
        ``Near miss/incident'' &
        ``An event, not necessarily defined under national laws and regulations, that could have caused harm to persons at work or to the public, e.g. a brick that falls off scaffolding but does not hit anyone.'' \cite[p. iv]{ilo_investigation_2015} 
        \\ 

        &
        &
        ``Occupational accident'' &
        ``An occurrence arising out of, or in the course of, work which results in a fatal or non-fatal injury, e.g. a fall from a height or contact with moving machinery.'' \cite[p. iv]{ilo_investigation_2015} 
        \\ 

        &
        &
        ``Hazard'' &
        ``Anything with the potential to cause harm, such as chemicals, electricity, work on ladders, an unguarded machine, an open drawer, demanding and stressful work, etc.'' \cite[p. iv]{ilo_investigation_2015} 
        \\ 

        &
        &
        ``Risk'' &
        ``The risk is the chance, high or low, that somebody could be harmed by these and other hazards, together with an indication of how serious the harm could be.'' \cite[p. iv]{ilo_investigation_2015} 
        \\ 

        \midrule
    \end{tabularx}
    \caption[]{Examples of Different Issue and Incident Definitions from Various Industries (cont'd)}
\end{table*}

\setcounter{table}{19}
\begin{table*}[p]
    \centering
    \small
    \begin{tabularx}{\textwidth}{ s s s X }
        \toprule
        
        \textbf{Field} & 
        \textbf{Source} & 
        \textbf{Term} & 
        \textbf{Definition} 
        \\ 
        \midrule 

        &
        \multirow{2}{=}{Occupational Safety and Health Administration (OSHA)} & 
        ``Close Call'' &
        ``An incident that could have caused serious injury or illness but did not; also called a `near miss.'\,'' \cite[p. 2]{osha_incident_2015} 
        \\ 

        &
        &
        ``Hazard'' &
        ``Any object, situation, or behavior that has the potential to cause injury, illness, or damage to property or the environment (hazards can be present anywhere including off-site, offices, communal areas, labs, etc.).'' \cite{osha_chapter_2020} 
        \\ 
        
        &
        &
        ``Incident'' &
        ``A work‐related event in which an injury or ill‐health (regardless of severity) or fatality occurred, or could have occurred.'' \cite[p. 2]{osha_incident_2015}\textsuperscript{1}

        \,
        
        \textsuperscript{1} ``In the past, the term `accident' was often used when referring to an unplanned, unwanted event. To many, `accident' suggests an event that was random, and could not have been prevented. Since nearly all worksite fatalities, injuries, and illnesses are preventable, OSHA suggests using the term `incident' investigation.'' \cite{osha_incident_nodate} 
        \\ 

        &
        &
        ``Incident'' &
        ``An event that causes an injury, illness, near miss, or report of a hazard that affects OSHA personnel.'' \cite{osha_chapter_2020} 
        \\ 

        &
        &
        ``Near miss'' &
        ``A near miss is a potential hazard or incident in which no property was damaged and no personal injury was sustained, but where, given a slight shift in time or position, damage or injury easily could have occurred. Near misses also may be referred to as close calls, near accidents, or injury-free events.'' \cite{osha_template_nodate} 
        \\ 
        
        &
        &
        ``Near miss'' &
        ``A close call or narrow escape, an unintentional incident that could have resulted in human injury, illness, environmental or equipment damage, or an interruption to normal operation.'' \cite{osha_chapter_2020} 
        \\ 
        \midrule

        Pesticides &
        Environmental Protection Agency (EPA) & 
        ``Pesticide incident'' &
        ``A simple way to define it is that an incident is any effect of a pesticide’s use that is not expected or intended. For relatively rare incidents adversely affecting humans, such as spills, leaks, pesticide movement off site, or chronic exposure to sensitive people or those with pre existing diseases, symptoms may include head aches, nausea, a breathing problem, or more severe symptoms. In extreme situations, systematic effects, serious illness, or death may occur. For an incident affecting the environment, impacts may include fish or bird kills or damage to plants. In addition, incidents may occur to domestic animals or wildlife. Incidents may be occupational, residential, or ecological.'' \cite[p. 3]{epa_opp_nodate} 
        \\

        \bottomrule
        
    \end{tabularx}
    \caption[]{Examples of Different Issue and Incident Definitions from Various Industries (cont'd)}
\end{table*}

\clearpage

\section{Appendix: Policy Initiatives for AI Incident Reporting} \label{sec:Appendix_Policy_Initiatives}

This Appendix provides examples of policy initiatives related to incident reporting for general-purpose AI systems (Tables \ref{tab:Policy_Initiatives_Enacted} and \ref{tab:Policy_Initiatives_Proposed}). Note that these lists were compiled in May 2024 and are not meant to be up-to-date, complete, or exhaustive; rather, we intend only to provide a sense of various initiatives around AI incident reporting. 

Table \ref{tab:Policy_Initiatives_Enacted} includes policy initiatives that have already been implemented and enacted---e.g., voluntary guidelines or codes of conduct that are currently in place, or legislation that is currently in effect (even though in some cases, regulatory requirements are in a buffer period). Table \ref{tab:Policy_Initiatives_Proposed} includes proposed (but not yet enacted as of the time of compilation in May 2024) initiatives---e.g., draft guidelines, proposed legislation, or policy papers from various governments. 

We included mostly national-level legislation, as well as some legislative proposals from US state governments. Policy initiatives were primarily sourced from \citet{iapp_global_2024} or Google Search using combinations of the keywords in Table \ref{tab:Review_Keywords}.

\begin{table*}[p]
    \centering
    \small
    \begin{tabularx}{\textwidth}{ s s s X }
        \toprule
        
        \textbf{Jurisdiction} & 
        \textbf{Initiative} & 
        \textbf{Type} &
        \textbf{Text} 
        \\ 
        
        \midrule 

        Canada & 
        Voluntary Code of Conduct on the Responsible Development and Management of Advanced Generative AI Systems & 
        Voluntary Guidelines &
        ``Developers'' of ``[a]ll advanced generative systems'' and ``[d]evelopers'' of ``[a]dvanced generative systems available for public use'' committed to ``[m]aintain a database of reported incidents after deployment, and provide updates as needed to ensure effective mitigation measures.''
        \cite{ised_voluntary_2023} 
        \\
        
        \midrule

        China & 
        Interim Measures for the Management of Generative Artificial Intelligence Services & 
        Regulation &
        ``Where providers discover illegal content they shall promptly employ measures to address it such as stopping generation, stopping transmission, and removal, employ measures such as model optimization training to make corrections and report to the relevant departments in charge.

        \,

        Where providers discover that users are using generative AI services to engage in illegal activities, they shall employ measures in accordance with laws and agreements to address it, including warnings, limiting functions, and suspending or concluding the provision of services, and store the relevant records and report to the relevant departments in charge.'' \cite[Art. 14]{cac_translation_2023}\textsuperscript{1}

        \,
        
        ``Providers shall establish and complete mechanisms for making complaints and reports, setting up easy complaint and reporting portals, disclosing the process for handling them and the time limits for giving responses, and promptly accepting and handling complaints and reports from the public and giving feedback on the outcome.'' \cite[Art. 15]{cac_translation_2023}
        
        \,
        
        ``Where users discover that generative AI services do not comply with laws, administrative regulations, or these Measures, they have the right to make a complaint or report to the relevant departments in charge.'' \cite[Art. 18]{cac_translation_2023}\textsuperscript{2} 

        \,

        \textsuperscript{1} ``Providers'' refers to ``generative AI service providers,'' defined as ``organizations and individuals that use generative AI technology to provide generative AI services (including providing generative AI services through programmable interfaces and other means).'' ``Generative AI technology'' is defined as ``models and relevant technologies that have the ability to generate content such as texts, images, audio, or video.'' \cite[Art. 23]{cac_translation_2023}  
        
        \textsuperscript{2} ``Users'' refers to ``generative AI service users'', defined as ``organizations and individuals that use generative AI services to generate content.'' \cite[Art. 23]{cac_translation_2023}
        \\

        \midrule
        
        China & 
        Provisions on the Administration of Deep Synthesis Internet Information Services &
        Regulation &
        ``Where deep synthesis service providers discover illegal or negative information, they shall employ measures to address it in accordance with law, store related records, and promptly make a report to the telecommunications department or relevant departments in charge; and take measures in accordance with laws and agreements against the related deep synthesis service users, such as giving warnings, restricted functions, service suspensions, and account closures.'' \cite[Art. 10]{cac_translation_2022}\textsuperscript{3} 
        \\
        
        \midrule
    \end{tabularx}
    \caption{Policy Initiatives for General-Purpose AI Incident Reporting, Enacted or Implemented}\label{tab:Policy_Initiatives_Enacted}
\end{table*}

\setcounter{table}{20}
\begin{table*}[p]
    \centering
    \small
    \begin{tabularx}{\textwidth}{ s s s X }
        \toprule
        
        \textbf{Jurisdiction} & 
        \textbf{Initiative} & 
        \textbf{Type} &
        \textbf{Text} 
        \\ 

        \midrule

        &
        &
        &

        ``Deep synthesis service providers shall establish and complete mechanisms for dispelling rumors, and where it is discovered that deep synthesis information services were used to produce, reproduce, publish, or transmit false information, they shall promptly employ measures to dispel the rumors, store related records, and make a report to the internet information departments and relevant departments in charge. '' \cite[Art. 11]{cac_translation_2022} 

        \,
        
        ``Deep synthesis service providers shall set up convenient portals for user appeals, public complaints, and reports, and shall publish the process for handling them and the time limits for responses, promptly accepting and handling them, and giving feedback on the outcome.'' \cite[Art. 12]{cac_translation_2022}

        \,
        
        \textsuperscript{3} ``Providers'' refers to ``deep synthesis service providers'', which are defined as ``organizations and individuals providing deep synthesis services.'' ``Deep synthesis technology'' is defined as ``use of technologies such as deep learning and virtual reality, that use generative sequencing algorithms to create text, images, audio, video, virtual scenes, or other information''. \cite[Art. 23]{cac_translation_2022} 
        \\
        
        \midrule 

        China & 
        Basic Safety Requirements for Generative Artificial Intelligence Services & 
        Technical Standard &
        ``[Corpus content safety r]equirements for providers are as follows . . . [b)] 3)Channels for reporting complaints on IPR [Intellectual Property Rights] issues shall be established. . . . 6) The following IPR measures should be in place . . . Support in complaint reporting channels for third-party inquiries about corpus usage and related IPR circumstances.'' \cite[Sec. 5.2]{tc_260_translation_2024}\textsuperscript{4}

        \,

        ``[Safety measure r]equirements for providers are as follows . . . f) Acceptance of complaints and reports from the public or users: 1) Ways for accepting complaints and reports from the public or users, as well as feedback methods, shall be provided, including but not limited to one or more methods such as telephone, email, interactive windows, and text messages; 2) The rules for handling complaints and reports from the public or users and the time limit for said handling shall be established.'' \cite[Sec. 7]{tc_260_translation_2024}

        \,

        \textsuperscript{4} ``Providers'' refers to ``service provider[s]'', defined as ``organization[s] or individual[s] that provides generative AI services in the form of interactive interfaces, programmable interfaces, etc.'' \cite[Sec. 3.2]{tc_260_translation_2024}. ``Generative AI services'' are defined as ``[t]he use of generative AI technology to provide text, graphics, audio, video, and other content generation services to the public within [the borders of] the People’s Republic of China.'' \cite[Sec. 3.1]{tc_260_translation_2024} 
        \\
        
        \midrule
    \end{tabularx}
    \caption[]{Policy Initiatives for General-Purpose AI Incident Reporting, Enacted or Implemented (cont'd)}
\end{table*}

\setcounter{table}{20}
\begin{table*}[p]
    \centering
    \small
    \begin{tabularx}{\textwidth}{ s s s X }
        \toprule
        
        \textbf{Jurisdiction} & 
        \textbf{Initiative} & 
        \textbf{Type} &
        \textbf{Text} 
        \\ 

        \midrule

        Colorado (US state) & 
        Consumer Protections for Artificial Intelligence (SB24-205) &
        Legislation &
        ``If a deployer deploys a high-risk artificial intelligence system on or after February 1, 2026, and subsequently discovers that the high-risk artificial intelligence system has caused algorithmic discrimination, the deployer, without unreasonable delay, but no later than ninety days after the date of the discovery, shall send to the Attorney General, in a form and manner prescribed by the Attorney General, a notice disclosing the discovery.'' \cite[6-1-1601(7)]{rodriguez_sb24-205_2024}
        \\
        
        \midrule 

        European Union & 
        Artificial Intelligence Act (AI Act or AIA) &
        Legislation &
        ``\,`[S]erious incident’ means an incident or malfunctioning of an AI system that directly or indirectly leads to any of the following:

        \begin{enumerate}
            \item [(a)] the death of a person, or serious harm to a person’s health;
            \item [(b)] a serious and irreversible disruption of the management or operation of critical infrastructure.
            \item [(c)] the infringement of obligations under Union law intended to protect fundamental rights;
            \item [(d)] serious harm to property or the environment[.]'' \citepalias[Art. 3(49)]{AIA}
        \end{enumerate} 
        
        \,
        
        ``Deployers shall monitor the operation of the high-risk AI system on the basis of the instructions for use and, where relevant, inform providers in accordance with Article 72.  . . . Where deployers have identified a serious incident, they shall also immediately inform first the provider, and then the importer or distributor and the relevant market surveillance authorities of that incident. If the deployer is not able to reach the provider, Article 73 shall apply mutatis mutandis.'' \citepalias[Art. 26(5)]{AIA}

        \,

        ``In addition to the obligations listed in Articles 53 and 54, providers of general-purpose AI models with systemic risk shall:  . . . (c) keep track of, document, and report, without undue delay, to the AI Office and, as appropriate, to national competent authorities, relevant information about serious incidents and possible corrective measures to address them;'' \citepalias[Art. 55(1)]{AIA}

        \,

        ``Any serious incident identified in the course of the testing in real world conditions shall be reported to the national market surveillance authority in accordance with Article 73 [reproduced in part below]. The provider or prospective provider shall adopt immediate mitigation measures or, failing that, shall suspend the testing in real world conditions until such mitigation takes place, or otherwise terminate it. The provider or prospective provider shall establish a procedure for the prompt recall of the AI system upon such termination of the testing in real world conditions.'' \citepalias[Art. 60(7)]{AIA}

        \,

        ``Providers of high-risk AI systems placed on the Union market shall report any serious incident to the market surveillance authorities of the Member States where that incident occurred.'' \citepalias[Art. 73(1)]{AIA}

        \\
        
        \midrule
    \end{tabularx}
    \caption[]{Policy Initiatives for General-Purpose AI Incident Reporting, Enacted or Implemented (cont'd)}
\end{table*}

\setcounter{table}{20}
\begin{table*}[p]
    \centering
    \small
    \begin{tabularx}{\textwidth}{ s s s X }
        \toprule
        
        \textbf{Jurisdiction} & 
        \textbf{Initiative} & 
        \textbf{Type} &
        \textbf{Text} 
        \\ 
        
        \midrule 

        &
        &
        &
        The report referred to in paragraph 1 [immediately above] shall be made immediately after the provider has established a causal link between the AI system and the serious incident or the reasonable likelihood of such a link, and, in any event, not later than 15 days after the provider or, where applicable, the deployer, becomes aware of the serious incident. The period for the reporting referred to in the first subparagraph shall take account of the severity of the serious incident.'' \citepalias[Art. 73(2)]{AIA}

        \,
        
        ``Notwithstanding paragraph 2 of this Article [Article 73(2), reproduced above], in the event of a widespread infringement or a serious incident as defined in Article 3, point (49) (b) [reproduced above; 'an incident or malfunctioning of an AI system that directly or indirectly leads to . . .  serious and irreversible disruption of the management or operation of critical infrastructure'], the report referred to in paragraph 1 of this Article shall be provided immediately, and not later than two days after the provider or, where applicable, the deployer becomes aware of that incident.'' \citepalias[Art. 73(3)]{AIA}
        
        \,

        ``Notwithstanding paragraph 2 [Article 73(2), reproduced above], in the event of the death of a person [see Article 3(49)(b), reproduced above], the report shall be provided immediately after the provider or the deployer has established, or as soon as it suspects, a causal relationship between the high-risk AI system and the serious incident, but not later than 10 days after the date on which the provider or, where applicable, the deployer becomes aware of the serious incident.'' \citepalias[Art. 73(4)]{AIA}
        
        \,

        ``Where necessary to ensure timely reporting, the provider or, where applicable, the deployer, may submit an initial report that is incomplete, followed by a complete report.'' \citepalias[Art. 73(5)]{AIA}
        
        \,

        ``Following the reporting of a serious incident pursuant to paragraph 1 [of Article 73, reproduced above], the provider shall, without delay, perform the necessary investigations in relation to the serious incident and the AI system concerned. This shall include a risk assessment of the incident, and corrective action. The provider shall cooperate with the competent authorities, and where relevant with the notified body concerned, during the investigations referred to in the first subparagraph, and shall not perform any investigation which involves altering the AI system concerned in a way which may affect any subsequent evaluation of the causes of the incident, prior to informing the competent authorities of such action.'' \citepalias[Art. 73(6)]{AIA}
        
        \\
        
        \midrule
    \end{tabularx}
    \caption[]{Policy Initiatives for General-Purpose AI Incident Reporting, Enacted or Implemented (cont'd)}
\end{table*}

\setcounter{table}{20}
\begin{table*}[p]
    \centering
    \small
    \begin{tabularx}{\textwidth}{ s s s X }
        \toprule
        
        \textbf{Jurisdiction} & 
        \textbf{Initiative} & 
        \textbf{Type} &
        \textbf{Text} 
        \\ 
        
        \midrule 

        &
        &
        &
        
        ``Upon receiving a notification related to a serious incident referred to in Article 3, point (49)(c) [reproduced above; `an incident or malfunctioning of an AI system that directly or indirectly leads to . . . the infringement of obligations under Union law intended to protect fundamental rights'], the relevant market surveillance authority shall inform the national public authorities or bodies referred to in Article 77(1). The Commission shall develop dedicated guidance to facilitate compliance with the obligations set out in paragraph 1 of this Article [Article 73(1), reproduced above]. That guidance shall be issued by . . . [12 months after the entry into force of this Regulation], and shall be assessed regularly.'' \citepalias[Art. 73(7)]{AIA}
        
        ``The market surveillance authority shall take appropriate measures, as provided for in Article 19 of Regulation (EU) 2019/1020, within seven days from the date it received the notification referred to in paragraph 1 of this Article [Article 73(1), reproduced above], and shall follow the notification procedures as provided in that Regulation.'' \citepalias[Art. 73(8)]{AIA}
        
        \,

        ``For high-risk AI system referred to in Annex III that are placed on the market or put into service by providers that are subject to Union legislative instruments laying down reporting obligations equivalent to those set out in this Regulation, the notification of serious incidents shall be limited to those referred to in Article 3, point (49)(c) [reproduced above; `an incident or malfunctioning of an AI system that directly or indirectly leads to . . . the infringement of obligations under Union law intended to protect fundamental rights'].'' \citepalias[Art. 73(9)]{AIA}
        
        \,

        ``For high-risk AI systems which are safety components of devices, or are themselves devices, covered by Regulations (EU) 2017/745 and (EU) 2017/746, the notification of serious incidents shall be limited to those referred to in Article 3, point (49)(c) of this Regulation [reproduced above; `an incident or malfunctioning of an AI system that directly or indirectly leads to . . . the infringement of obligations under Union law intended to protect fundamental rights'], and shall be made to the national competent authority chosen for that purpose by the Member States where the incident occurred.'' \citepalias[Art. 73(10)]{AIA}
        
        \,

        ``National competent authorities shall immediately notify the Commission of any serious incident, whether or not they have taken action on it, in accordance with Article 20 of Regulation (EU) 2019/1020.'' \citepalias[Art. 73(11)]{AIA}

        \,

        ``Where a market surveillance authority [when supervising `testing in real-world conditions' per Article 60] has been informed by the prospective provider, the provider or any third party of a serious incident or has other grounds for considering that the conditions set out in Articles 60 and 61 are not met, it may take either of the following decisions on its territory, as appropriate:

        \begin{enumerate}
            \item [(a)] to suspend or terminate the testing in real world conditions;
            \item [(b)] to require the provider or prospective provider and the deployer or prospective deployer to modify any aspect of the testing in real world conditions.''
        \end{enumerate}
        \citepalias[Art. 76(3)]{AIA} 
        \\
        
        \midrule
    \end{tabularx}
    \caption[]{Policy Initiatives for General-Purpose AI Incident Reporting, Enacted or Implemented (cont'd)}
\end{table*}

\setcounter{table}{20}
\begin{table*}[p]
    \centering
    \small
    \begin{tabularx}{\textwidth}{ s s s X }
        \toprule
        
        \textbf{Jurisdiction} & 
        \textbf{Initiative} & 
        \textbf{Type} &
        \textbf{Text} 
        \\ 
        
        \midrule 

        Japan &
        AI Guidelines for Business Ver1.0 & 
        Voluntary Guidelines &
        ``[In `Common Guiding Principles for AI Business actors involved in advanced AI systems'] II) Identify and mitigate vulnerabilities, and, where appropriate, incidents and patterns of misuse, after deployment including placement on the market. (`5) Ensuring security' and `7) Accountability')

        \,
        
        Use, as and when appropriate commensurate to the level of risk, AI systems as intended and monitor for vulnerabilities, and take appropriate action to address these.

        \,
        
        encouraged [sic] to maintain appropriate documentation of reported incidents and to mitigate the identified risks and vulnerabilities, in collaboration with other stakeholders.'' \cite[pt. 2.D]{mic_draft_2024}

        \,
        
        ``[In `Common Guiding Principles for AI Business actors involved in advanced AI systems'] IV) Work towards responsible information sharing and reporting of incidents among organizations developing advanced AI systems including with industry, governments, civil society, and academia. (`5) Ensuring security', `6) Transparency', `7) Accountability', `10) Innovation')

        \,
        
        These include reports on monitoring results and documents related to security and safety risks.'' \cite[pt. 2.D]{mic_draft_2024}

        \,  
        
        ``[In `Matters Related to AI Developers'] When an unexpected incident occurs in an AI system, any party in the AI value chain may be required to explain that incident. Bearing this in mind, it is important for AI developers to leave records that help them reasonably explain how they were involved in the AI system.'' \cite[pt. 3]{mic_draft_2024}

        \,

        ``[In `Matters Related to AI Providers''] Upon provision, AI providers are expected to pay attention to prevent violations of stakeholders' rights and the occurrence of social drawbacks, etc., and share information of incidents and the like within reasonable extent to provide safer, more secure and reliable AI systems and services. . . . 

        \,
        
        P-6) ii. Providing relevant stakeholders with information
        
        \,
        
        Provide information on the AI system or service to be provided (for example, the items listed below) in a timely and appropriate manner so that it can be easily understood and accessed (`6) Transparency'). . . . 

        \,
        
        Information on the operational status of the AI systems and services, causes of failures, status of actions against them, incidents, etc. (`2) Safety').'' \cite[pt. 4]{mic_draft_2024} 
        \\

        \midrule
        
    \end{tabularx}
    \caption[]{Policy Initiatives for General-Purpose AI Incident Reporting, Enacted or Implemented (cont'd)}
\end{table*}

\setcounter{table}{20}
\begin{table*}[tp]
    \centering
    \small
    \begin{tabularx}{\textwidth}{ s s s X }
        \toprule
        
        \textbf{Jurisdiction} & 
        \textbf{Initiative} & 
        \textbf{Type} &
        \textbf{Text} 
        \\ 
        
        \midrule 

        United States &
        Artificial Intelligence Risk Management Framework v1.0 (NIST AI RMF) &
        Voluntary Guidelines &
        ``GOVERN 4.3: Organizational practices are in place to enable AI testing, identification of incidents, and information sharing.

        \,
        
        GOVERN 6.2: Contingency processes are in place to handle failures or incidents in third-party data or AI systems deemed to be high-risk.

        \,

        MANAGE 4.1: Post-deployment AI system monitoring plans are implemented, including mechanisms for capturing and evaluating input from users and other relevant AI actors, appeal and override, decommissioning, incident response, recovery, and change management.

        \,
        
        MAP 5.1: Likelihood and magnitude of each identified impact (both potentially beneficial and harmful) based on expected use, past uses of AI systems in similar contexts, public incident reports, feedback from those external to the team that developed or deployed the AI system, or other data are identified and documented.

        \,

        MANAGE 4.3: Incidents and errors are communicated to relevant AI actors, including affected communities. Processes for tracking, responding to, and recovering from incidents and errors are followed and documented.'' \cite{nist_ai_2023}
        \\
        
        \midrule

        United States & 
        Executive Order on the Safe, Secure, and Trustworthy Development and Use of Artificial Intelligence & 
        Executive Order &
        ``Within 365 days of the date of this order [on October 30, 2023], the Secretary of HHS [the Department of Health and Human Services] shall, in consultation with the Secretary of Defense and the Secretary of Veterans Affairs, establish an AI safety program that, in partnership with voluntary federally listed Patient Safety Organizations . . . establishes a common framework for approaches to identifying and capturing clinical errors resulting from AI deployed in healthcare settings as well as specifications for a central tracking repository for associated incidents that cause harm, including through bias or discrimination, to patients, caregivers, or other parties[.]'' \cite{biden_executive_2023}
        \\
        
        \bottomrule
    \end{tabularx}
    \caption[]{Policy Initiatives for General-Purpose AI Incident Reporting, Enacted or Implemented (cont'd)}
\end{table*}

\setcounter{table}{21}
\begin{table*}[p]
    \centering
    \begin{tabularx}{\textwidth}{ s s s X } 
        \toprule
        
        \textbf{Jurisdiction} & 
        \textbf{Initiative} & 
        \textbf{Type} &
        \textbf{Text} 
        \\ 
        
        \midrule 

        Brazil & 
        Draft AI Law, Bill No. 2338 & 
        Legislation &
        ``Artificial intelligence agents will report to the competent authority the occurrence of serious security incidents, including when there are risks posed to life and integrity of individuals, interruption of the functioning of critical infrastructure operations, serious damage to property or the environment, as well as serious violations of fundamental rights, in accordance with the applicable regulation.
        
        \begin{enumerate}
            \item The communication will be made within a reasonable timeframe, as defined by the competent authority.
            \item The competent authority will assess the incident's severity and may order the agent, whenever necessary, to take steps and measures to reverse or mitigate the effects of the incident.''
        \end{enumerate}
        \cite[Art. 31]{pacheco_bill_2023}
        \\

        \midrule

    \end{tabularx}
    \caption{Policy Initiatives for General-Purpose AI Incident Reporting, Proposed}
    \label{tab:Policy_Initiatives_Proposed}
\end{table*}

\setcounter{table}{21}
\begin{table*}[p]
    \centering
    \begin{tabularx}{\textwidth}{ s s s X } 
        \toprule
        
        \textbf{Jurisdiction} & 
        \textbf{Initiative} & 
        \textbf{Type} &
        \textbf{Text} 
        \\ 
        
        \midrule 

        Canada & 
        Artificial Intelligence and Data Act, Proposed Amendment 11 &
        Legislation &
        ``A person who manages the operations of a general-purpose system must . . . if there are reasonable grounds to suspect that the use of the system has resulted, directly or indirectly, in serious harm or that the mitigation measures are not effective in mitigating risks of serious harm that could result from the use of the system, assess whether the use of the system did actually result in serious harm or the measures are actually not effective in mitigating those risks and, if so, 

        \begin{enumerate}
            \item [(i)] cease the system’s operations until additional or modified measures are put in place that will mitigate risks of serious harm,
            \item [(ii)] prepare a report and provide it to the Commissioner, in accordance with the regulations, and
            \item [(iii)] notify the Commissioner when the system resumes operation;''
        \end{enumerate}

        \cite[Amd. 8.2(1).]{champagne_letter_2023} [Amd. 11.2 would impose the same requirement on operators of ``high-impact system[s].'']
        \\

        \midrule

        China & 
        Model AI Law v2.0 from the Chinese Academy of Social Sciences &
        Legislation &
        ``AI developers and providers are also obligated to promptly disseminate best practices for security, guiding users toward the safe and correct application of artificial intelligence. In compliance with the provisions of this law regarding record-keeping and technical documentation, AI developers and providers must ensure traceability, enabling prompt and accurate tracing and pinpointing of issues in the event of incidents, thereby ensuring AI safety.'' \cite[Art. 34]{zhou_model_2024}

        \,

        ``AI developers and providers must enhance risk monitoring and, upon identifying any safety defects or security vulnerabilities, must immediately take remedial actions. If AI developers discover any safety incidents with the AI they have developed, they must immediately implement corrective measures and notify AI providers. In turn, AI providers, in accordance with the third provision of this article, must fulfill their notification obligations. When AI providers identify any safety incidents with the AI they offer, they must immediately take action, comply with the notification obligations as outlined in the third provision of this article, and promptly inform the AI developers. 

        \,
        
        AI providers are required to take immediate corrective actions and, in accordance with national regulations, promptly inform users and report to the supervisory authorities regarding the following matters:

        \begin{enumerate}
            \item [a.] The occurrence and impact scope of the safety incident;
            \item [b.] The remedial actions already taken by the AI provider, as well as potential actions that users can take to mitigate harm;
            \item [c.] The contact information of the AI provider.'' (cont'd)
        \end{enumerate}
        \\
        
        \midrule
    \end{tabularx}
    \caption[]{Policy Initiatives for General-Purpose AI Incident Reporting, Proposed (Cont'd)}
\end{table*}

\setcounter{table}{21}
\begin{table*}[p]
    \centering
    \begin{tabularx}{\textwidth}{ s s s X } 
        \toprule
        
        \textbf{Jurisdiction} & 
        \textbf{Initiative} & 
        \textbf{Type} &
        \textbf{Text} 
        \\ 
        
        \midrule 

        &
        &
        &
        (cont'd) ``If the measures taken by the AI providers can effectively prevent substantial harm to users, notifying the users may be deemed unnecessary. However, if the National AI Administrative Authority believes the situation could be harmful, it reserves the authority to require the provider to notify affected users.
        
        \,
        
        Upon discovering a safety incident or receiving notification of such from the provider, AI developers should immediately perform an assessment. If the assessment reveals risks that originated during the development stage, the developer must immediately notify other AI providers and temporarily suspend or remove the affected AI products or services until the risks are resolved. Upon receiving such notification, other AI providers must immediately take remedial actions to manage the risks associated with their products and services.'' \cite[Art. 37]{zhou_model_2024}

        \,

        ``The National AI Administrative Authority shall exercise the following administrative responsibilities for AI in accordance with the law: . . . Establish an emergency response mechanism for AI safety incidents.'' \cite[Art. 54]{zhou_model_2024}

        \,

        ``If the National AI Administrative Authority and local AI administrative authorities at all levels, in the course of fulfilling their responsibilities, discover that activities related to the R\&D or provision of AI pose significant risks or have resulted in safety/security incidents, they may, in accordance with specified authorities and procedures, interview the relevant AI developers or providers and require them to take the following measures:

        \begin{enumerate}
            \item [a.] Implement corrective actions as specified to eliminate potential risks;
            \item [b.] Provide an appropriate explanation of their R\&D, and provisioning activities, clarify responsibilities related to the development, management, and operation of AI services, discuss measures taken to ensure fairness, safety, and stability, and assess the impact on stakeholders;
            \item [c.] Commission a professional organization to conduct a compliance audit of their AI R\&D and provisioning activities.
        \end{enumerate}
        
        If AI developers and providers commit to timely corrective actions to achieve compliance, and can effectively avoid causing harm through their AI R\&D, provision, or usage activities, they may not be required to suspend relevant activities. However, if the National AI Administrative Authority deems that potential harm may occur, it may order the suspension of such activities.'' \cite[Art. 58]{zhou_model_2024}
        
        \\
        
        \midrule
    \end{tabularx}
    \caption[]{Policy Initiatives for General-Purpose AI Incident Reporting, Proposed (Cont'd)}
\end{table*}

\setcounter{table}{21}
\begin{table*}[p]
    \centering
    \begin{tabularx}{\textwidth}{ s s s X } 
        \toprule
        
        \textbf{Jurisdiction} & 
        \textbf{Initiative} & 
        \textbf{Type} &
        \textbf{Text} 
        \\ 
        
        \midrule 

        &
        &
        &
        ``Where an artificial intelligence developer or provider violates the provisions of this Law during R\&D or provision activities, resulting in a significant safety/security incident, multiple safety/security incidents, or multiple administrative penalties, the National AI Administrative Authority may suspend the license and order rectification within a prescribed period; if rectification is not made upon the expiration of the period, or if a safety/security incident occurs again or an administrative penalty is imposed after the suspension of the license, the National AI Administrative Authority may revoke the license.'' \cite[Art. 67]{zhou_model_2024}
        \\

        \midrule

        China & 
        Model AI Law from the China University of Political Science and Law &
        Legislation &
        ``Reporting of Major Safety and Security Incidents[. ]AI developers and providers shall immediately take measures for handling AI safety and security incidents; in the event of a major AI safety or security incident, it shall be reported promptly to the main oversight department for AI in accordance with regulations. The administrative methods for reporting major AI safety and security incidents shall be formulated by the main oversight departments for AI under the State Council.''\textsuperscript{1} \cite[Art. 43]{zhang_artificial_2024}

        \,

        \textsuperscript{1} The same term is used to mean both ``safety'' and ``security'' in Chinese. Only sections identified in the English translation as relating to ``safety'' incidents are excerpted here, though other sections contain references to ``security'' incidents.
        \\

        \midrule

        G7 &
        Hiroshima Process International Code of Conduct for Organizations Developing Advanced AI Systems &
        Voluntary Guidelines &
        ``Identify and mitigate vulnerabilities, and, where appropriate, incidents and patterns of misuse, after deployment including placement on the market. 
        
        \,
        
        Organizations should use, as and when appropriate commensurate to the level of risk, AI systems as intended and monitor for vulnerabilities, incidents, emerging risks and misuse after deployment, and take appropriate action to address these. Organizations are encouraged to consider, for example, facilitating third-party and user discovery and reporting of issues and vulnerabilities after deployment such as through bounty systems, contests, or prizes to incentivize the responsible disclosure of weaknesses. Organizations are further encouraged to maintain appropriate documentation of reported incidents and to mitigate the identified risks and vulnerabilities, in collaboration with other stakeholders. Mechanisms to report vulnerabilities, where appropriate, should be accessible to a diverse set of stakeholders.'' \cite{g7_hiroshima_2023-1}

        \,
    
        ``Work towards responsible information sharing and reporting of incidents among organizations developing advanced AI systems including with industry, governments, civil society, and academia[.]'' (cont'd)
        \\
        
        \midrule
    \end{tabularx}
    \caption[]{Policy Initiatives for General-Purpose AI Incident Reporting, Proposed (Cont'd)}
\end{table*}

\setcounter{table}{21}
\begin{table*}[p]
    \centering
    \begin{tabularx}{\textwidth}{ s s s X } 
        \toprule
        
        \textbf{Jurisdiction} & 
        \textbf{Initiative} & 
        \textbf{Type} &
        \textbf{Text} 
        \\ 
        
        \midrule

        \,

        &
        &
        &
        (cont'd) ``This includes responsibly sharing information, as appropriate, including, but not limited to evaluation reports, information on security and safety risks, dangerous intended or unintended capabilities, and attempts by AI actors to circumvent safeguards across the AI lifecycle.

        \,
        
        Organizations should establish or join mechanisms to develop, advance, and adopt, whereappropriate, shared standards, tools, mechanisms, and best practices for ensuring thesafety, security, and trustworthiness of advanced AI systems.

        \,

        This should also include ensuring appropriate and relevant documentation and transparency across the AI lifecycle in particular for advanced AI systems that cause significant risks to safety and society.

        \,
        
        Organizations should collaborate with other organizations across the AI lifecycle to share and report relevant information to the public with a view to advancing safety, security and trustworthiness of advanced AI systems. Organizations should also collaborate and share the aforementioned information with relevant public authorities, as appropriate.

        \,

        Such reporting should safeguard intellectual property rights.'' \cite{g7_hiroshima_2023-1}
        \\

        \midrule

        G7 &
        Hiroshima Process International Guiding Principles for Organizations Developing Advanced AI system &
        General Principles &
        ``Patterns of misuse, after deployment including placement on the market. 
        
        \,
        
        Organizations should use, as and when appropriate commensurate to the level of risk, AI systems as intended and monitor for vulnerabilities, incidents, emerging risks and misuse after deployment, and take appropriate action to address these. Organizations are encouraged to consider, for example, facilitating third-party and user discovery and reporting of issues and vulnerabilities after deployment. Organizations are further encouraged to maintain appropriate documentation of reported incidents and to mitigate the identified risks and vulnerabilities, in collaboration with other stakeholders. Mechanisms to report vulnerabilities, where appropriate, should be accessible to a diverse set of stakeholders.'' \cite{g7_hiroshima_2023}

        \,

        ``[Organizations should] Work towards responsible information sharing and reporting of incidents among organizations developing advanced AI systems including with industry, governments, civil society, and academia. 
        
        \,
        
        This includes responsibly sharing information, as appropriate, including, but not limited to evaluation reports, information on security and safety risks, dangerous, intended or unintended capabilities, and attempts AI actors to circumvent safeguards across the AI lifecycle.'' \cite{g7_hiroshima_2023}
        \\
        
        \midrule
    \end{tabularx}
    \caption[]{Policy Initiatives for General-Purpose AI Incident Reporting, Proposed (Cont'd)}
\end{table*}

\setcounter{table}{21}
\begin{table*}[p]
    \centering
    \begin{tabularx}{\textwidth}{ s s s X } 
        \toprule
        
        \textbf{Jurisdiction} & 
        \textbf{Initiative} & 
        \textbf{Type} &
        \textbf{Text} 
        \\ 
        
        \midrule 

        New York (US state) & 
        Advanced Artificial Intelligence Licensing Act (A8195) &
        Legislation &
        ``A licensee shall have the duty to notify the department and, if applicable, a relevant law enforcement agency or governmental entity where the licensee's system  fails to operate as intended for any significant period of time. A period of time is deemed `significant' for purposes of this section where the period of time that the malfunction occurred had the capacity to or has harmed a person or persons.  . . . A licensee shall have the duty to notify a relevant law enforcement agency or governmental entity of a malfunction where designated by the department upon receipt of a license. The secretary shall issue such a requirement upon the licensee where such systems interact with law enforcement systems or the systems of a government agency, engage in law enforcement functions or the functions of a government agency, or where such systems operate, in whole or in part, or are, a weapon.'' \cite[Sec. 420]{hyndman_a8195_2023}
        \\

        \midrule

        Singapore &
        Proposed Model AI Governance Framework for Generative AI &
        Voluntary Guidelines &
        ``Even with the most robust development processes and safeguards, no software that we use today is foolproof. The same applies to AI. Incident reporting is an established practice, including in critical domains such as telecommunications, finance and cybersecurity. It allows for timely notification and remediation. Establishing the structures and processes to enable incident reporting is therefore key. This in turn supports continuous improvement of AI systems through insights, remediation and patching. 
        
                \,
                
        . . . After incidents happen, organisations need internal processes to report the incident for timely notification and remediation. Depending on the impact of the incident and how extensively AI was involved, this could include notifying both the public as well as governments. Defining `severe AI incidents' or setting the materiality threshold for formal reporting is therefore key. Borrowing from cybersecurity, AI incidents can be reported to the equivalent of `Information Sharing and Analysis Centres', which are trusted entities to foster information sharing and good practices, as well as to relevant authorities where required by law.

        \,

        Reporting should be proportionate, which means striking a balance between comprehensive reporting and practicality. This will need to be calibrated to suit the specific local context. In this regard, the impending EU AI Act provides one reference point for legal reporting requirements (see box below).'' \cite[Sec. 4]{aivf_proposed_2024}
        \\
        
        \midrule
    \end{tabularx}
    \caption[]{Policy Initiatives for General-Purpose AI Incident Reporting, Proposed (Cont'd)}
\end{table*}

\setcounter{table}{21}
\begin{table*}[p]
    \centering
    \begin{tabularx}{\textwidth}{ s s s X } 
        \toprule
        
        \textbf{Jurisdiction} & 
        \textbf{Initiative} & 
        \textbf{Type} &
        \textbf{Text} 
        \\ 

        \midrule

        United Kingdom &
        ``Emerging processes for frontier AI safety'' from the Department for Science, Innovation \& Technology &
        Policy Paper &
        ``Report any details of security or safety incidents or near misses to relevant government authorities. This includes any compromise to the security of the organisation or its systems, or any incident where an AI system -- deployed or not -- causes substantial harm or is close to doing so. This will enable government authorities to build a clear picture of when safety and security incidents occur and make it easier to anticipate and mitigate future risks. Incident reports could include a description of the incident, the location, start and end date, details of any parties affected and harms occurred, any specific models involved, any relevant parties responsible for managing and responding to the incident, as well as ways in which the incident could have been avoided. It is important that incidents indicative of more severe risks are reported as soon as possible after they occur. High-level details of safety and security incidents - with sensitive information removed - could also be made public, such as have been shared in the AI incident database.'' \cite{dsit_emerging_2023} [The document also discusses reporting of security incidents and ``vulnerabilities,'' which is broadly defined and appears to include some safety or rights incidents.]
        \\

        \midrule
        
        United Nations &
        ``Interim Report: Governing AI for Humanity'' from the UN High-Level Advisory Body on Artificial Intelligence &
        Policy Paper &
        ``Institutional Function 6: Monitor risks, report incidents, coordinate emergency response

        \,
        
        The borderless nature of AI tools, which can proliferate across the globe at the stroke of a key, pose new challenges to international security and global stability. AI models could lower the barriers for access to weapons of mass destruction. AI-enabled cyber tools increase the risk of attacks on critical infrastructure and dual-use AI can be used to power lethal autonomous weapons which could pose a risk to international humanitarian law and other norms. Bots can rapidly disseminate harmful information, with increasingly human characteristics, in a manner that can cause significant damage to markets and public institutions. The possibility of rogue AI escaping control and posing still larger risks cannot be ruled out. Given these challenges, capabilities must be created at a global level to monitor, report, and rapidly respond to systemic vulnerabilities and disruptions to international stability (subfunctions 13, 14 in Table 1).

        \,
        
        For example, a techno-prudential model, akin to the macro-prudential framework used to increase resilience in central banking and bringing together those developed at the national level, may help to similarly insulate against AI risks to global stability. Such a model must be grounded in human rights principles.
        
        \,
        
        Reporting frameworks can be inspired by existing practices of the IAEA for mutual reassurance on nuclear safety and nuclear security, as well as the WHO on disease surveillance.'' \cite[para. 70--72]{UN_AI_Report}
        \\
        
        \midrule
    \end{tabularx}
    \caption[]{Policy Initiatives for General-Purpose AI Incident Reporting, Proposed (Cont'd)}
\end{table*}

\setcounter{table}{21}
\begin{table*}[p]
    \centering
    \begin{tabularx}{\textwidth}{ s s s X } 
        \toprule
        
        \textbf{Jurisdiction} & 
        \textbf{Initiative} & 
        \textbf{Type} &
        \textbf{Text} 
        \\ 
        
        \midrule

        United States &
        ``Artificial Intelligence Accountability Report'' from the National Telecommunications and Information Administration &
        Policy Paper &
        ``The federal government should strengthen its capacity to address cross-sectoral risks and practices related to AI. . . . Such cross-sectoral horizontal capacity, wherever housed, would be useful for creating accountability inputs such as . . . A national AI adverse incidents reporting database and platform for receiving reports[.]'' \cite{goodman_ai_2024}
        \\

        \midrule

        United States &
        Bipartisan Framework for US AI Act from US Senators Blumenthal and Hawley &
        Policy Framework &
        ``The new oversight body should establish a public database and reporting so that consumers and researchers have easy access to A.I. model and system information, including when significant adverse incidents occur or failures in A.I. cause harms.'' \cite{blumenthal_bipartisan_2023} 
        \\

        \midrule

        United States &
        ``Driving US Innovation in Artificial Intelligence'' from the Bipartisan Senate AI Working Group &
        Policy Paper & 
        ``The AI Working Group also encourages the relevant committees to . . . Explore whether there is a need for an AI-focused Information Sharing and Analysis Center (ISAC) to serve as an interface between commercial AI entities and the federal government to support monitoring of AI risks.'' \cite{schumer_driving_2024}
        \\

        \midrule

        United States &
        ``Managing Misuse Risk for Dual-Use Foundation Models'' draft from the US AI Safety Institute &
        Voluntary Guidelines & 

        ``Practice 7.3: Report incidents and hazards related to the foundation model to AI incident databases.

        \,
        
        Recommendations to effectively implement Practice 7.3:

        \,
        
        1. Based on existing best practices and adequate review of the benefits and risks of disclosing certain information, define the category of misuse events to report.

        \,
        
        2. Collate verified reports of misuse in a commonly used and machine-readable format.

        \,
        
        3. Share verified reports of misuse with relevant third parties, such as AI incident databases.
        
        \,
        
        The following documentation can help provide transparency about how Practice 7.3 is
        implemented:
        
        \,
        
        a. A description of the incident reporting process, including an example of a type of incident that might be reported and to whom it would be reported.'' \cite[pp. 17]{us_ai_safety_institute_nist_2024}
        \\

        \midrule

        United States &
        Secure A.I. Act of 2024 &
        Legislation &
        ``SEC. 3. Voluntary tracking and processing of security and safety incidents and risks associated with artificial intelligence.

        \,

        (a) Processes and procedures for vulnerability management.---Not later than 180 days after the date of the enactment of this Act, the Director of the National Institute of Standards and Technology shall---'' (cont'd)
        \\
        
        \midrule
    \end{tabularx}
    \caption[]{Policy Initiatives for General-Purpose AI Incident Reporting, Proposed (Cont'd)}
\end{table*}

\setcounter{table}{21}
\begin{table*}[p]
    \centering
    \begin{tabularx}{\textwidth}{ s s s X } 
        \toprule
        
        \textbf{Jurisdiction} & 
        \textbf{Initiative} & 
        \textbf{Type} &
        \textbf{Text} 
        \\ 

        \midrule

        &
        &
        &
        (cont'd) ``(1) initiate a process to update processes and procedures associated with the National Vulnerability Database of the Institute to ensure that the database and associated vulnerability management processes incorporate artificial intelligence security vulnerabilities to the greatest extent practicable; and

        \,
        
        (2) identify any characteristics of artificial intelligence security vulnerabilities that make utilization of the National Vulnerability Database inappropriate for their management and develop processes and procedures for vulnerability management for those vulnerabilities.

        \,
        
        (b) Voluntary tracking of artificial intelligence security and artificial intelligence safety incidents.---
        
        \,
        
        (1) VOLUNTARY DATABASE REQUIRED.---Not later than 1 year after the date of the enactment of this Act, the Director of the Institute, in coordination with the Director of the Cybersecurity and Infrastructure Security Agency, shall---
    
        \,
        
        (A) develop and establish a comprehensive, voluntary database to publicly track artificial intelligence security and artificial intelligence safety incidents; and
    
        \,
            
        (B) in establishing the database under subparagraph (A)---
    
        \,
        
        (i) establish mechanisms by which private sector entities, public sector organizations, civil society groups, and academic researchers may voluntarily share information with the Institute on confirmed or suspected artificial intelligence security or artificial intelligence safety incidents, in a manner that preserves confidentiality of any affected party;
    
        \,
        
        (ii) leverage, to the greatest extent possible, standardized disclosure and incident description formats;
    
        \,
            
        (iii) develop processes to associate reports pertaining to the same incident with a single incident identifier;
    
        \,
        
        (iv) establish classification, information retrieval, and reporting mechanisms that sufficiently differentiate between artificial intelligence security incidents and artificial intelligence safety incidents; and
    
        \,
        
        (v) create appropriate taxonomies to classify incidents based on relevant characteristics, impact, or other relevant criteria.'' (cont'd)
        
        \\
        
        \midrule
    \end{tabularx}
    \caption[]{Policy Initiatives for General-Purpose AI Incident Reporting, Proposed (Cont'd)}
\end{table*}

\setcounter{table}{21}
\begin{table*}[p]
    \centering
    \begin{tabularx}{\textwidth}{ s s s X } 
        \toprule
        
        \textbf{Jurisdiction} & 
        \textbf{Initiative} & 
        \textbf{Type} &
        \textbf{Text} 
        \\ 

        \midrule

        &
        &
        &
        (cont'd) ``(2) IDENTIFICATION AND TREATMENT OF MATERIAL ARTIFICIAL INTELLIGENCE SECURITY OR ARTIFICIAL INTELLIGENCE SAFETY RISKS.---
    
        \,
        
        (A) IN GENERAL.---Upon receipt of relevant information on an artificial intelligence security or artificial intelligence safety incident, the Director of the Institute shall determine whether the described incident presents a material artificial intelligence security or artificial intelligence safety risk sufficient for inclusion in the database developed and established under paragraph (1).
    
        \,
        
        (B) PRIORITIES.---In evaluating a reported incident pursuant to paragraph (1), the Director shall prioritize inclusion in the database cases in which a described incident---
    
        \,
        
        (i) describes an artificial intelligence system used in critical infrastructure or safety-critical systems;

        \,
        
        (ii) would result in a high-severity or catastrophic impact to the people or economy of the United States; or

        \,
        
        (iii) includes an artificial intelligence system widely used in commercial or public sector contexts.

        \,
        
        (3) REPORTS AND ANONYMITY.---The Director shall populate the voluntary database developed and established under paragraph (1) with incidents based on public reports and information shared using the mechanism established pursuant to subparagraph (B)(i) of such paragraph, ensuring that any incident description sufficiently anonymizes those affected, unless those who are affected have consented to their names being included in the database.'' \cite[Sec. 3]{warner_s_2024}
        \\
        
        \midrule
    \end{tabularx}
    \caption[]{Policy Initiatives for General-Purpose AI Incident Reporting, Proposed (Cont'd)}
\end{table*}

\end{document}